\documentclass[usenatbib,fleqn]{mnras}

\usepackage[T1]{fontenc}
\usepackage{ae,aecompl}
\usepackage{graphicx, amssymb}
\usepackage[fleqn]{amsmath}
\usepackage{epsfig}
\usepackage{color}
\usepackage{times}

\title[Radiation pressure in super star cluster formation]
{Radiation pressure in super star cluster formation}

\author[B.~T.-H. Tsang \& M. Milosavljevi\'{c}]{Benny T.-H. Tsang and Milo\v{s} Milosavljevi\'{c}\\
Department of Astronomy, University of Texas at Austin, Austin, TX 78712, USA}

\begin{document}

\maketitle
\topmargin-1cm

\begin{abstract}
The physics of star formation at its extreme, in the nuclei of the densest and the most massive star clusters in the universe---potential massive black hole nurseries---has for decades eluded scrutiny.  Spectroscopy of these systems has been scarce, whereas theoretical arguments suggest that radiation pressure on dust grains somehow inhibits star formation.
Here, we harness an accelerated Monte Carlo radiation transport scheme to report a
radiation hydrodynamical simulation of super star cluster formation in
turbulent clouds. We find that
radiation pressure reduces the global star formation efficiency
by 30--35\%, and the star formation rate by 15--50\%, both relative to a
radiation-free control run.
Overall, radiation pressure does not
terminate the gas supply for star formation and the final stellar mass of the most massive cluster is $\sim 1.3\times10^6\,M_\odot$.
The limited impact as compared to in idealized theoretical models is attributed to a radiation-matter anti-correlation in the supersonically turbulent, gravitationally collapsing medium.
In isolated regions outside massive clusters,
where the gas distribution is less disturbed, radiation pressure
is more effective in limiting star formation.
The resulting stellar density at the cluster core is
$\ge 10^{8}$\,$M_{\odot}$\,pc$^{-3}$,
with stellar velocity dispersion $\gtrsim 70\,\text{km}\,\text{s}^{-1}$.
We conclude that the super star cluster nucleus is propitious to the formation of very massive stars
via dynamical core collapse and stellar merging. We speculate that the very massive star may avoid the claimed catastrophic mass loss by continuing to accrete dense gas condensing from a gravitationally-confined ionized phase.
\newline

\end{abstract}


\section{Introduction}

Very massive star clusters, also known as super star clusters (SSCs),
have extreme stellar mass and star formation rate (SFR)  densities.
They are found in dusty star-forming galaxies at high redshifts
\citep{Blain02,Schinnerer08,Casey17}
and in local starbursts and interacting galaxies
\citep{Johnson03,Whitmore10,Herrera17}.
The SFRs in single clusters that are mere parsecs across can be as high as
$\sim10^3\,M_{\odot}$\,yr$^{-1}$ and the final cluster masses as high as $\sim 10^{8}\,M_{\odot}$
\citep{PortegiesZwart10,Maraston04,Pollack07}.
Since the nearest SSC-forming systems are distant and the immediate formation sites are heavily obscured by dust,
direct astronomical constraints on SSC formation
are scarce.
Observations have been historically limited to
optical and near-infrared wavelengths at which stars emit after most
of the natal gas clouds has been removed
\citep{Bastian14,Hollyhead15}.
The early embedded state requires longer wavelengths.
Free-free radio emission from the embedded ultra-dense H\,II regions shows
that these dense stellar systems are born at much higher pressures than typical
in the Galactic interstellar medium
\citep[ISM;][]{Elmegreen97,Kobulnicky99,AZ01}.
The high angular resolution of ALMA also enables the mapping of dense
cluster-forming gas in the molecular lines
\citep{Herrera11,Herrera12,Johnson15,Ginsburg17,Turner17}.

The lifetimes of massive O and B stars are longer
than the dynamical time of the densest molecular clouds from which SSCs form. The pre-supernova energy and momentum input from stars, and in particular the radiation pressure, have been claimed to be an important factor influencing massive cluster formation
\citep{KM09,Fall10,MQT10,AT11,Kim16}. The radiation pressure force could potentially slow down and even truncate star formation in a newborn cluster. However, the quoted studies attributing dynamical importance to radiation pressure modeled star forming systems with idealized geometries.
Here, we present a realistic simulation that allows us to assess the radiation's importance in SSC formation.  We follow in the footsteps of \citet[hereafter SO15]{SO15} who, to our knowledge, carried out the first radiation-hydrodynamical simulations of massive cluster formation from turbulent initial conditions.

Computational radiation hydrodynamics (RHD) of star forming molecular clouds is technically challenging. Recent proof-of-concept simulations have highlighted sensitivities to numerical discretizations and closure schemes.
The first to be treated had idealized setups in which radiation injected
at the base of a computational box accelerated gas against gravity. \citet{KT12,KT13} and \citet{RT15} applied the flux-limited diffusion
(FLD) approximation and the M1 closure, respectively, and found
suppression of momentum transfer from radiation to gas after radiation-driven Rayleigh-Taylor instability (RTI) has fragmented the layer into clumps.
\citet{Davis14}, \citet{ZD17}, and \citet[hereafter Paper I]{TM15}, applied the variable Eddington tensor (VET) and implicit Monte Carlo (IMC) methods to find weaker suppression of momentum transfer and argued that the FLD and M1 closures
underestimated radiation-matter coupling in clumpy media.
Whether RTIs develop in massive star formation itself seems to depend on
the implementation of radiating sources and the spatial resolution
\citep{Krumholz09,Kuiper11,Kuiper12,Rosen16,Harries17}.

% Revised.

SO15 carried out RHD simulations of gravitational collapse in turbulent molecular clouds.  Injecting the dust-reprocessed infrared (IR)
radiation and transporting it in the M1 closure, they found that for radiation pressure to significantly reduce the star formation efficiency (SFE) in turbulent gas clouds,
the dust opacity must be rather high
($\kappa_{\rm IR} \sim$15\,cm$^{2}$\,g$^{-1}$).
\citet{ROS16} targeted a regime with lower density in which the direct ultraviolet
(UV) radiation pressure is expected to dominate over the IR.
In the presence of radiation pressure the SFE increases with
gas surface density and decreases with the virial parameter.
In these 3D RHD simulations the effect of radiation is evident. The simulations show that the coupling of radiation to turbulent
gas is quantitatively different than in spherically symmetric models.

%Similar to SO15, they found that
%the degree of turbulent compression of the gas, rather than the non-ionizing
%radiation pressure, is the principal parameter influencing the SFE.

In Paper I we demonstrated the
accuracy of the IMC radiation transport scheme in strongly compressible RHD simulations.
In local thermodynamical equilibrium (LTE), the net exchange of energy between
radiation and gas is small. IMC replaces a fraction of absorption and
immediate re-emission with an \emph{effective scattering} process.
The reduction in radiation-matter coupling permits longer radiation transport
time steps. The IMC scheme was originally introduced by \citet{FC71},
and was recently generalized by \citet{Abdikamalov12}.
The IMC method has been extended to Lagrangian meshes and multi-frequency transport as required for
transporting photons in thermonuclear supernovae \citep{Wollaeger13} and has been improved with
variance-reduction estimators for the radiation source terms \citep{Roth15}.

The IMC method differs from the methods based on closures of the moment equations. It is a particle-based scheme that discretizes the full angle-dependent radiation field with Monte Carlo particles (MCPs) and then transports the MCPs so as to approximate solutions of the radiation transfer equation.
Each MCP represents a collection of photons sharing a common position,
velocity, and frequency. In the limit of an increasing MCP number, IMC converges to the exact solution of the radiation transport equation on the finite volume mesh. However the scheme is intrinsically noisy. Historically, noise reduction with brute MCP numbers has been deemed prohibitively computationally expensive.  We introduce practical improvements that make the Monte Carlo approach to astrophysical RHD competitive in its computational cost and accuracy with moment-based methods.

At high gas densities, the photon mean free path can be extremely short. IMC's effective scatterings are local and rapidly isotropizing, a clear overkill.
The inefficiency of IMC in this regime can be alleviated by selectively applying
the diffusion approximation to the MCPs themselves \citep{FC84,Gentile01,Densmore07}.
The \emph{modified random walk} approach of \citet{FC84} replaces
multiple effective scatterings with a single spatial excursion within the same finite volume cell.  The excursion is sampled consistent with an approximate solution of the radiation diffusion closure.
We find that in the finite volume context, because of the breakdown of the diffusion approximation in the vicinity of cell boundaries, this approach provides only a limited speedup in
three-dimensional simulations.
An improved \emph{partially gray random work} (PGRW) algorithm by \citet{KC17} selectively accelerates the
transport of optically-thick MCPs when their individual opacities are strongly dependent on frequency.
This has offered speedups of 2--4 over the standard modified random walk.

The discrete diffusion Monte Carlo (DDMC) method of \citet{Densmore07,Densmore12}, on the other hand,
is extremely effective in speeding up the transport of MCPs
without compromising accuracy.
DDMC approximately solves the diffusion equation by implementing  discrete MCP jumps across cells.
Each jump replaces many effective scatterings but the MCP retains a physically self-consistent time variable.
The continuous time treatment in DDMC allows easy hybridization with
the IMC that is applied in less optically thick cells. \citet{Abdikamalov12} have generalized the gray, static matter DDMC scheme of
\citet{Densmore07} to frequency-dependent transport in moving matter.
While the PGRW algorithm is slower than DDMC, it can be more easily
adopted to unstructured meshes;
a hybridization between PGRW and DDMC may therefore be optimal for future
simulations \citep{KC17}.

Here, we keep the radiation transport gray and choose
an isothermal equation of state for the gas to demonstrate, in a controlled setting, the
capabilities of the Monte Carlo methods in research-scale simulations. The specific contributions of this paper are:
\begin{enumerate}
\item Demonstration of DDMC's superior speedup of radiation transport at high optical depths.
\item Demonstration of
seamless hybridization with IMC on adaptively-refined finite volume meshes.
\item Demonstration of the reliability of radiation-to-gas momentum deposition in the IMC-DDMC hybrid.
\item Proof of the feasibility of simulating clustered star formation in turbulent clouds with the hybrid method.
\item A quantitative test of the claims that radiation pressure plays a decisive dynamical role in massive star cluster formation.
\item The first simulation of the formation of $\sim10^6\,M_\odot$ SSCs employing accurate RHD and resolving the cluster nuclei.
\end{enumerate}

This paper is organized as follows. In Section \ref{sec:nm} we review the IMC
formalism and provide the mathematical basis and implementation
strategies for the DDMC scheme.
In Section \ref{sec:tests} we assess the accuracy of the hybrid
IMC-DDMC scheme and demonstrate its
efficiency gain.
In Section \ref{sec:sim_setup} we describe the massive-cluster-forming setup and the details of our implementation.
In Section \ref{sec:results} we present the results and
in Section \ref{sec:prior_art} we compare our results to prior art and discuss observational implications.
Finally, in Section
\ref{sec:conclusions} we summarize our conclusions.

\section{Monte Carlo radiation transport schemes}
\label{sec:nm}

In Paper I, we implemented the IMC radiation transport
scheme in the hydrodynamic code \textsc{flash} \citep{Fryxell00},
and applied it to RHD simulations of radiative driving of gas clouds
in a gravitational field.
There, the optical depth of absorption across a single cell was
$\tau \lesssim 10$, a regime in which the basic Monte Carlo procedure is
effective.
However, in high-density regions where massive star clusters form,
the optical thickness to IR radiation from dust can reach
$\tau \gtrsim 100 - 1000$.
Tracking MCPs with short mean free paths is extremely inefficient.
In this section, we summarize the DDMC extension of the IMC radiation transfer
equation and discuss the associated Monte Carlo solution procedure.
Our presentation closely follows \citet{Densmore07} and the
energy-dependent generalization in \citet{Abdikamalov12}.
We note that we have concurrently developed a generalization of the DDMC
scheme for resonant line transfer \citep{Smith17b}.

We adopt the gray approximation and an isothermal
equation of state for the gas.
These assumptions will further simplify the DDMC equation.
The radiation source terms are accumulated during
the Monte Carlo transfer step and are coupled to gas dynamics in an
operator-split  manner (cf.\ Steps ii{\it a} and ii{\it b} of Paper I).

\subsection{Implicit Monte Carlo (IMC)}

The radiation transfer equation for elastic and isotropic physical scattering in LTE reads
\begin{eqnarray}
  \label{eqn:RTE-original}
  \frac{1}{c} \frac{\partial I(\epsilon,\mathbf{n})}{\partial t} +
  \mathbf{n} \cdot \nabla I(\epsilon,\mathbf{n}) &=&
  k_{\rm a}(\epsilon) B(\epsilon) + j_{\rm s}(\epsilon) \nonumber \\
  &-& [k_{\rm a}(\epsilon) + k_{\rm s}(\epsilon, \mathbf{n})]
  I(\epsilon,\mathbf{n}) ,
\end{eqnarray}
where $I(\epsilon, \mathbf{n})$ is the specific intensity,
$\epsilon$ and $\mathbf{n}$ are the photon energy and direction of
propagation,
$c$ is the speed of light,
$k_{\rm a}$ and $k_{\rm s}$ are the absorption and scattering
coefficients, $B$ is the Planck function, and
$j_{\rm s}$ is the emission coefficient for physical scattering $j_{\rm s} (\epsilon) = \int  k_{\rm s}(\epsilon, \mathbf{n})\,I(\epsilon,\mathbf{n})d\Omega = \int (4\pi)^{-1} k_{\rm s}(\epsilon)\,I(\epsilon,\mathbf{n}) d\Omega= k_{\rm s}(\epsilon) J(\epsilon)$.
Here, the factor of $(4\pi)^{-1}$ arises from the per-solid-angle normalization of
the isotropic scattering kernel and $J(\epsilon) = (4\pi)^{-1} \int I(\epsilon,\mathbf{n}) d\Omega$ is the mean intensity.

The IMC scheme integrates the transfer equation over a
time step $\Delta t = t^{n+1} - t^{n}$ in the following alternative form given in Equation (29) of Paper I
\begin{align}
  \label{eqn:RTE-IMC}
&  \frac{1}{c} \frac{\partial I(\epsilon,\mathbf{n})}{\partial t} +
    \mathbf{n} \cdot \nabla I(\epsilon,\mathbf{n})  =\notag\\
     &\ \ \ \ \ \ \ \ \ \ \ \ \ \  \tilde{k}_{\rm ea}(\epsilon) \tilde{b}  (\epsilon) c \tilde{u}_{\rm r}  - [\tilde{k}_{\rm ea}(\epsilon) + \tilde{k}_{\rm es}(\epsilon) + \tilde{k}_{\rm s}(\epsilon)] I(\epsilon,\mathbf{n}) \notag \\
     &\ \ \ \ \ \ \ \ \ + 4\pi \frac{\tilde{k}_{\rm a}(\epsilon) \tilde{b}(\epsilon) }{\tilde{k}_{\rm p}}
          \int_{0}^{\infty} \tilde{k}_{\rm es} (\epsilon') J(\epsilon') d\epsilon' + k_{\rm s}(\epsilon) J(\epsilon).
\end{align}
The tildes denote the corresponding time-centered quantities within the time step.
In practice, an explicit choice of evaluating the quantities at the initial time $t^{n}$ is made.
A few auxiliary variables have been introduced in the derivation:
the normalized Planck function $b(\epsilon)=(4\pi)^{-1}B(
\epsilon)/\int_0^\infty B(\epsilon) d\epsilon$,
the effective radiation energy density if radiation and gas are at equilibrium
$u_{r}$,
the Planck mean absorption opacity $k_{\rm P}$,
the effective absorption and scattering coefficients $\tilde{k}_{\rm ea}(\epsilon) = f \,\tilde{k}_{\rm a} (\epsilon)$ and $\tilde{k}_{\rm es}(\epsilon) = (1 - f) \,\tilde{k}_{\rm a} (\epsilon)$,
where $f= (1+\Delta t \tilde{\beta} c \tilde{k}_{p})^{-1}$ is the Fleck factor
and the radiation-gas coupling nonlinearity factor $\beta=\partial u_\mathrm{r}/\partial u_\mathrm{g}$,
where $u_\mathrm{g}$ is the internal energy density of the gas.
We encourage the interested reader to consult Paper I for further detail.

\subsection{Discrete diffusion Monte Carlo (DDMC)}

At high densities, the $k_{\rm P}$ term is large and the
Fleck factor approaches zero.
The effective scattering term then dominates and
the extremely short mean-free paths make the MCP flights extremely
costly to follow. This is precisely the radiative diffusion regime. DDMC adaptively identifies this regime and accelerates transport by
recognizing the diffusive nature of random walks.

To derive a diffusion scheme from the IMC scheme, we first define
the flux $\mathbf{F}(\epsilon) = \int I(\epsilon,\mathbf{n}) \, \mathbf{n}\, d\Omega$.
In the diffusion regime, radiation is isotropized by scattering and one considers the isotropic moment of the
IMC radiation transfer equation (Equation \ref{eqn:RTE-IMC})
\begin{align}
  \label{eqn:ang_int_RTE}
&  \frac{1}{c} \frac{\partial J(\epsilon)}{\partial t} +
    \nabla \cdot \mathbf{F}(\epsilon)  = \tilde{k}_{\rm ea}(\epsilon) \tilde{b}  (\epsilon) c \tilde{u}_{\rm r}  - [\tilde{k}_{\rm ea}(\epsilon)
                         + \tilde{k}_{\rm es}(\epsilon)] J(\epsilon) \notag \\
     &\ \ \ \ \ \ \ \ \ + 4 \pi \frac{\tilde{k}_{\rm a}(\epsilon) \tilde{b}(\epsilon) }{\tilde{k}_{\rm p}}
          \int_{0}^{\infty}\,\tilde{k}_{\rm es} (\epsilon') J(\epsilon') d\epsilon',
\end{align}
where the source and sink terms for isotropic scattering cancel.
Indeed the gain in computational efficiency is the most
profound if an isotropic and elastic physical scattering is present.
In this case, as we shall see, the spatial excursion of photons due to physical scattering
is treated together with effective scattering via the introduction of a `transport opacity'
in Fick's law.

For the remainder of this section, we present formalism for transport in
the $x$-direction. The transport in the $y$- and $z$-directions
 is equivalent.
Similar to the IMC scheme, we discretize the spatial domain
into cells indexed by $j$ that have boundaries at $[x_{j-1/2}, x_{j+1/2}]$.
To specify our IMC-DDMC interfacing scheme, without a loss of generality we assume that the cells $j\geq 2$ are DDMC \emph{interior cells} where DDMC applies. The cell $j = 1$ is a DDMC \emph{interface cell} that is adjacent to IMC cells with $j\leq 0$.
Due to the slight difference in formulation, we will discuss the interior and interface cells separately below.

\subsubsection{Interior cells}
To derive an equation that enables fast transport in the diffusive regime,
we finite-difference Equation (\ref{eqn:ang_int_RTE}) over space,
\begin{align}
  \label{eqn:fd_RTE}
&  \frac{1}{c} \frac{\partial J_{j}(\epsilon)}{\partial t} +
    \frac{F_{j+1/2}(\epsilon) - F_{j-1/2}(\epsilon)}{\Delta x_{j}}  =\notag\\
     &\ \ \ \ \ \ \ \ \ \ \ \ \ \  \tilde{k}_{{\rm ea},j} \tilde{b}_{j} c \tilde{u}_{\rm r} - [\tilde{k}_{{\rm ea},j}(\epsilon)
                         + \tilde{k}_{{\rm es},j}(\epsilon)] J_{j}(\epsilon) \notag \\
     &\ \ \ \ \ \ \ \ \ + 4 \pi \frac{\tilde{k}_{{\rm a}, j}(\epsilon) \tilde{b}_{j}(\epsilon) }{\tilde{k}_{{\rm p},j}}
          \int_{0}^{\infty}\,\tilde{k}_{{\rm es},j}(\epsilon') J_{j}(\epsilon') d\epsilon',
\end{align}
where $J_{j}= \frac{1}{\Delta x_{j}} \int_{x_{j-1/2}}^{x_{j + 1/2}} J(x)dx$
and $F_{j \pm 1/2} = F(x=x_{j \pm 1/2})$ are the radiative fluxes
across cell boundaries.
The subscript $j$ refers to the corresponding cell-centered quantities.
The mean intensity $J_{j}(\epsilon)$
can be more intuitively understood as the radiation energy density
$u_{{\rm rad},j}(\epsilon) = 4\pi J_{j}(\epsilon)/c$ in cell $j$.

To complete the numerical scheme,
we express $F_{j \pm 1/2}$ in terms of certain known quantities and
the mean intensity.
By finite-differencing Fick's law of radiative diffusion:
\begin{align}
  F_{j \pm 1/2} = \frac{1}{3 k_{\rm T}} \left.\frac{\partial J}{\partial x}
                      \right|_{x=x_{j \pm 1/2}},
\end{align}
where $k_{\rm T} = k_{\rm a} + k_{\rm s}$ is the `transport opacity',
we obtain an approximate flux at face $j+1/2$ of cell $j$
\begin{align}
  \label{eqn:F_1}
  F_{j+1/2} = -\frac{1}{3\tilde{k}^{-}_{{\rm T}, j+1/2}}
                   \frac{J_{j+1/2} - J_{j}}{\Delta x_{j}/2}.
\end{align}
The flux at the same face can be obtained similarly by considering cell $j+1$
\begin{align}
  \label{eqn:F_2}
  F_{j+1/2} = -\frac{1}{3\tilde{k}^{+}_{{\rm T}, j+1/2}}
                   \frac{J_{j+1} - J_{ j+1/2}}{\Delta x_{j+1}/2} .
\end{align}
Here the superscripts $+$ and $-$ pertain to the immediate right and left,
respectively, of the cell boundary $j+1/2$.
From Equations (\ref{eqn:F_1}) and (\ref{eqn:F_2})
we can solve for the mean intensity at the interface,
\begin{align}
  \label{eqn:J_face}
  J_{j+1/2} = \frac{\tilde{k}^{+}_{{\rm T}, j+1/2} \Delta x_{j+1} J_{j}
                       +\tilde{k}^{-}_{{\rm T}, j+1/2} \Delta x_{j} J_{j+1}}
                       {\tilde{k}^{-}_{{\rm T}, j+1/2} \Delta x_{j}
                        + \tilde{k}^{+}_{{\rm T}, j+1/2} \Delta x_{j+1}} .
\end{align}
By substituting this into Equation (\ref{eqn:F_1}) or (\ref{eqn:F_2}) we finally have
\begin{align}
  \label{eqn:F_face}
  F_{j+1/2} = - \frac{2}{3}
                    \frac{J_{j+1} - J_{j}}
                    {\tilde{k}^{-}_{{\rm T}, j+1/2} \Delta x_{j}
                     + \tilde{k}^{+}_{{\rm T}, j+1/2} \Delta x_{j+1}}.
\end{align}
Substituting Equation (\ref{eqn:F_face}) into (\ref{eqn:fd_RTE})
we obtain the DDMC version of the radiation transfer equation
\begin{align}
  \label{eqn:DDMC_RTE}
&  \frac{1}{c} \frac{\partial J_{j}(\epsilon)}{\partial t} =
     \tilde{k}_{{\rm ea},j}(\epsilon) \tilde{b}_{j}(\epsilon) c \tilde{u}_{\rm r} \notag \\
     &\ \ \ \ \ \ \ \  -[\tilde{k}_{{\rm ea},j}(\epsilon) + \tilde{k}_{{\rm es},j}(\epsilon)
                             + \tilde{k}_{{\rm L},j}(\epsilon) + \tilde{k}_{{\rm R},j}(\epsilon) ] J_{j}(\epsilon) \notag \\
     &\ \ \ \ \ \ \ \ + \left( \frac{\Delta x_{j+1}}{\Delta x_{j}} \right)
                            \tilde{k}_{{\rm L},j+1}(\epsilon) J_{j+1}(\epsilon)\notag\\
         &\ \ \ \ \ \ \ \  +\left( \frac{\Delta x_{j-1}}{\Delta x_{j}} \right)
                              \tilde{k}_{{\rm R},j-1}(\epsilon) J_{j-1}(\epsilon) \notag \\
     &\ \ \ \ \ \ \ \ \ + 4 \pi \frac{\tilde{k}_{{\rm a}, j}(\epsilon) \tilde{b}_{j}(\epsilon)} {\tilde{k}_{{\rm p},j}}
          \int_{0}^{\infty}\,\tilde{k}_{{\rm es},j}(\epsilon') J_{j}(\epsilon') d\epsilon'.
\end{align}
Here we have introduced the left and right \emph{leakage opacity} that are crucial for the DDMC speedup,
\begin{align}
  \label{eqn:k_L}
  \tilde{k}_{{\rm L},j} = \frac{2}{3\Delta x_{j}}
                        \frac{1}{\tilde{k}^{+}_{{\rm T}, j-1/2} \Delta x_{j}
                        + \tilde{k}^{-}_{{\rm T}, j-1/2} \Delta x_{j-1}},
\end{align}
and
\begin{align}
  \label{eqn:k_R}
  \tilde{k}_{{\rm R},j} = \frac{2}{3\Delta x_{j}}
                        \frac{1}{\tilde{k}^{-}_{{\rm T}, j+1/2} \Delta x_{j}
                        + \tilde{k}^{+}_{{\rm T}, j+1/2} \Delta x_{j+1}}.
\end{align}
The leakage opacities quantify the radiation's diffusive `shortcuts'
to adjacent cells.
At high optical densities this shortcut  replaces a large
number of scatterings. This is is how DDMC accelerates the transport scheme.
The leakage opacities vary inversely with the transport opacity
$\propto k^{-1}_{\rm T}$, therefore the leakage `mean free path'
is longer at higher optical densities.

\citet{Abdikamalov12} were the first to generalize the DDMC scheme
into an energy-dependent transport scheme.
Equation (\ref{eqn:DDMC_RTE}) is discretized into
 separate energy \emph{groups} corresponding to intervals of photon energy,
\begin{align}
  \label{eqn:DDMC_mg}
&  \frac{1}{c} \frac{\partial J_{j,k}}{\partial t} =
     \tilde{k}_{{\rm ea},j,k} \tilde{b}_{j,k} c \tilde{u}_{\rm r} \notag \\
     &\ \ \ \ \ \ \ \  - [\tilde{k}_{{\rm ea},j,k} + \tilde{k}_{{\rm es},j,k}
                             + \tilde{k}_{{\rm L},j,k} + \tilde{k}_{{\rm R},j,k} ] J_{j,k} \notag \\
     &\ \ \ \ \ \ \ \  + \left( \frac{\Delta x_{ j+1}}{\Delta x_{j}} \right)
                              \tilde{k}_{{\rm L},j+1,k} J_{j+1,k}
                             +\left( \frac{\Delta x_{j-1}}{\Delta x_{j}} \right)
                              \tilde{k}_{{\rm R},j-1,k} J_{j-1,k} \notag \\
     &\ \ \ \ \ \ \ \ \ + 4 \pi \frac{\tilde{k}_{{\rm a}, j,k} \tilde{b}_{j,k}} {\tilde{k}_{{\rm p},j}}
          \sum_{l}\,\tilde{k}_{{\rm es},j,l} J_{j,l} \Delta\epsilon_{l}.
\end{align}
We have replaced the energy parameter $\epsilon$ with an energy group subscript $k$.
The last term on the RHS is a sum over the effective scatterings from all groups into group $k$.

The latter form can be further simplified to invite a more transparent
Monte Carlo interpretation.
Extracting the $l=k$ term from the sum and combining it with the corresponding
sink term $\tilde{k}_{{\rm es},j,k} J_{j,k}$ yields
\begin{align}
  \label{eqn:DDMC_full}
&  \frac{1}{c} \frac{\partial J_{j,k}}{\partial t} =
     \tilde{k}_{{\rm ea},j,k} \tilde{b}_{j,k} c \tilde{u}_{\rm r}\notag \\
     &\ \ \ \ \ \ \ \  - \left[\tilde{k}_{{\rm ea},j,k} + \tilde{\sigma}_{{\rm es},j,k}
                             + \tilde{k}_{{\rm L},j,k} + \tilde{k}_{{\rm R},j,k} \right] J_{j,k} \notag \\
     &\ \ \ \ \ \ \ \  + \left( \frac{\Delta x_{j+1}}{\Delta x_{j}} \right)
                              \tilde{k}_{{\rm L},j+1,k} J_{j+1,k}
                             +\left( \frac{\Delta x_{j-1}}{\Delta x_{j}} \right)
                              \tilde{k}_{{\rm R},j-1,k} J_{j-1,k} \notag \\
     &\ \ \ \ \ \ \ \ \ + 4 \pi \frac{\tilde{k}_{{\rm a}, j,k} \tilde{b}_{j,k}} {\tilde{k}_{{\rm p},j}}
          \sum_{l\neq k}\,\tilde{k}_{{\rm es},j,l} J_{j,l} \Delta\epsilon_{l} ,
\end{align}
where the revised effective scattering coefficient becomes
\begin{align}
  \label{eqn:sigma_es}
  \tilde{\sigma}_{{\rm es},j,k} = \left( 1 - \frac{\tilde{k}_{{\rm a},j,k} \tilde{B}_{j,k} \Delta \epsilon_{k}}{\int_{0}^{\infty}\tilde{k}_{{\rm a}, j}(\epsilon) \tilde{B}_{j}(\epsilon) d\epsilon} \right) \tilde{k}_{{\rm es},j,k} .
\end{align}
From the last definition it is clear that
$\tilde{\sigma}_{{\rm es},j,k} < \tilde{k}_{{\rm es},j,k}$.
This trick removes effective scattering events without energy group changes
and further improves the efficiency of DDMC.
This property is unique to DDMC---the corresponding IMC equation
(Equation \ref{eqn:RTE-IMC}) retains angular dependence of the
specific intensity which prohibits such simplification.

Having reviewed the multigroup formulation of DDMC,
we consider a gray setup.
By restricting to one energy group, the term given in Equation (\ref{eqn:sigma_es}) vanishes.
The effective
scattering terms therefore cancel each other out.
If we further assume an isothermal equation of state, because $u_\text{g}$ is constant, we have $\beta\rightarrow\infty$, and thus
the Fleck factor vanishes.
This simplifies the DDMC radiation transfer equation
\begin{align}
  \label{eqn:DDMC_gray_iso}
&  \frac{1}{c} \frac{\partial J_{j}}{\partial t} = - [\tilde{k}_{{\rm L},j} + \tilde{k}_{{\rm R},j} ] J_{j} \notag \\
     &\ \ \ \ \ \ \ \ \  + \left( \frac{\Delta x_{j+1}}{\Delta x_{j}} \right)
                              \tilde{k}_{{\rm L},j+1} J_{j+1}
                             +\left( \frac{\Delta x_{j-1}}{\Delta x_{j}} \right)
                              \tilde{k}_{{\rm R},j-1} J_{j-1} .
\end{align}

\subsubsection{The Monte Carlo scheme}
This DDMC transfer equation, Equation (\ref{eqn:DDMC_gray_iso}), is solved approximately with a Monte Carlo
scheme. Recall that $J_{j}$ is proportional to the radiation energy density in cell $j$.
Equation (\ref{eqn:DDMC_gray_iso}) therefore governs the decrease in
MCP number in cell $j$ due to the leakage into neighboring cells
(the first term on the RHS)
and the gain in MCPs from neighboring cells (the remaining terms on the RHS).
The Monte Carlo solution procedure for DDMC differs from that for IMC in
at least two ways.
The DDMC does not keep track of the MCP location inside
a cell but only of the host cell's identity. The spatial position of an MCP inside a cell is re-sampled uniformly once it has
entered a DDMC region.
Also, the leakage terms are new in DDMC.

During a hydrodynamic time step $\Delta t=t^{n+1}- t^{n}$, an MCP in the DDMC region always keeps its current time $t_{i}$.
It can either remain in the cell
or undergo leakage events to transfer to neighboring cells.
We transport each MCP in a DDMC region as follows:
\begin{enumerate}
  \item Calculate the free-streaming distance until the end of the time step
        $d_{\rm t} = c (t^{n+1} - t_{i})$.
  \item Calculate the distance to the next leakage event $d_{\rm L}$
        using the total leakage opacity
        $k_{\rm tot} = k_{\rm L} + k_{\rm R}$
        of the cell\footnote{In the general multi-group, non-isothermal case, the `distance to collision'
        is computed instead. The total leakage opacity is then replaced
        by the total opacity in the first term on the RHS of
        Equation (\ref{eqn:DDMC_full}), which also contains contributions
        from effective absorption and effective scattering.}
\begin{align}
  d_{\rm L} = -\frac{\ln \xi}{k_{\rm tot}}
            = -\frac{\ln \xi}{k_{\rm L} + k_{\rm R}},
\end{align}
        where $k_{\rm L} = k_{{\rm L},x}+k_{{\rm L},y}+k_{{\rm L},z}$ and
              $k_{\rm R} = k_{{\rm R},x}+k_{{\rm R},y}+k_{{\rm R},z}$ are
        the total three-dimensional left and right leakage opacities.
  \item Execute the event that corresponds to the minimum of the distances
  $d_{\rm min} = \min(d_{\rm t}, d_{\rm L})$.
  If $d_{\rm t}$ is shorter, the MCP remains in the cell.
  If $d_{\rm L}$ is shorter, the leakage direction is sampled
  probabilistically consistent with the relative magnitudes of the leakage opacities.
  The three directions are treated independently.
  MCP is then moved to the neighboring cell in
 the direction of leakage.
  \item Update the current time of the MCP according to $t_{i} \rightarrow t_{i} + d_{\rm min}/c$
   or advance to
  the end of time step $t^{n+1}$, whichever comes first.
  \item Accumulate the momentum deposited onto the gas (the details of momentum deposition are provided in Section
        \ref{sec:md} below).
  \item Repeat the above operations until all MCPs have advanced to the end of the time
        step $t^{n+1}$.
\end{enumerate}
The above particle transport scheme approximately solves the radiative transfer
equation in the diffusion limit.
In Section \ref{sec:tests} below we present solutions of test problems that demonstrate high accuracy and extraordinary efficiency of this scheme.

\subsubsection{Interface cells}
In interface cells, the DDMC transport equation
takes a slightly different form.
For the cell $j = 1$
Equation (\ref{eqn:fd_RTE}) reads
\begin{align}
&  \frac{1}{c} \frac{\partial J_{\rm 1}(\epsilon)}{\partial t} =
                  \tilde{k}_{\rm ea,1} \tilde{b}_{\rm 1} c \tilde{u}_{\rm r} +
                  \frac{1}{\Delta x_{\rm 1}} F_{\rm 1/2}\notag \\
  &\ \ \ \ \ \ \ \ \ \ - [\tilde{k}_{\rm ea,1}(\epsilon)
                         + \tilde{k}_{\rm es,1}(\epsilon)
                         + \tilde{k}_{\rm R,1}(\epsilon) ] J_{\rm 1}(\epsilon)\notag \\
     &\ \ \ \ \ \ \ \ \ +  \left( \frac{\Delta x_{\rm 2}}{\Delta x_{\rm 1}} \right)
                           \tilde{k}_{{\rm L},2}(\epsilon) J_{\rm 2}(\epsilon)\notag \\
     &\ \ \ \ \ \ \ \ \ + 4 \pi \frac{\tilde{k}_{\rm a, 1}(\epsilon) \tilde{b}_{\rm 1}(\epsilon) }{\tilde{k}_{\rm p,1}}
          \int_{0}^{\infty}\,\tilde{k}_{\rm es,1}(\epsilon') J_{\rm 1}(\epsilon') d\epsilon'.
  \label{eqn:interface_RTE}
\end{align}
The only unknown besides $J_{\rm 1}$ is $F_{\rm 1/2}$,
the flux at the IMC-DDMC interface $x=x_{1/2}$.
Following \citet{Densmore07} and \citet{Abdikamalov12}, we adopt the
\emph{asymptotic diffusion limit} as the interface boundary condition, namely,
\begin{align}
  \label{eqn:adl}
  \int^{1}_{0}W(\mu) I_{\rm b}(\mu,t)d\mu = \left[J -
                  \frac{\lambda}{\tilde{k}^{+}_{{\rm T},1/2}}
                  \frac{\partial J}{\partial x}\right]_{x=x_{\rm 1/2}},
\end{align}
where $I_{\rm b}(\mu, t)$ is the radiation intensity incident on the boundary
at $x=x_{\rm 1/2}$,
$\mu$ is the cosine between the direction of photon propagation
and the IMC-DDMC interface normal vector pointing into the DDMC cell ($\hat{x}$ in the
current setup),
$W(\mu)$ is a transcendental function that can be approximated as $W(\mu) \simeq \mu + \frac{3}{2}\mu^2$,
and $\lambda \simeq 0.7104$ is a constant extrapolation distance
\citep{Habetler75}.
\citet{Densmore06} studied three interfacing methods and found that among them the
asymptotic interface is accurate regardless of the actual MCP directional distribution at the interface.

Finite-differencing the derivative term in
Equation (\ref{eqn:adl}) gives
\begin{align}
  J_{\rm 1/2} &= \frac{\tilde{k}^{+}_{\rm T,1/2} \Delta x_{\rm 1}}
                     {\tilde{k}^{+}_{\rm T,1/2} \Delta x_{\rm 1} + 2 \lambda}
                \int_{0}^{1} W\left( \mu \right) I_{\rm b}\left(\mu\right) d\mu \notag \\
                &+ \frac{2 \lambda}{\tilde{k}^{+}_{\rm T,1/2} \Delta x_{\rm 1} + 2 \lambda}
                J_{\rm 1}.
\end{align}
This can be solved for $F_{\rm 1/2}$ by eliminating $J_{\rm 1/2}$ in the above
equation with Equation (\ref{eqn:F_2}),
\begin{align}
  \label{eqn:f_interface}
  F_{\rm 1/2} &= -\frac{2}{3 \tilde{k}^{+}_{\rm T,1/2}\Delta x_{\rm 1} + 6\lambda}
                 \left( J_{\rm 1} - \int^{1}_{0} W\left(\mu\right)
                                        I_{\rm b}\left(\mu \right) d\mu\right) .
\end{align}
If instead the IMC region lies to the right of the DDMC region,
the normal vector used to define $\mu$ would become $-\hat{x}$.
While this will introduce a minus sign in the $\partial J/ \partial x$ term of
equation (\ref{eqn:adl}) as well as in the corresponding $J_{\rm 1}$ term of equation (\ref{eqn:f_interface}), the derivation that follows is identical.

Substituting Equation (\ref{eqn:f_interface}) back into (\ref{eqn:interface_RTE}) we obtain the DDMC
transport equation for the interface cell,
\begin{align}
  \label{eqn:DDMC_interface}
    \frac{1}{c} \frac{\partial J_{\rm 1}(\epsilon)}{\partial t} &=
                 \tilde{k}_{\rm ea,1} \tilde{b}_{\rm 1} c \tilde{u}_{\rm r} \notag \\
                 &- [\tilde{k}_{\rm ea,1}(\epsilon)
                         + \tilde{k}_{\rm es,1}(\epsilon)
                         + \tilde{k}_{\rm L,if}(\epsilon)
                         + \tilde{k}_{\rm R,1}(\epsilon) ] J_{\rm 1}(\epsilon) \notag \\
     &+  \left( \frac{\Delta x_{\rm 2}}{\Delta x_{\rm 1}} \right)
                           \tilde{k}_{{\rm L},2}(\epsilon) J_{\rm 2}(\epsilon)\notag \\
     &+ \frac{1}{\Delta x_{\rm 1}} \int^{1}_{0}P\left( \mu\right)  \mu I_{\rm b}
                           \left(\mu\right) d\mu \notag \\
     &+ 4 \pi \frac{\tilde{k}_{\rm a, 1}(\epsilon) \tilde{b}_{\rm 1}(\epsilon) }{\tilde{k}_{\rm p,1}}
          \int_{0}^{\infty}\,\tilde{k}_{\rm es,1}(\epsilon') J_{\rm 1}(\epsilon') d\epsilon',
\end{align}
where the left DDMC-to-IMC interfacial leakage opacity is redefined to be
\begin{align}
  \tilde{k}_{\rm L, if} =\frac{1}{\Delta x_{\rm 1}}
                        \frac{2}{3 \tilde{k}^{+}_{\rm T,1/2}\Delta x_{\rm 1} + 6\lambda},
\end{align}
and we have rearranged the terms in the integral to endow $P(\mu)$ with the physical
interpretation of an IMC-to-DDMC `conversion probability',
\begin{align}
  \label{eqn:conv_prob}
  P(\mu) = \frac{4}{3 \tilde{k}^{+}_{\rm T,1/2}\Delta x_{\rm 1} + 6\lambda}
          \left(1 + \frac{3}{2}\mu \right).
\end{align}
With this interpretation $\mu I_{b}$ can be understood as the flux of radiation
 at the interface.

%%% Show results in a plot of urad
\begin{figure*}
   \begin{center}
   \includegraphics[width=0.33 \textwidth]{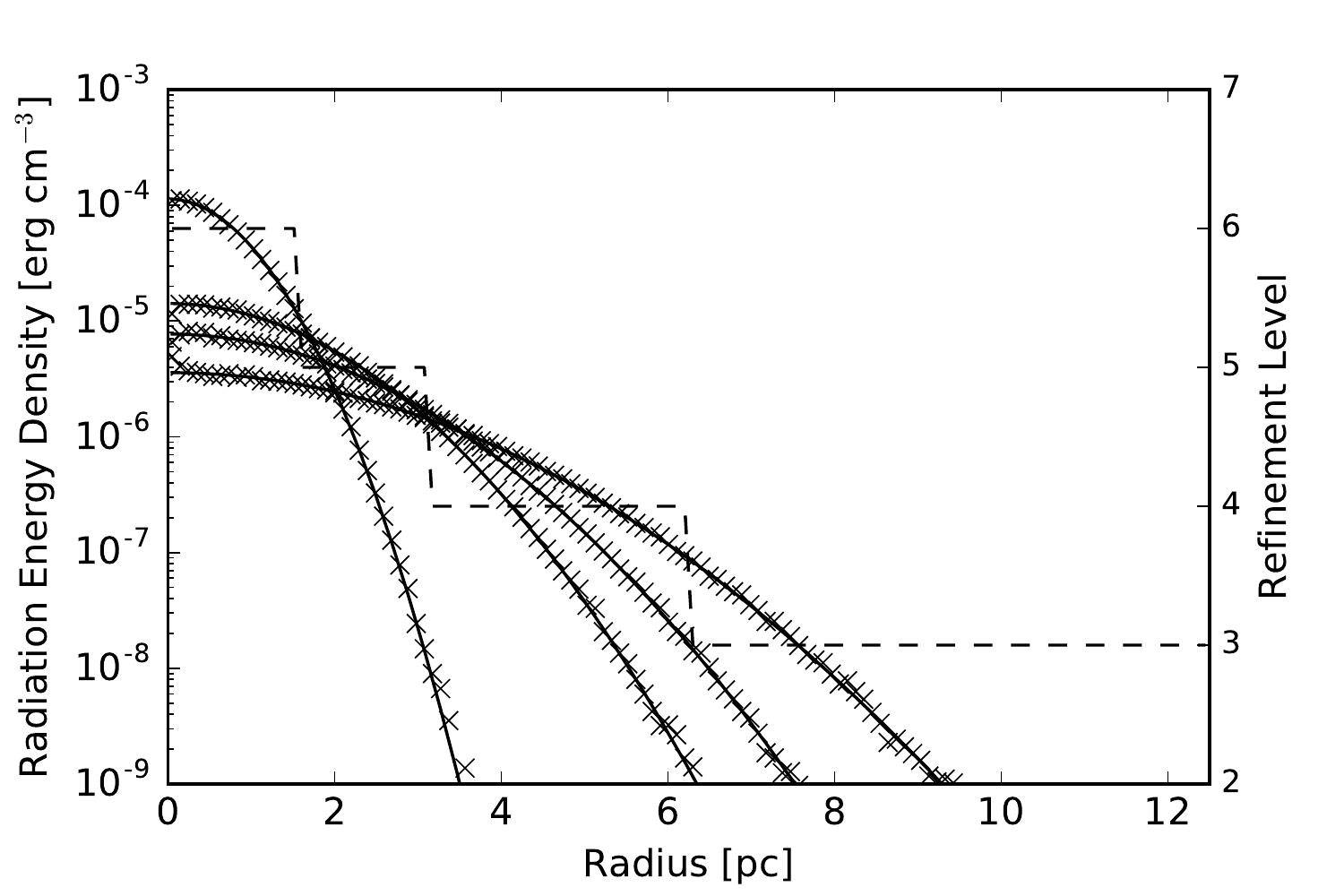}
      \includegraphics[width=0.33 \textwidth]{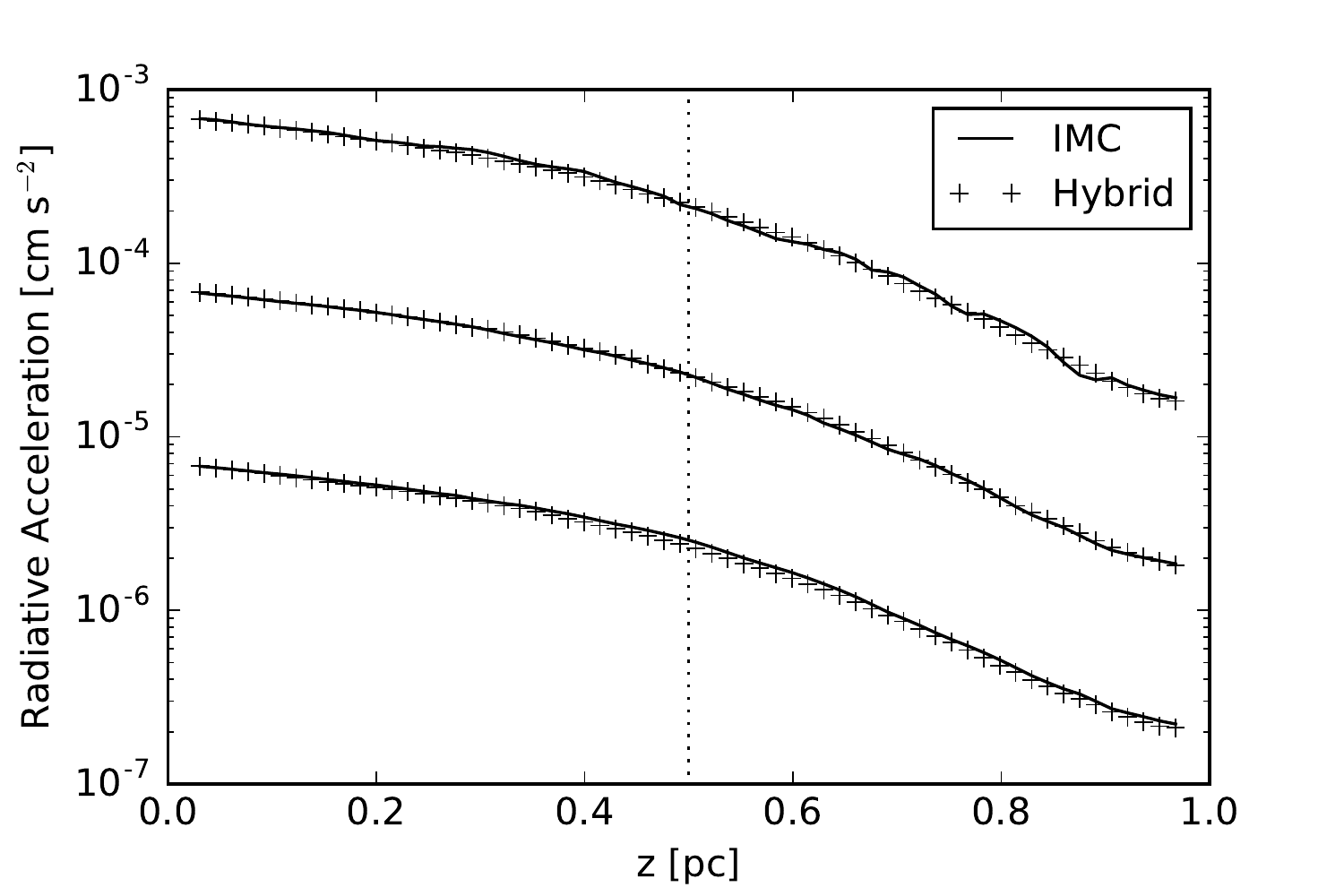}
   \includegraphics[width=0.33 \textwidth]{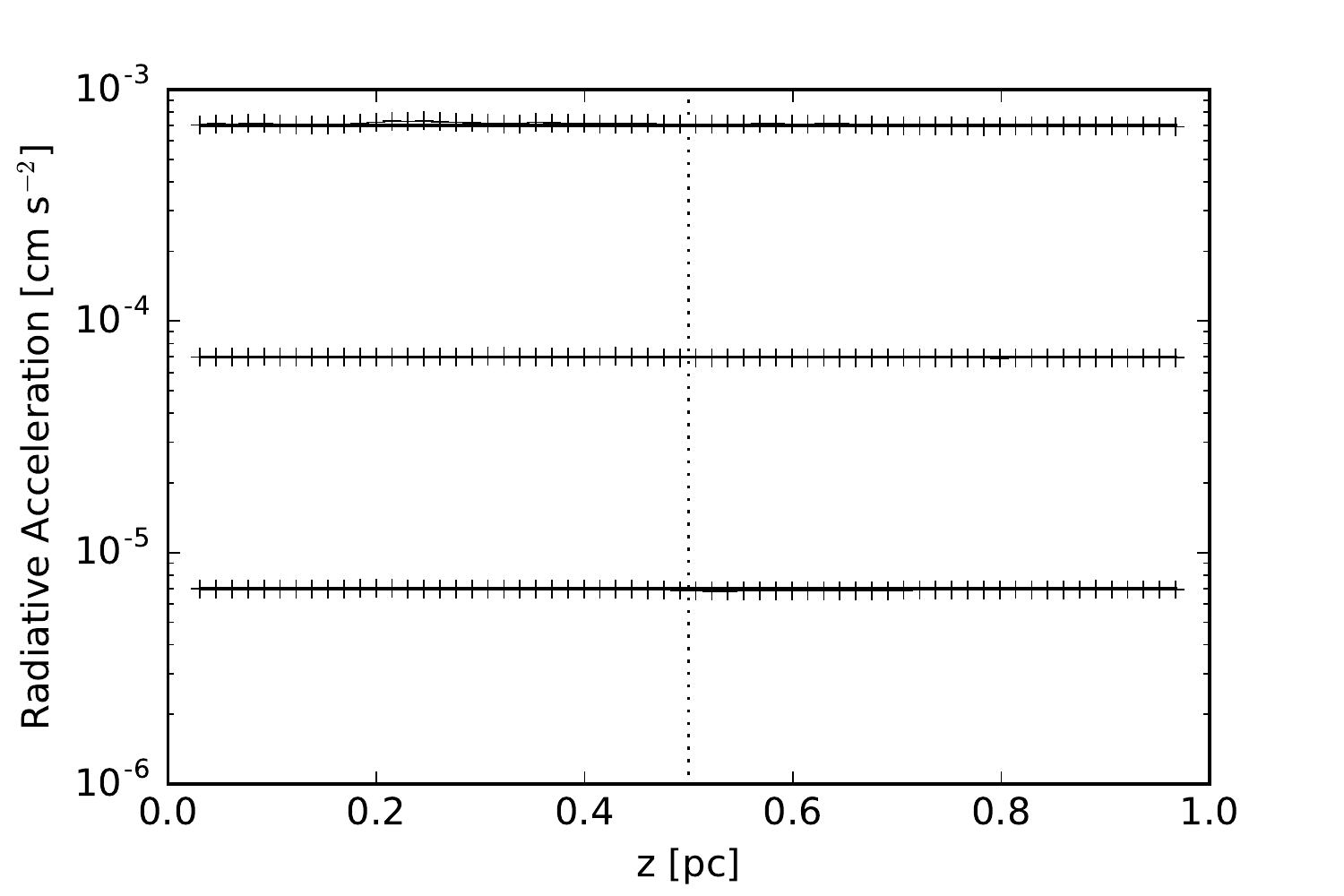}
   \end{center}
   \caption
   {Left panel: Spherically-averaged radiation energy density profiles in the
    radiative diffusion test in Section \ref{sec:rad_diff}.
    The analytical (solid lines) and the numerical
    (crosses) solutions are shown at four different times,
    $(1,\,4,\,6,\,10)\times10^{10}\,\mathrm{s}$.
    The dashed line and the right axis show the refinement level.  The middle and right panels show radiative acceleration profiles in the $z$-direction for the plane-parallel
    slab in Section \ref{sec:ss_pp}.
    The solid lines and crosses are the results with and without DDMC enabled, respectively.
    The IMC-DDMC interface is marked with the dotted line at $z=0.5$\,pc.
    Note that no artifact of the IMC-DDMC transition is observed across the interface.
    Middle panel: Profiles at the optical-thickness-dependent time 0.5~$t_{\rm slab}$ with,
    from top to bottom, $\tau_{\rm slab} = 80, 320, 3200$.
    The profiles for $\tau_{\rm} = 320$ and 3200 are scaled by factors of 10 and 100
    for visual clarity.
    Right panel: Same as in the middle panel but at steady state.
    }   \label{fig:ddmc_diff}
\end{figure*}

Similarly, we can divide the equation into energy groups,
\begin{align}
  \label{eqn:DDMC_mg_interface}
    \frac{1}{c} \frac{\partial J_{1,k}}{\partial t} &=
                 \tilde{k}_{{\rm ea},1,k} \tilde{b}_{1,k} c \tilde{u}_{\rm r} \notag \\
                 &- [\tilde{k}_{{\rm ea},1,k}
                         + \tilde{\sigma}_{{\rm es},1,k}
                         + \tilde{k}_{{\rm L,if},k}
                         + \tilde{k}_{{\rm R},1,k}] J_{1,k} \notag \\
     &+  \left( \frac{\Delta x_{2}}{\Delta x_{1}} \right)
                           \tilde{k}_{{\rm L},2,k} J_{2,k} \notag \\
     &+ \frac{1}{\Delta x_{1}} \int^{1}_{0}P_{k}\left( \mu\right) \mu I_{{\rm b},k}
                           \left(\mu\right) d\mu \notag \\
     &+ 4 \pi \frac{\tilde{k}_{{\rm a,}1,k} \tilde{b}_{1,k} }{\tilde{k}_{{\rm p},1}}
          \sum_{l\ne k}\,\tilde{k}_{{\rm es},1,l} J_{1,l} \Delta \epsilon_{l},
\end{align}
where the reduced effective coefficient $\tilde{\sigma}_{{\rm es},1,k}$ is
defined as in Equation (\ref{eqn:sigma_es}).
This transfer equation for the interface cell differs from the version
for interior cells (Equation \ref{eqn:DDMC_full}) in two ways:
it contains a revised DDMC-to-IMC leakage opacity, and a new radiation
source term that describes the addition of radiation due to IMC-to-DDMC conversion at the boundary of the interface cell.

As in interior cells, we adopt the gray approximation and an isothermal equation of state,
the equation can be simplified,
\begin{align}
  \label{eqn:DDMC_gray_iso_interface}
    \frac{1}{c} \frac{\partial J_{1,k}}{\partial t} =
                 &- [\tilde{k}_{{\rm L,if},k}
                         + \tilde{k}_{{\rm R},1,k} ] J_{1,k} \notag \\
     &+  \left( \frac{\Delta x_{2}}{\Delta x_{1}} \right)
                           \tilde{k}_{{\rm L},2,k} J_{2,k} \notag \\
     &+ \frac{1}{\Delta x_{\rm 1}} \int^{1}_{0}P_{k}\left( \mu\right) \mu I_{{\rm b},k}
                           \left(\mu\right) d\mu,
\end{align}
where again, the effective absorption and scattering terms have canceled out.

\subsubsection{The IMC-DDMC hybridization}

To decide whether radiation transport in a cell should be performed with IMC or DDMC,
we compute optical depths of all cells and flag those above a threshold
value $\tau_{\rm DDMC}$ as DDMC-active.
For Equation (\ref{eqn:conv_prob}) to have a valid probabilistic
interpretation, we must require $0 \leq P(\mu) \leq 1$ with
$0 < \mu \le 1$. These conditions impose a minimum threshold optical depth of
$\tau_{\rm DDMC} \ge 2.0$.
In Section~\ref{sec:tests} we will show that the exact value of
$\tau_{\rm DDMC}$ does not affect the accuracy of the scheme as long as the
above criterion is satisfied.

When a DDMC MCP undergoes a leakage event into an IMC cell,
it is placed randomly at the cell boundary with a direction sampled
isotropically toward the interior of the IMC cell.
Once the MCP has entered the IMC cell, it is tracked with the IMC procedure described in Paper I.  When an IMC MCP reaches a DDMC cell interface,
it is converted into a DDMC MCP with probability $P(\mu)$.
In the DDMC cell the converted photon is transported with the DDMC scheme.
The non-converted MCPs are returned to the IMC region isotropically
and their IMC transport continues.
This reflection of MCPs according the chosen form of $W(\mu)$ guarantees
that the radiation flux across the IMC-DDMC interface is consistent with the asymptotic diffusion limit, which is an approximate solution
of the radiation transfer equation at the interface with radiation impinging on a diffusive,
isotropic scattering medium.

\begin{figure*}
   \centering
   \includegraphics[width=0.285\textwidth]{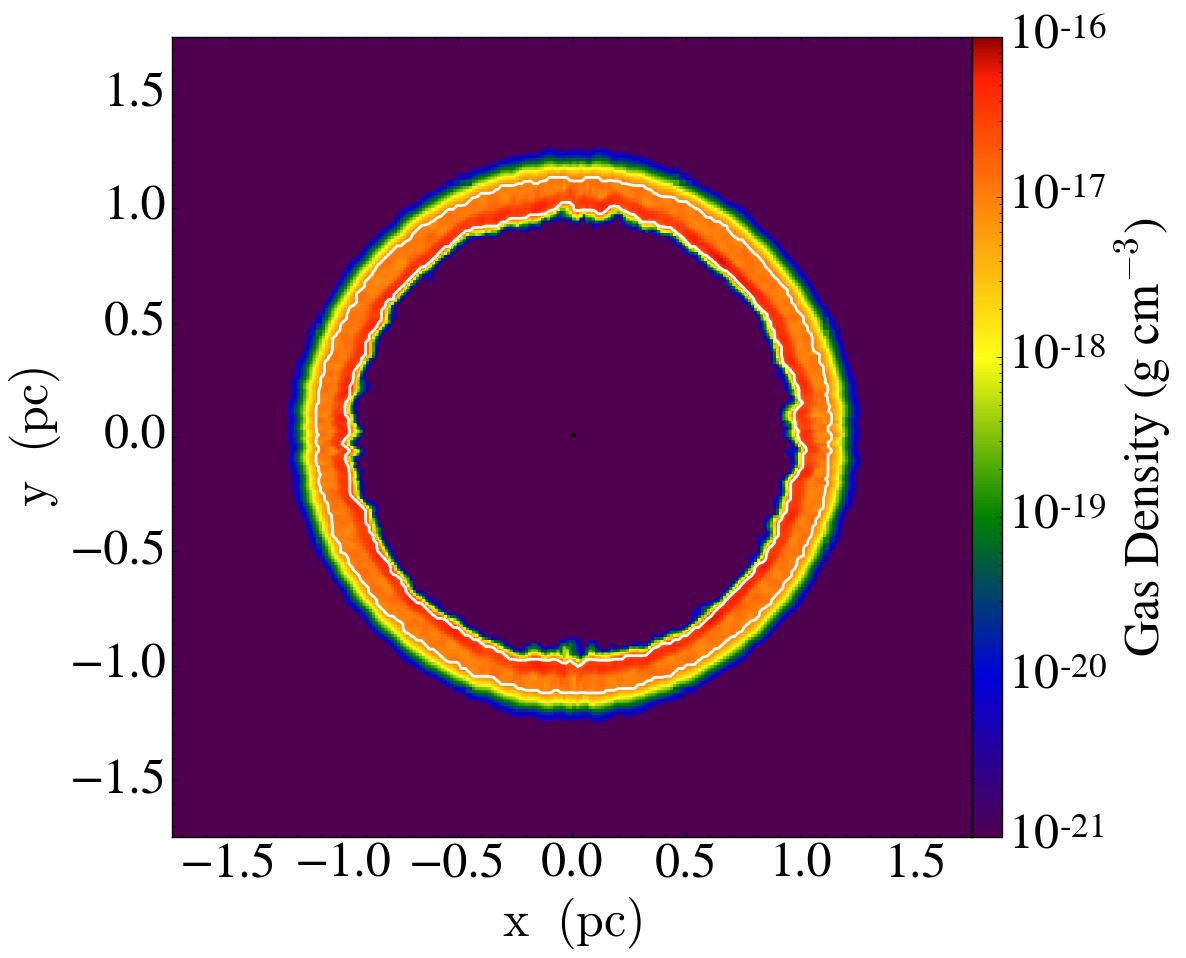}
   \includegraphics[width=0.345\textwidth]{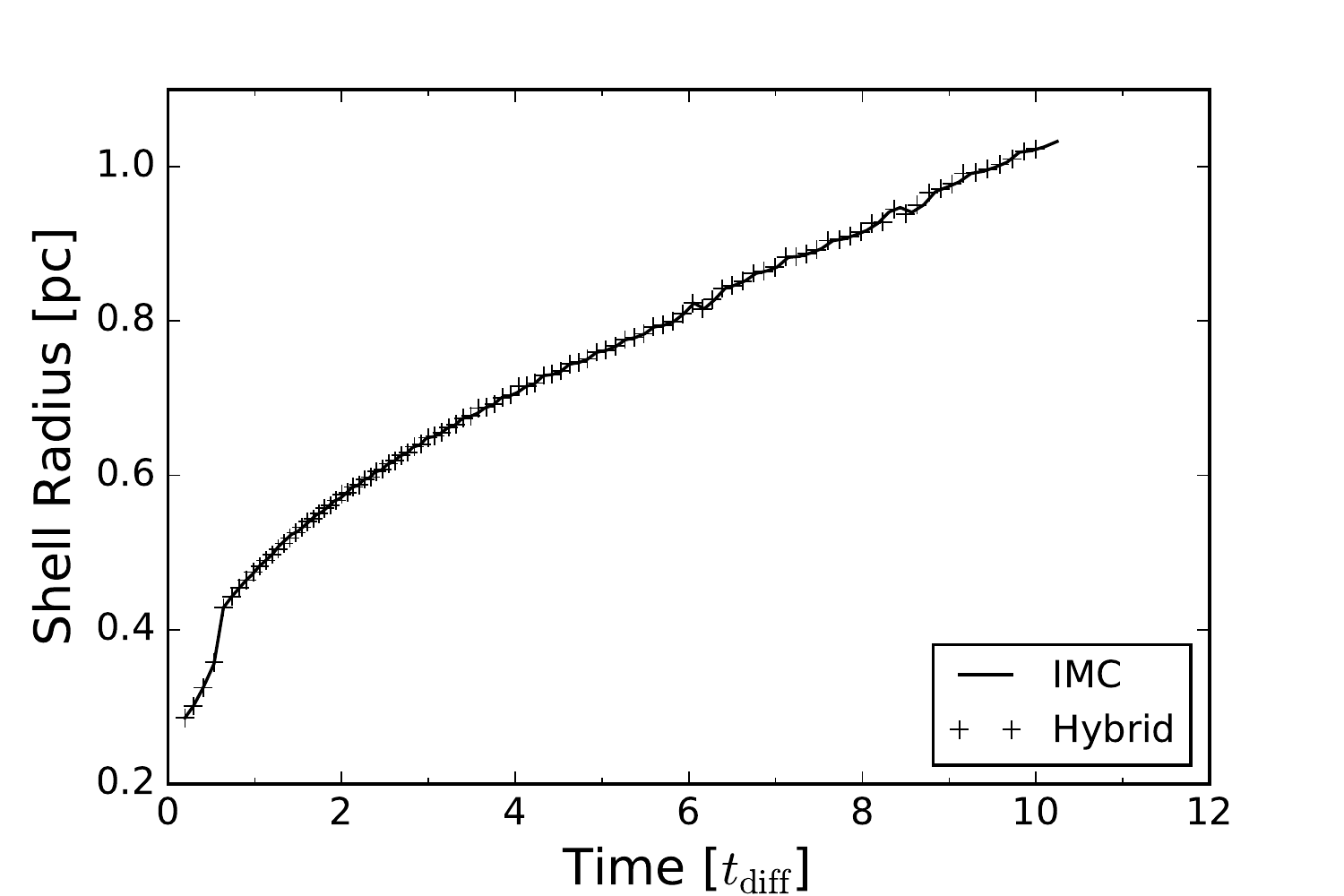}
   \includegraphics[width=0.345\textwidth]{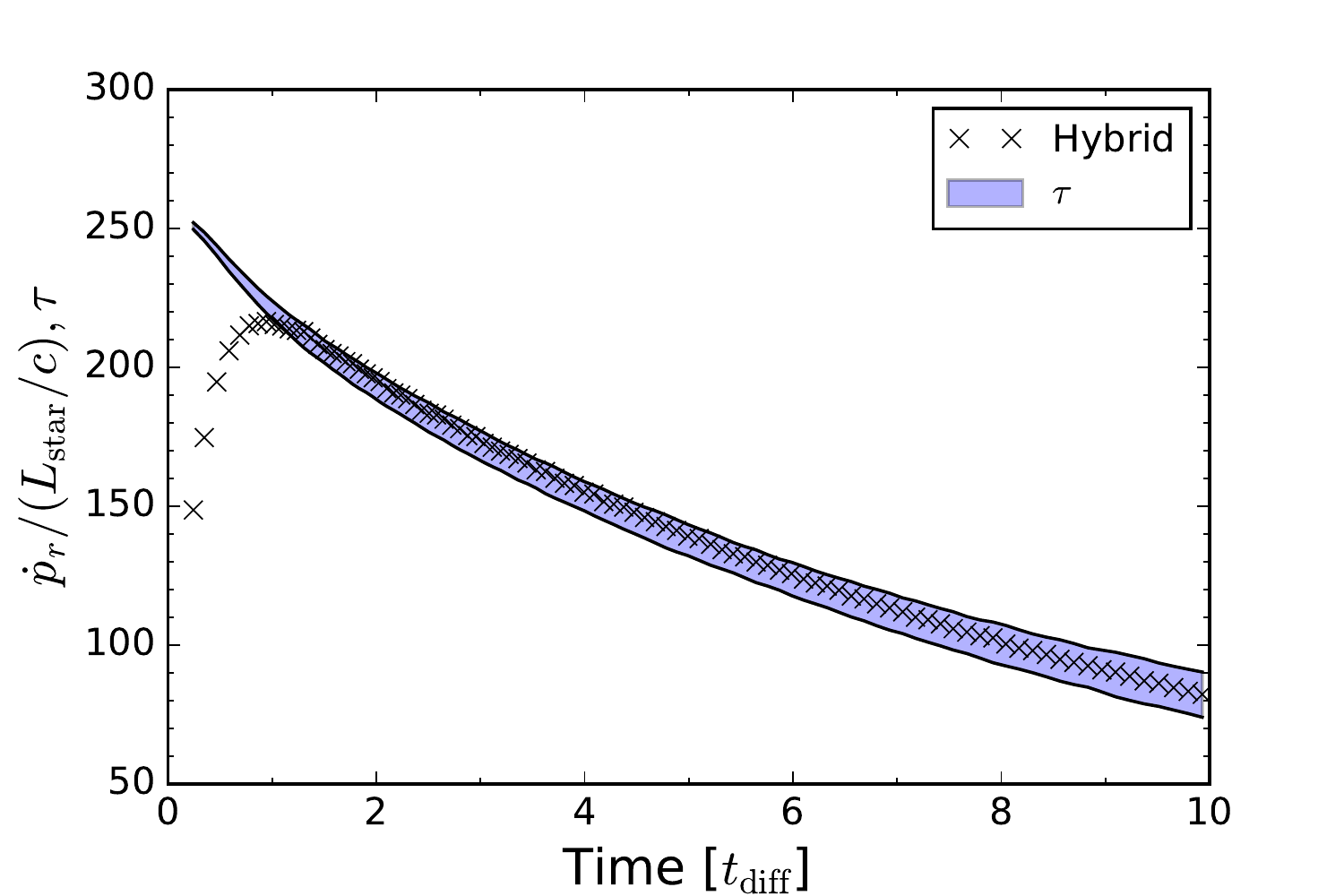}
   \caption
   {Left panel: Gas density slice at $t=10\,t_{\rm diff}$ in the setup in Section \ref{sec:rd_shell} in which radiation sweeps up a dense shell from the center of a spherical cloud.
    The shell has retained spherical symmetry. The region between the thin white
    contours, dynamically allocated in the course of the simulation, is DDMC-active.
    Middle panel: Evolution of the radiation-pressure-driven shell's radius.
    The slight fluctuations are instability in the numerical extraction of the shell radius from Cartesian data. The plot shows that activating DDMC yields consistent shell dynamics.
    Right panel: Evolution of the effective radiation force and the optical depth across the dense shell. The blue band represents the $1\sigma$ line-of-sight spread of the
    optical depth from the discretization of the spherical shell on the AMR grid.
    At $t\gtrsim t_{\rm diff}$, the diffusive radiation couples closely with the shell.
    }   \label{fig:shell}
\end{figure*}

\subsubsection{Momentum deposition}
\label{sec:md}
Here we present our scheme for momentum deposition from radiation on gas.
Because we treat the gas as isothermal, energy exchange reduces to momentum deposition.
In Paper I, we tallied the momentum deposition from individual effective
scattering events within a cell,
\begin{align}
  \label{eqn:dp_scat}
  \Delta \mathbf{p}_{\rm es} &= - \sum_{\text{eff.\,scatt.}(i)\in\text{cell}} p_{i} (\mathbf{n}'_{i} -\mathbf{n}_{i})
      \simeq \sum_{\text{eff.\,scatt.}(i)\in\text{cell}}  p_{i} \mathbf{n}_{i},
\end{align}
where $p_{i}$ denotes the magnitude of the momentum of MCP $i$
and the sum is over all scattering events that occur inside the cell during
$\Delta t$.  Each event randomizes the direction of the MCP $\mathbf{n}_{i} \rightarrow \mathbf{n}'_{i}$.
The approximate equality follows from the statistical isotropy of $\mathbf{n}'_i$.

In optically thin regions, this event-based momentum exchange estimator is noisy.
To reduce the noise in IMC regions we adopt a commonly used path-length-based estimator for
momentum deposition \citep{Lucy99,Hykes09,Harries15,Roth15,Smith17}
\begin{align}
  \label{eqn:dp_path}
  \Delta \mathbf{p}_{\rm es,IMC} &= \frac{k_{\rm es}}{c}
                                \sum \epsilon_{i} c \delta t_{i} \mathbf{n}_{i},
\end{align}
where the sum is over all the piecewise paths MCPs execute within the cell.

In DDMC regions where MCPs are not transported with piecewise paths
Equation (\ref{eqn:dp_path}) does not apply.
We follow \citet{Densmore07} to evaluate the momentum deposition by
averaging the radiation fluxes across cell faces in each grid direction,
\begin{align}
  \Delta \mathbf{p}_{\rm es,DDMC} &= \frac{1}{2c} k_{\rm es}
                                      \left( F_{-1/2} + F_{+1/2} \right)
                              \Delta t \Delta V,
\end{align}
where $\Delta V$ is the volume of the cell and $F_{\pm 1/2}$ are
the net interfacial radiation fluxes due to
leakage and IMC-to-DDMC conversion (`crossing') events,
\begin{align}
  F_{\pm 1/2} = \sum_{\text{crossing}(i)\in\pm\text{face}(j)} \frac{d_{i} \epsilon_{i}}{\Delta A_{j} \Delta t} .
\end{align}
Here, $\epsilon_{i}$ is the energy of the face-crossing MCP,
$d_{i} = \pm1$ for
leakage/conversion towards to the right (upper sign) and left (lower sign) of the cell,
$\Delta A_{j}$ is the area of the cell face, and
the sum includes all leakage/conversion events across the face.
Similar to Paper I, the momentum deposition tallied during the radiation
transport is used to construct the source terms for the hydrodynamical updates
\begin{align}
  \mathbf{S} = \frac{ \Delta \mathbf{p}_{\rm es,IMC} +\Delta \mathbf{p}_{\rm es,DDMC}}{\Delta t \Delta V} .
\end{align}
and
\begin{align}
  c S_{0} = \mathbf{v} \cdot \mathbf{S}.
\end{align}
These source terms are deposited as gas kinetic energy and momentum at the
end of the hydrodynamic time step according to the same procedure reported in
Paper I.

\section{Performance tests}
\label{sec:tests}
To validate the hybrid IMC-DDMC radiation transport scheme we solved a suite of test problems.
The radiative diffusion setup in Section \ref{sec:rad_diff} validates the reliability of tracking the diffusion of radiation across mesh refinement jumps.
The plane-parallel setups in Section \ref{sec:ss_pp}
demonstrates the robustness of momentum deposition within IMC and DDMC regions
as well as across IMC-DDMC interfaces.
The radiation pressure-driven shell sweeping setup in Section \ref{sec:rd_shell}
puts the coupled RHD to test, showing the consistency of
the DDMC scheme with the IMC scheme.

\subsection{Radiative diffusion across mesh refinement jumps}

\label{sec:rad_diff}
To test optically thick radiative diffusion across mesh refinement jumps, we simulate a $25\,\text{pc}$ cube
filled with a uniform, purely scattering medium with density $\rho=10^{-17}$g\,cm$^{-3}$.
The absorption opacity is set to zero and the scattering opacity to a constant value $\kappa_{\rm s} = 4$\,cm$^{2}$\,g$^{-1}$.  Optical depth across a half of the box width is $\tau_{\rm box} = 1540$.
Such setup is the standard for validating the diffusion limit of any radiation transport scheme \citep{Commercon11,Harries11,Noebauer12}.
Since the objective is to test the transport of MCPs on a nonuniformly refined grid, we disable momentum deposition and gas dynamics in this test. The radiation does not reach the boundary of the box and thus boundary condition is immaterial.

At $t = 0$\,s, we insert
$E_{\rm init} = 2\times10^{52}$\,erg of radiative energy in the form of 320,000 isotropically directed MCPs, all initially at the center
of the grid.
The grid has the base resolution of $32^{3}$. The center of the box is refined three levels higher and has an effective resolution of $256^3$.
The smallest cell width is $\sim0.1$\,pc which corresponds to a cell optical depth of
$\tau_{\rm cell} =12$.
The simulation is run for $10^{11}$\,s (5\% of the half-domain diffusion time) at a constant time step of $\Delta t = 2\times10^{9}$\,s.
In this test, DDMC is active in the entire simulation domain.

The left panel of Figure \ref{fig:ddmc_diff} shows spherically-averaged radiation energy
density profiles at four different times. The radial
dependence of the mesh refinement level shown on the right axis.
Note the excellent agreement with the analytical solution (solid lines in the figure; see Paper I).
The test also demonstrates that DDMC attains the same level of
accuracy with fewer MCPs and on a coarser mesh than the basic
Monte Carlo method of Paper I.

\begin{table*}
  \centering
  \begin{tabular}{ccccccc}
  \hline \hline
  $\kappa_{\rm s}$
  (cm$^{2}$\,g$^{-1}$) & $\tau_{\rm box}$ & $\tau_{\rm cell, min}$ & $\tau_{\rm cell, max}$ & IMC runtime (s) & DDMC runtime (s) & Speedup \\
  \hline
        1    & 385  & 3  & 24  &  9.6$\times10^{3}$    & 1.9$\times10^{3} $     & 5.1 \\
        4    & 1540 & 12 & 96  &  3.4$\times10^{4}$    & 1.5$\times10^{3}$      & 22.4 \\
        16   & 6160 & 48 & 385 &  1.4$\times10^{5}$   & 1.0$\times10^{3} $     & 137.0 \\
    \hline
  \end{tabular}
   \caption{Parameters and timing results for the timing test presented in Section \ref{sec:timing}.}
\label{tab:diff_speedup}
\end{table*}

\subsection{Radiative acceleration of a plane-parallel slab}
\label{sec:ss_pp}

To test the accuracy of momentum deposition under IMC-DDMC hybridization, we construct a
three-dimensional plane-parallel setup with sharp transitions in optical thickness.
The objective is twofold, to validate that momentum deposition
from radiation to gas is accurate in both the IMC and DDMC regions and to check that there are no spurious discontinuities at
IMC-DDMC interfaces.

The simulation domain is a cube with side $L_{\rm slab}=1\,\text{pc}$,
with a low density $\rho_{\rm low} = 2.6\times10^{-22}\,\text{g}\,\text{cm}^{-3}$
in one half ($z < 0.5$\,pc) of the cube
and a higher constant density in the other half ($z \ge 0.5$\,pc).
Specifically, we carry out runs with three choices of the higher density of
$\rho_{\rm high}=2.6\times10^{-18}$, $1.03\times10^{-17}$,
and $1.03\times10^{-16}$\,g\,cm$^{-3}$.
Assuming a purely scattering medium with an opacity
$\kappa_{\rm s} =20\,\text{cm}^{2}\,\text{g}^{-1}$,
the low-density half is an optically thin layer with
$\tau(z<0.5\,\rm pc) = 0.008$ and the high density half has optical depth
$\tau(z \ge 0.5\,\rm pc) = 80$, 320, and 3200, respectively.
The grid resolution is fixed at 64$^{3}$.
As a result, the single-cell optical depth jumps from $2.5\times10^{-4}$ to
2.5, 10, and 100 across the $z=0.5\,\text{pc}$ interface in the three runs.
A constant vertical radiation flux of
$2.54\times10^{7}\,L_{\odot}\,\text{pc}^{-2} \hat{z}$
is introduced at the $z=0$ face.
To test the accuracy of momentum deposition,
gas dynamics is again disabled; we simply tally the momentum deposited by the MCPs in each cell.

The middle panel of Figure \ref{fig:ddmc_diff} shows the $z$-profiles of the radiative
acceleration of gas at the half-diffusion time of the high-density regions
$0.5\,t_{\rm slab} = 0.5\,\tau_{\rm slab} (L_{\rm slab}/2) / c$.  The DDMC is enabled but
the low-density region defaults to pure IMC.
The profiles agree with each other and with the pure IMC runs (in the figure they are arbitrarily rescaled for visual clarity).
We do not observe any artifacts at the IMC-DDMC interface.

In the $t\rightarrow\infty$ steady state limit, the radiative acceleration of a plane-parallel slab transporting a constant flux is uniform in space.
The right panel of Figure \ref{fig:ddmc_diff} shows the steady-state radiative acceleration profiles.
The solid lines denote the analytical constant value.
In all runs, the accelerations agree with the analytical value within $\le$~3\%.
We have also repeated the simulations with an additional
refinement level transition at the IMC-DDMC interface (not shown).
Our hybrid scheme handles the radiation transfer equally well with a
coincident jump in cell size and optical thickness.

\subsection{Radiation pressure driven shell}
\label{sec:rd_shell}
To test how well the IMC-DDMC hybrid scheme couples to the unsplit HLLC Riemann solver we use for the hydrodynamic update,
we let the pressure of radiation injected at the center of an optically thick spherical cloud drive the expansion of a shell of gas.
The setup is similar to the radiation pressure test in \citet[][see their Sec.\ 4.2]{Rosen17}, except that here we focus on the multi-scattered IR
radiation instead of the singly-scattered (i.e., absorbed) UV radiation.
Our test also bears similarities with the simulation of \citet{Costa17}. We employ the test to validate three key aspects of the hybrid scheme:
proper handling of radiation and momentum deposition in non-plane-parallel
geometries,
stability of radiation transport with dynamically activated DDMC regions,
and accuracy of radiative momentum deposition on gas.

A spherical gas cloud with mass $M_{\rm cl} = 1.25\times10^{6} M_{\odot}$,
radius $R_{\rm cl} = 1$\,pc, and uniform density
$\rho_{0} = 10^{-17}$\,g\,cm$^{-3}$ is placed at the
center of a (3.5\,pc)$^{3}$ domain.
The cloud is embedded in a uniform low-density $\rho_{\rm bg} = 10^{-22}$\,g\,cm$^{-3}$ background.
A constant dust opacity $\kappa_{\rm d} = 8.4$\,cm$^{2}$\,g$^{-1}$ is set throughout.
The resulting center-to-edge optical depth of the gas cloud is
$\tau_{\rm cl} = \kappa_{\rm d} \rho_{0} R_{\rm cl} = 264.2$.
As in our research-scale simulations in the next section,
the equation of state is isothermal.  Gravity is disabled in the test.

A radiation source with a constant luminosity
$L_{\rm star} = 5.6\times10^{42}$\,erg\,s$^{-1}$ is placed at the center.
The stellar radiation exerts a net outward
radiation pressure. The resulting supersonic expansion sweeps up a dense shell.
We adopt a base resolution of (32)$^{3}$ and allow two levels of
adaptive refinement activated with the standard gas density second derivative criterion. The minimum cell width is
$\Delta x \simeq 0.04$\,pc. We apply DDMC
to cells with optical depths exceeding the threshold $\tau_{\rm thres}=2$.
The left panel of Figure \ref{fig:shell} shows the slice of gas density through
the center at time $t=10$\,$t_{\rm diff}$,
where $t_{\rm diff}$ is the radiative diffusion timescale across the initial
uniform gas cloud $t_{\rm diff} = \tau_{\rm cl} R_{\rm cl} / c$.

The spherical symmetry of the swept-up shell is well-preserved. The small asymmetry arose at the beginning of the simulation when the shell was resolved by only a couple cells.
Throughout the simulation, the shell is refined at the highest level.
The DDMC-active region is not fixed in space but
is allowed to adapt to the local optical density and travels with the expanding shell.
The region between the white lines in the left panel of Figure \ref{fig:shell}
marks where DDMC is active.

To validate the shell dynamics, we repeat the simulation with
DDMC disabled and compare the results.
The middle panel of Figure \ref{fig:shell} shows the evolution of the dense
shell's position in both the IMC-only and the hybrid runs.
The shell's radius is computed as the mass-weighted average of the radius of the gas lying above $25\%$ of the peak density.
The position of the shell shows excellent agreement between the IMC-DDMC hybrid run and the pure IMC control run.

\begin{table}
  \centering
  \begin{tabular}{cccc}
  \hline \hline
  Quantity & Value & Units \\
  \hline \hline
  $M_{\rm box}$ & 10 & $10^{6} M_{\odot}$ \\
  $L_{\rm box}$ & 25 & pc                 \\
  $\rho_{\rm init}$ & 4.4 & 10$^{-20}$\,g\,cm$^{-3}$ \\
  $\Sigma_{\rm init}$ & 1.6 & $10^{4}\,M_{\odot}$\,pc$^{-2}$ \\
  $t_{\rm ff}$ & 0.32 & Myr \\
  ${\cal M}$ & 7.5  & -- \\
  $\Delta x_{\rm min}$ & 0.02 & pc \\
  $r_{\rm soft}$ & 0.06 & pc \\
  $t_{\rm max} / t_{\rm ff}$ & 1.0 & --  \\
    \hline
  \end{tabular}
  \caption{Simulation parameters used in the cluster formation runs.}
\label{tab:sim_par}
\end{table}

\begin{figure*}
\begin{center}
\includegraphics[width=0.8\textwidth]{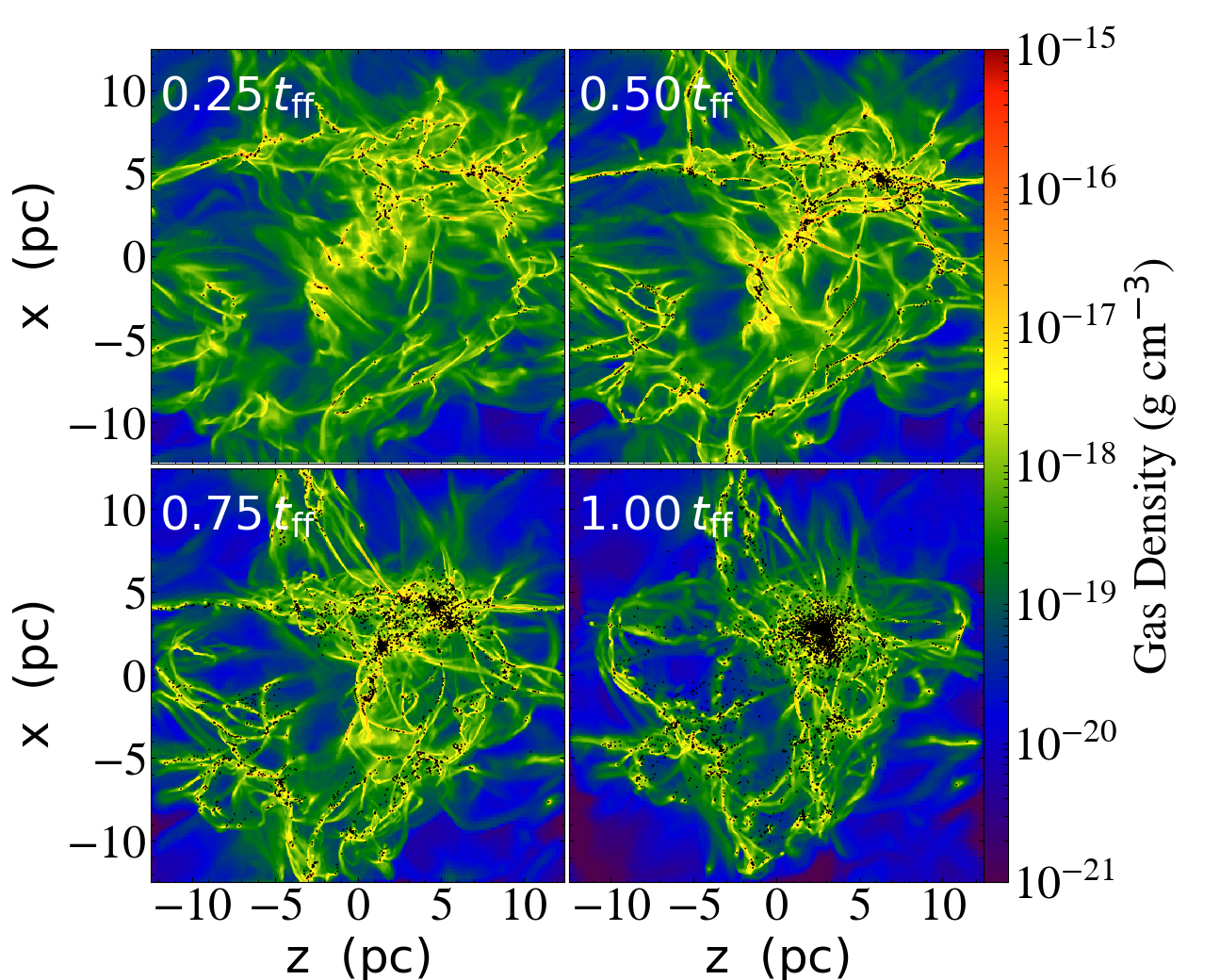}
\includegraphics[width=0.8\textwidth]{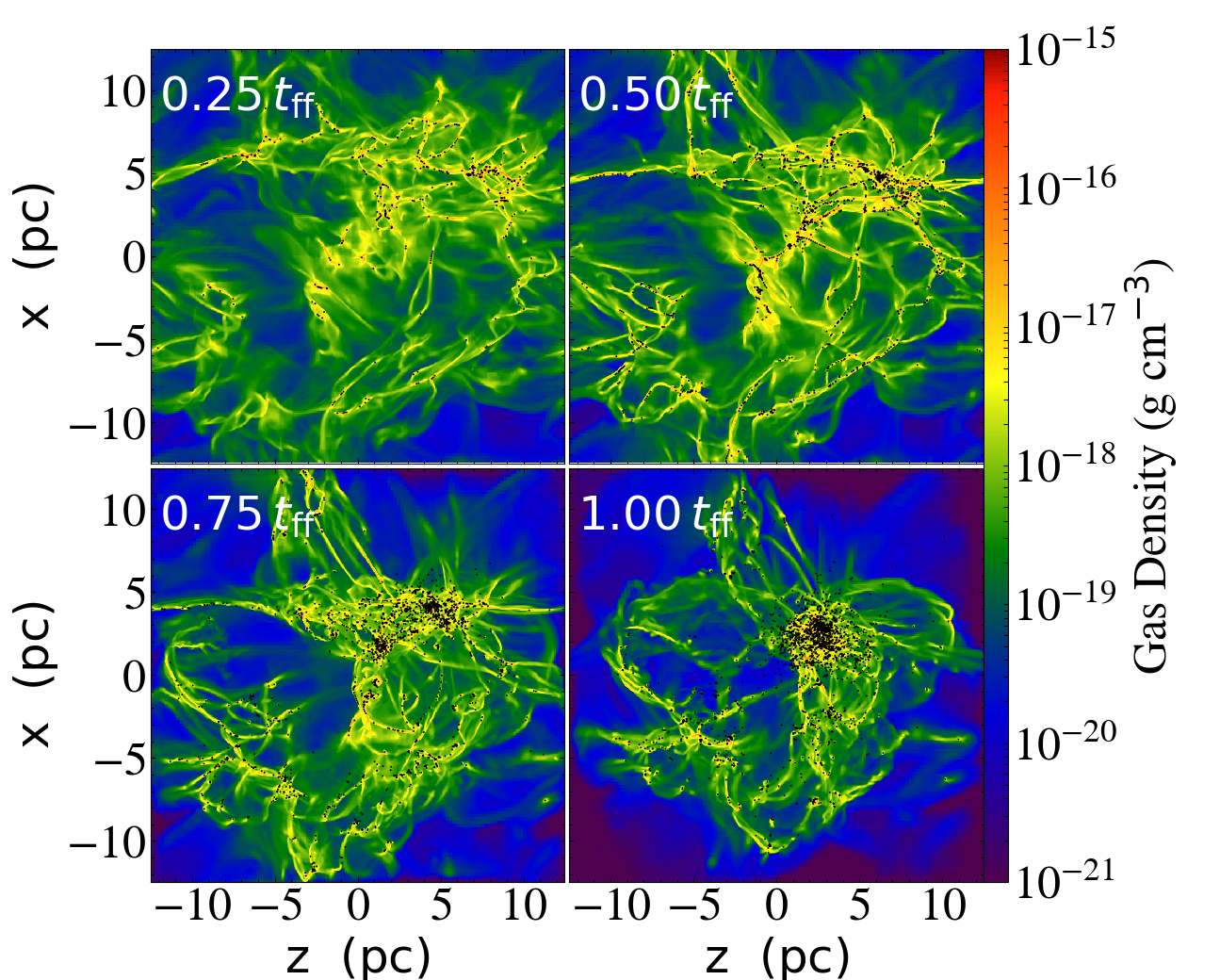}
\end{center}
\caption{Density-weighted density projections ($\int \rho^2 dy / \int \rho dy$)
of the entire simulation box for the RHD (top panel) and HD runs (bottom panel)
at 0.25, 0.5, 0.75, and 1.0\,$t_{\rm ff}$,
where the free fall time is $t_{\rm ff}= 0.3\,\text{Myr}$.
Black dots show the projected locations of all sink particles in the simulation.}
\label{fig:dens_projs}
\end{figure*}

\begin{figure*}
\begin{center}
\includegraphics[width=0.8\textwidth]{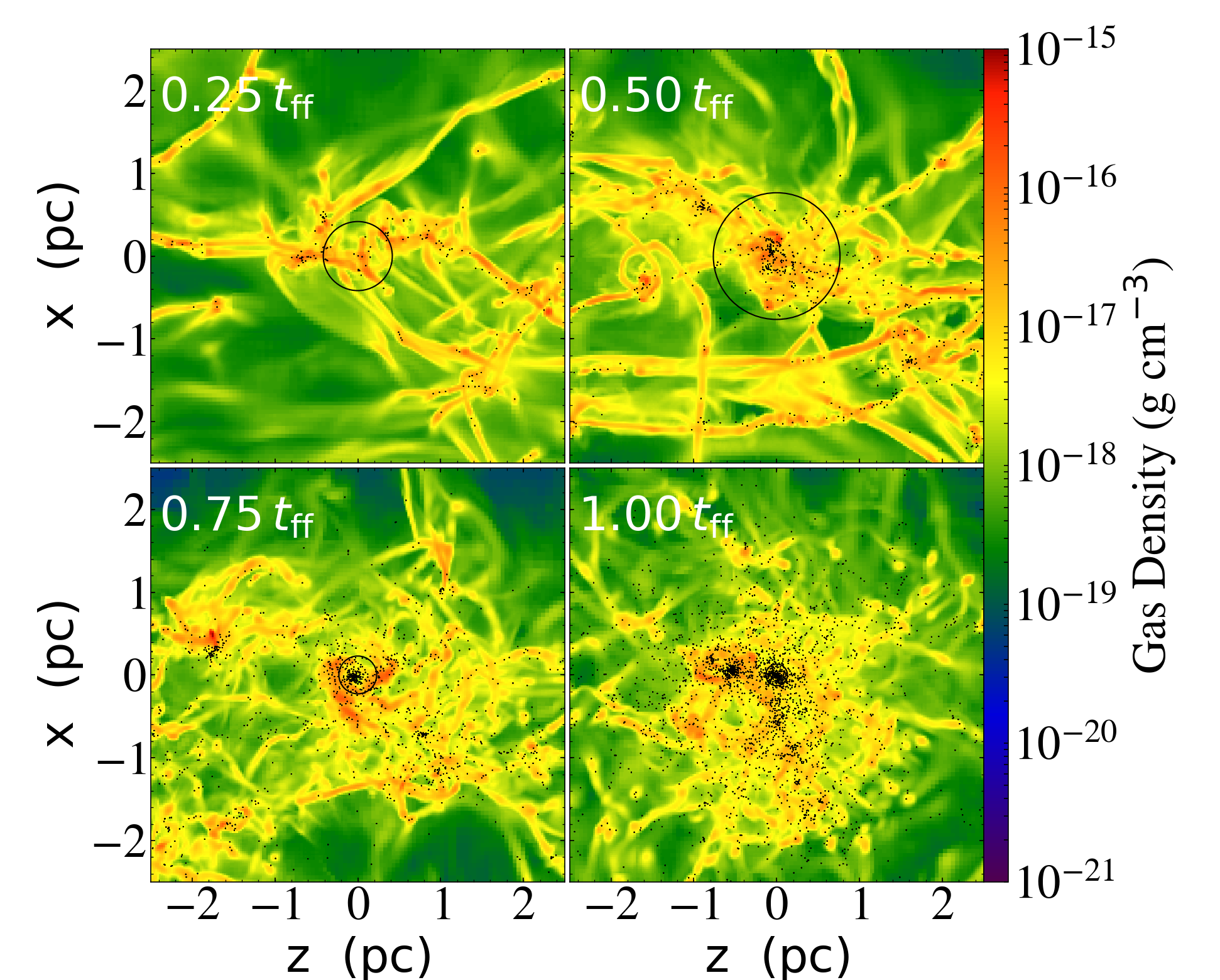}
\includegraphics[width=0.8\textwidth]{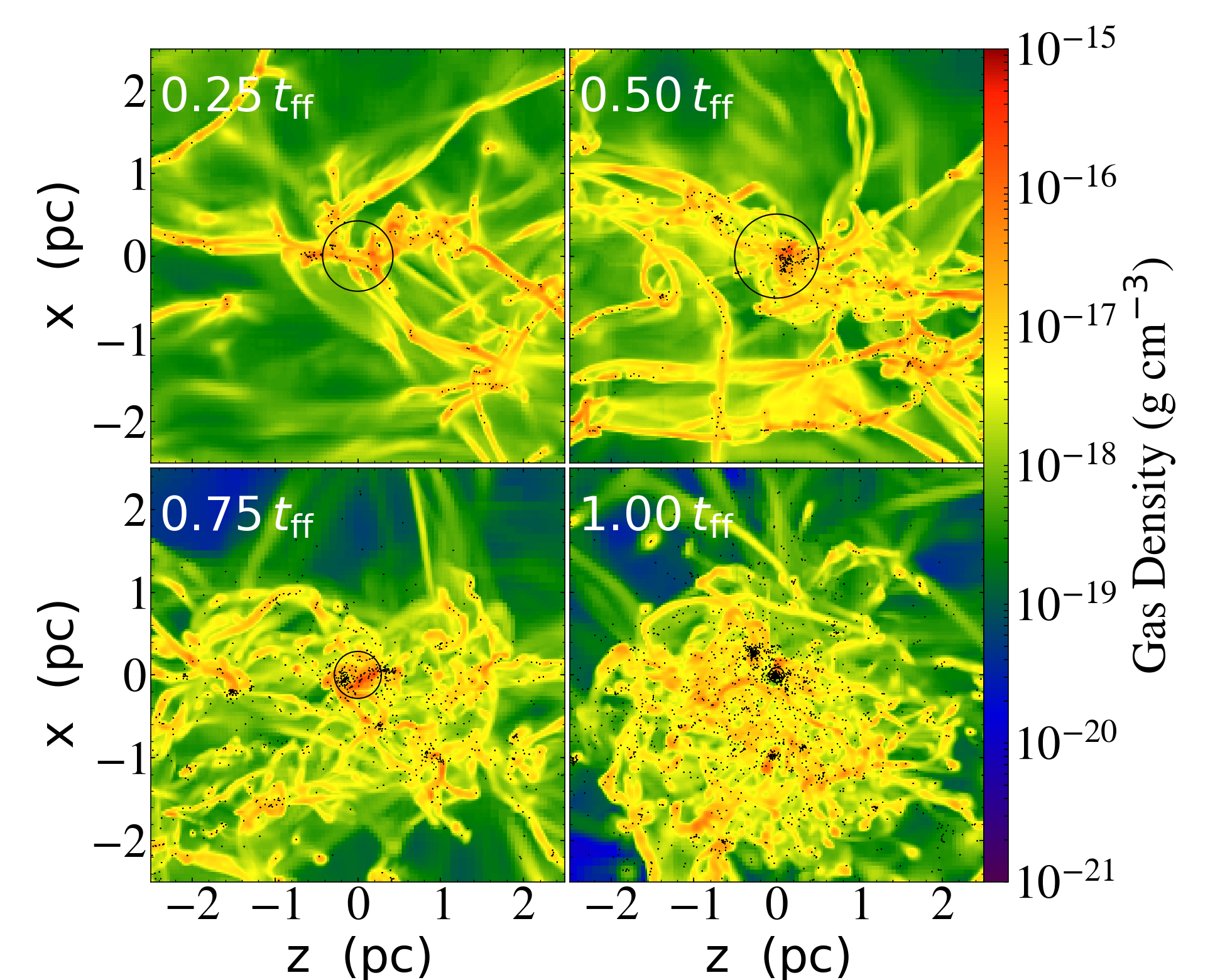}
\end{center}
\caption{The corresponding (5\,pc)$^{2}$ zoomed-in view of Figure \ref{fig:dens_projs}
         centered at the center of mass of the most massive clusters.
         They show the details of gas and sink distributions around the bases
         of the global gravitational potentials.
         Black circles show the size of the cluster's half-mass radius.
         }
\label{fig:dens_projs_zoomed}
\end{figure*}

The radiation pressure force is deposited over a layer of finite thickness.
To verify accuracy of the momentum coupling, we compare the effective radiative force
$\dot{p}_{\rm r}(t) / (L_{\rm star} / c)$
with the instantaneous shell optical thickness $\tau_{\rm sh}(t)$ in the right panel of
Figure \ref{fig:shell}.
The blue shaded region shows the $1\sigma$ spread of $\tau_{\rm sh}$ arising from the
Cartesian discretization of the spherical shell.
The agreement at $t \ge t_{\rm diff}$ is excellent.
At $t \leq t_{\rm diff}$, the radiative force has not yet reached its quasi-steady-state level because radiation has not yet diffused through the entire thickness of the shell. This limited momentum flux due to finite travel time of radiation is in line with the results of \citet{Costa17} who also observed limited momentum coupling
when $\tau_{\rm cl} \geq 50$.

In the non-inertial comoving frame of the dense shell, the net outward acceleration by the
radiation pressure can be viewed as an effective gravitational acceleration with magnitude $g_{\rm eff} = \dot{p}_{\rm r} / M_{\rm cl}$, where the assumption is that the entire cloud has been swept up into the shell.
RTI is expected to grow on the timescale of
$t_{\rm RT} = \sqrt[]{\lambda_{\rm RT} / 2 \pi g_{\rm eff}}$, where $\lambda_{\rm RT}$
is the length scale on which the instability structure grows
\citep{Chandrasekhar61}.
We expect to see RTIs if the dynamical time $t_{\rm dyn} = R_{\rm sh} / v_{\rm sh}$ is longer than
$t_{\rm RT}$, where $v_{\rm sh}$ is the shell's radial velocity.
At the end of the test simulation, $\lambda_{\rm RT} \leq 3.6\text{\,pc}$, with a corresponding
growth time of $t_{\rm RT} = 20$\,$t_{\rm diff}$.
The length scale is comparable to the size of the system, but we only run the test to
10\,$t_{\rm diff}$. On small scales RTI may be suppressed by radiative diffusion,
while on larger spatial scales the setup does not yet have enough time to develop significant RTI.

\subsection{Timing}
\label{sec:timing}

To assess the efficiency of the DDMC schem, we repeated the test in Section \ref{sec:rad_diff} in which radiation is deposited at the grid center
at $t = 0$\,s and allowed to spread radially.
With all other parameters fixed, we vary the scattering opacity
$\kappa_{\rm s}=1,\,4,\,16$.  The wall-clock runtimes required for the simulation to reach $t = 10^{11}$\,s are recorded.
Table \ref{tab:diff_speedup} summarizes the grid parameters and the
timing results.
The speedup is defined as the ratio of IMC runtime to the DDMC runtime.
DDMC consistently outperforms
IMC and the speedup increases at higher opacities.
This is a unique advantage of the leakage formalism of DDMC and is reflected in the inverse relation of the leakage opacities to  the physical opacities (Equations \ref{eqn:k_L} and \ref{eqn:k_R}).
At a fixed spatial resolution, leakage of MCPs across cells replaces a larger number of scattering events in regions with higher densities. This enhanced computational efficiency.

\section{Simulation methods}
\label{sec:sim_setup}

In this section we introduce a Monte Carlo simulation of SSC formation to serve
as a testbed for our hybrid radiation transport scheme.
The simulation is intended to demonstrate the feasibility and effectiveness
of the
method in research-scale applications.
The astrophysical motivation is to examine the effects of the pressure of the reprocessed stellar radiation in the assembly of SSCs.  Does the radiation pressure force become competitive with the gravitational force?  Does it temper star formation?  Does it affect the structure of the resulting SSC?

To simulate SSC formation, we perform a radiation hydrodynamical (RHD) run,
and also a radiation-free, purely hydrodynamical (HD) control run in which stellar radiation is disabled.
Comparison of the RHD with the HD run allows us to isolate the effects of
stellar radiation.
Other than for the absence of radiation in the HD run, the two runs have identical parameters and initial conditions. The parameters are explained in what follows and summarized
in Table~\ref{tab:sim_par}.

\begin{figure}
   \begin{center}
   \includegraphics[width=\columnwidth]{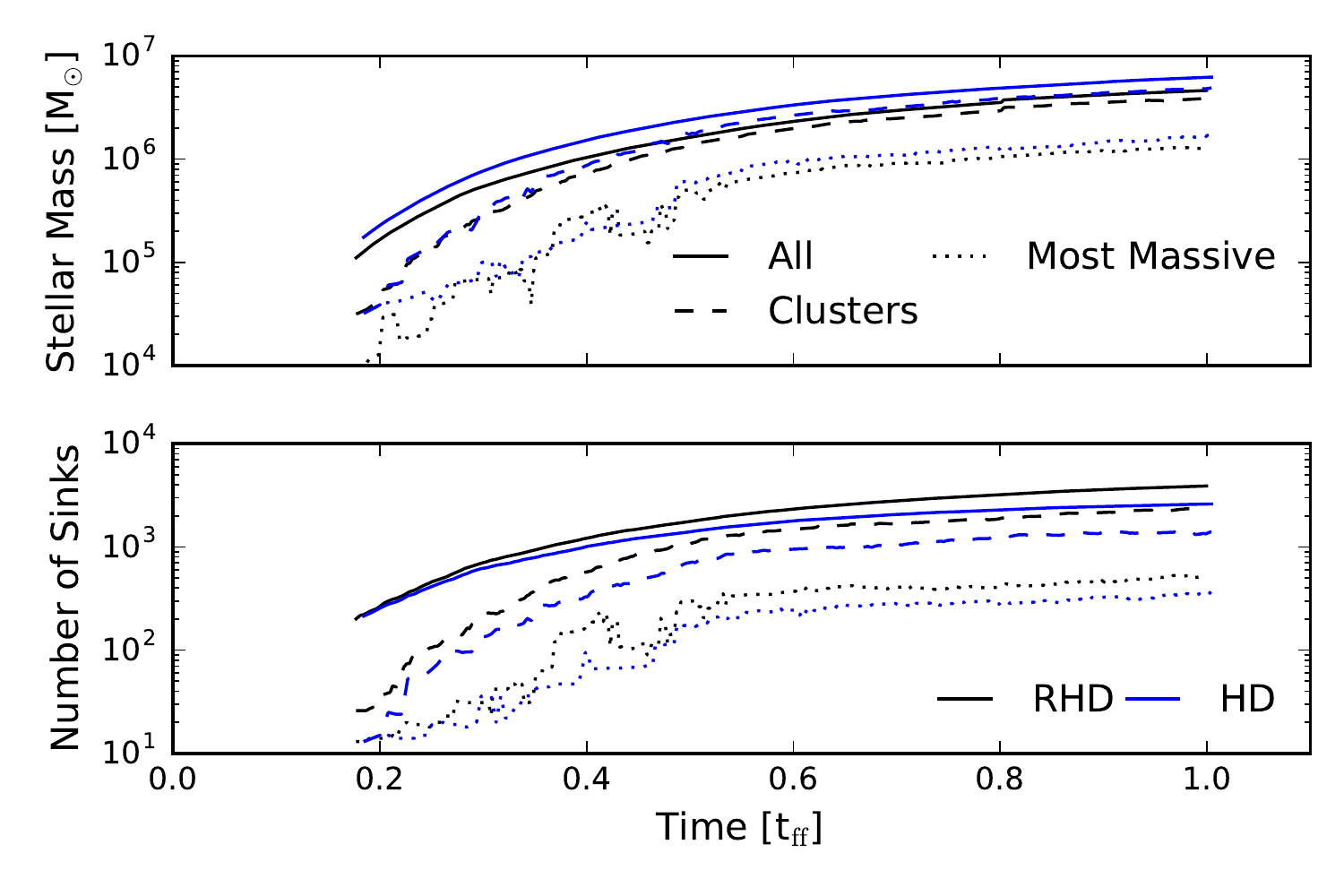}
   \end{center}
   \caption
   { Upper panel: Evolution of the total stellar mass (solid), total stellar mass in
    clusters (dashed) and in the most massive cluster
   for the RHD (black) and HD (blue) runs. Lower panel: Evolution of the total sink particle count.
    }   \label{fig:stellar_mass}
\end{figure}

\subsection{Preparation of turbulent initial conditions}
The gas and gravitational dynamics is computed with the adaptive mesh
refinement code \textsc{flash} \citep[][]{Fryxell00}, version 4.4.
We used the unsplit HLLC Riemann solver for the Euler subsystem.
Following \citet{Federrath08} we emulated an isothermal EOS at temperature $T_{\rm EOS}=480\,\text{K}$ via
a source term that cancels the $-p\nabla\cdot\mathbf{v}$ term in the
internal energy equation.
The corresponding speed of sound is
$c_{\rm s}= (k_\text{B}T_{\rm EOS}/m_p)^{1/2}\approx2\,\text{km}\,\text{s}^{-1}$.
The isothermal temperature and speed of sound are chosen to match those in
SO15 to permit direct comparison.

We prepare the initial density and velocity fields by forcing gas stochastically
with gravity disabled.  The forcing drives supersonic
turbulent fluctuations in the gas.
The initial condition preparation via stochastic driving should produce a more realistic initial density field, at least at the high Mach numbers expected in SSC-forming cloud complexes, than the common approach by imposing turbulent velocity fluctuations onto initially spherical, uniform-density clouds.

We start with a total gas mass of $10^7\,M_\odot$ at uniform
density $\bar{\rho}\approx4.4\times10^{-20}$\,g\,cm$^{-3}$
in a cubic periodic $L_{\rm box} = 25$\,pc domain.
Before turning on gravity, stochastic forcing containing
a statistical equipartition mix of solenoidal and compressive forcing modes is applied in the box
\citep[see, e.g.,][]{Federrath08}. The forcing is carried out
for five turbulence-crossing times; the crossing time is $T = \frac{1}{2} L_{\rm box} / V$, where
$V = {\cal M} c_{\rm s}$ and
${\cal M}=\langle v^2\rangle^{1/2} / c_{\rm s}$ is the mass-weighted average
Mach number of the gas flow. In this definition $\langle f \rangle \equiv \sum m_{i} f_{i} / \sum m_{i}$, where $m_{i}$ is the mass in cell $i$ and
and $f_{i}$ is
the cell-average of quantity $f$.
Although gravity is disabled, during the initial stochastic forcing the mesh is adaptively refined following the criterion of \citet{Truelove97}, namely, requiring that the nominal local Jeans length
$\lambda_\text{J}=c_\text{s}\sqrt{\pi/G\rho}$ be resolved by at least four
grid cells.
The base resolution is $64^3$ and we allow four levels of refinement,
which translate to a maximum and minimum cell size of
$\Delta x_{\rm max}\sim0.4\,\text{pc}$ and $\Delta x_{\rm min}\sim 0.02\,\text{pc}$, respectively.  Before gravity is activated, the \citet{Truelove97} criterion is satisfied
in the entire domain.

The density statistics and mass-weighted velocity dispersion of the turbulent flow reach a statistical steady state
after $\sim 2\,T$ when the Mach number is ${\cal M}\sim7.5$.
We verified that the steady-state density is approximately log-normally distributed as expected in isothermal turbulence.
The density statistics exhibits deviations from the log-normal shape that
are all consistent with \citet{Federrath08,Federrath10}.

\begin{figure*}
   \begin{center}
   \includegraphics[width=0.45\textwidth]{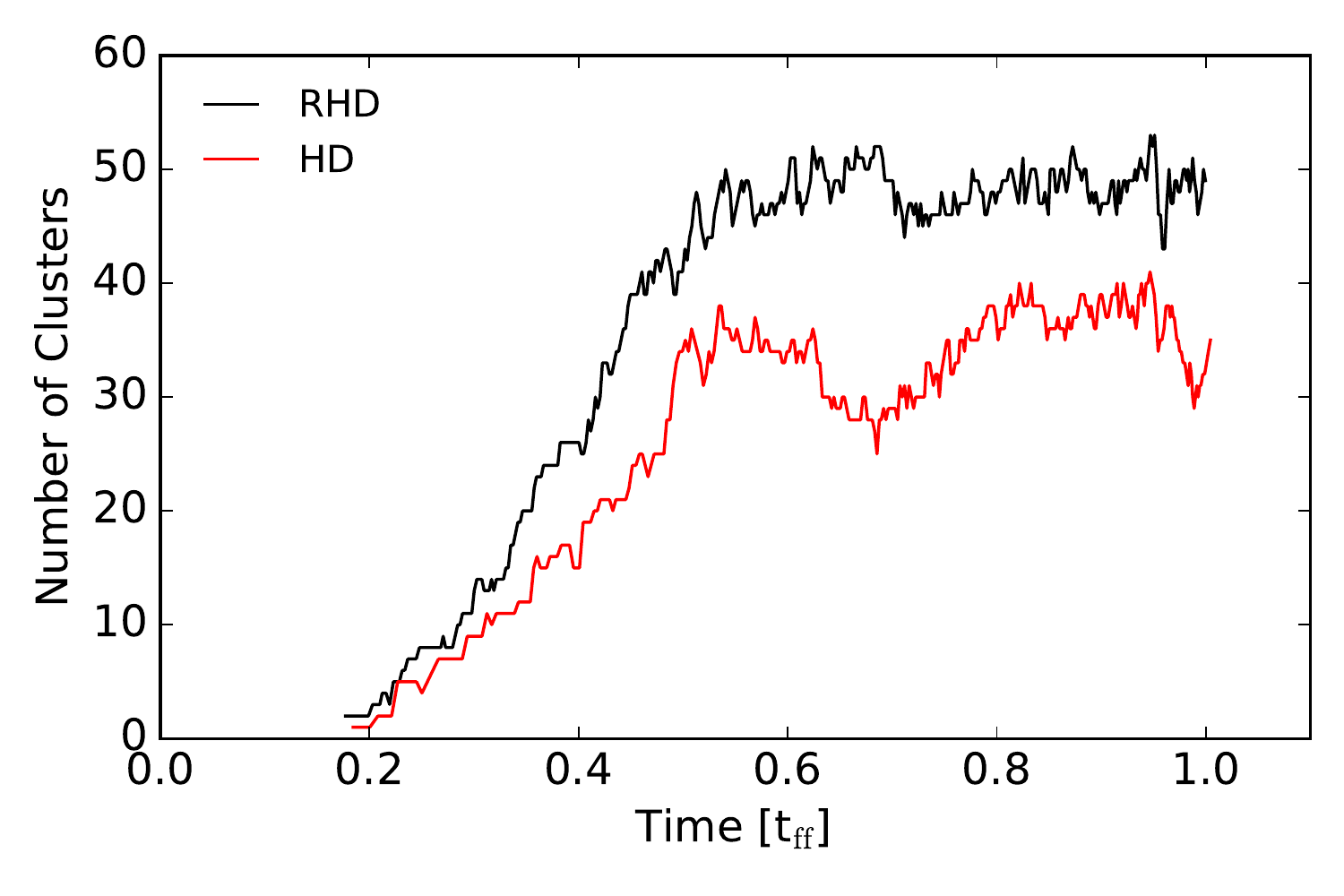}
   \includegraphics[width=0.45\textwidth]{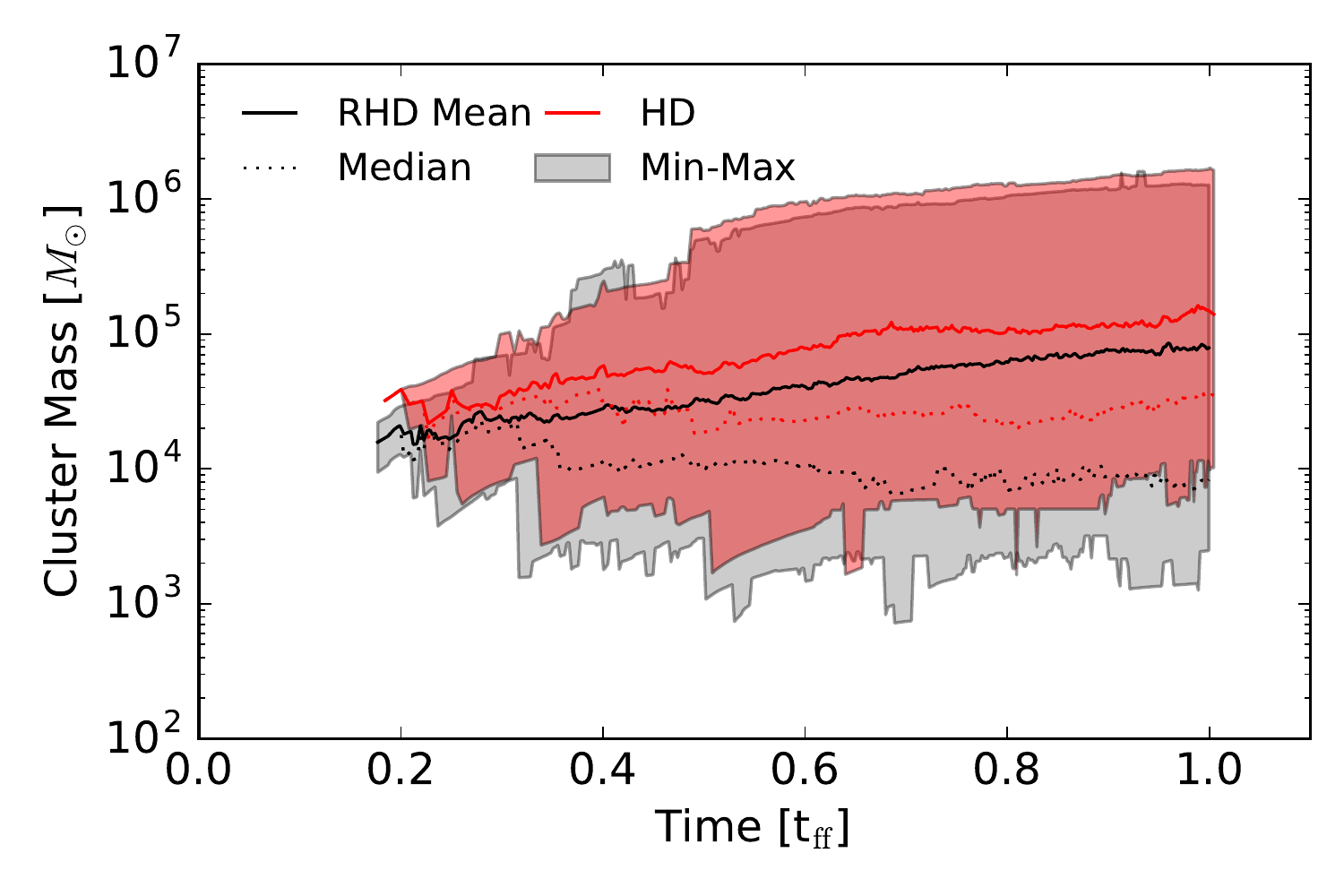}
   \end{center}
   \caption
   { Left panel: Evolution of the total number of star clusters returned from the
  DBSCAN detection routine in the RHD (black) and the HD (red) run.
     Right panel: Evolution of the range of cluster mass in the RHD (black) and the HD
   (red) run. The solid and dotted lines denote the mean and the median
   mass of all star clusters, and the shaded regions span the
   minimum and the maximum cluster mass.
    }   \label{fig:cluster_info}
\end{figure*}

\subsection{Gravitational hydrodynamics}
After the statistical steady state is identified at $5\,T$, we stop the stochastic forcing and activate gravity.  We compute the gas self-gravity with  the multigrid Poisson solver of \citet{Ricker08}.
After gravity is activated, inertial turbulent compression triggers distributed gravitational collapse.  The sites of collapse are organized into hierarchical, merging structures, just as in real observed star-forming galactic environments.
We track the formation of internally unresolved stellar groups by creating
gravitating sink particles from the gas that is too dense for its Jeans length
to be resolved.
Sink particles are introduced when the gas flow fulfills the joint criteria
for subgrid gravitational collapse \citep{Federrath10b}:
the local Jeans length is no longer resolved by $\geq 4$ grid cells even at the
highest level of refinement,
the flow is converging $\nabla\cdot\mathbf{v}<0$,
the gravitational potential has a local minimum,
absolute value of cell's gravitational energy exceeds the sum of the
internal and kinetic energies (therefore the cell is gravitationally bound),
and there are no pre-existing sink particles within the new particle's accretion radius
$r_{\rm acc} = 2.5\,\Delta x_{\rm min}$.

With our parameter choices the maximum density allowed by the \citet{Truelove97} condition is $\rho_{\rm th} = 1.3\times 10^{-17}$\,g\,cm$^{-3}$.
The sink particles are allowed to accrete gas subject to the same
criteria as for sink creation adding the requirement that the accreting gas be
gravitationally bound to the particle.

The gravitational force between the sinks and between the sinks and gas
is computed by direct summation.
Inside a small radius $r_{\rm soft}$, the sink-sink force is softened by spline kernel interpolation \citep{Federrath10b}.
The sink-gas force is softened linearly within the same radius. The gravitational acceleration on a cell due to a sink particle switches
continuously from the inverse-square law,
$a_{\rm grav}=-GM_{\rm sink}/r^2$, outside the softening radius to the linear scaling
$a_{\rm grav}=- GM_{\rm sink} r / r_\text{soft}^3$ inside.
We do not allow merging of sink particles; once formed, the sinks remain in
the simulation.
After about a few thousand sinks have formed, the cost of direct gravitational force summation
become significant.
To reduce this cost, we approximate the sink-gas gravitational
calculation by aggregating the force from the particles in distant blocks in the spirit of the \citet{Barnes86} approximation. (The blocks are the basic computational units aggregating 16$^{3}$ cells each.)
Each block is subjected to a solid-angle test comparing
$L^{2}_{\rm block}/d^{2}$, where $d$ is the particle to block center distance, to a constant value $1.53\times 10^{-3}$.  With this criterion the aggregation scheme is applied when a block at the highest refinement level is separated from a particle by a distance $d\ge5\,\text{pc}$.

\subsection{Sink particles as radiation sources}
\label{sec:sink_rad}
Each sink particle represents a group of stars sampling the entire stellar
initial mass function (IMF).
We implement a simple prescription for sink particle radiative output by fixing the
luminosity-to-mass ratio to $L_{\rm sink}/M_{\rm sink} = 3000$\,erg\,s$^{-1}$\,g$^{-1}$.
This ratio is chosen slightly higher than the value $\sim2000\,\text{erg}\,\text{s}^{-1}\,\text{g}^{-1}$
predicted by the Starburst99 models \citep{Starburst99} for the \citet{Kroupa01} IMF at solar metallicity.  The optimistic ratio is chosen to probe upper limits of dynamical forcing by radiation pressure.
The dust opacity is fixed at $\kappa_{\rm dust} = 20\,\text{cm}^{2}\,\text{g}^{-1}$,
about the value at which SO15 found radiation
pressure started to significantly influence star formation efficiency.  The opacity is consistent with the Planck mean opacity of stellar-radiation-heated,
warm ($100$--$500\,\text{K}$) dusty gas at $2$--$3$ times the solar metallicity \citep[e.g.,][]{Semenov03,Kuiper10}.
With these choices, the Eddington ratio is
$f_{\rm Edd, sph} = \kappa_{\rm dust} (L/M) / (4 \pi c G) \simeq 2.4$. A heuristic cell optical thickness threshold of $\tau_{\rm DDMC} = 5$ is set
to activate the DDMC acceleration scheme.

Radiation MCPs are transported across processors using the {\tt moveParticles}
subroutine in \textsc{flash}. The subroutine uses asynchronous, non-blocking message passing interface (MPI) calls to minimize communication overhead.
By default, \textsc{flash} estimates the total amount of work by assigning one unit of work to non-leaf
blocks and two units to leaf blocks.
It then distributes approximately equal amount of work to each processor. Sink particles are mapped to the processors hosting the blocks that contain them.
As a simulation proceeds, the clustering of sink particles can impose an asymmetric load on
isolated processors by encumbering them with unbalanced gravitational and radiation transport calculations.
To alleviate load imbalance, we modified the work estimate in the subroutine {\tt amr\_refine\_derefine} to add ten units of work to the hosting block for each sink particle.
This work allocation heuristic improved workload balance.

To avoid giving radiation pressure a non-physical
advantage over the artificially softened gravity, radiative acceleration is softened in the same way as gravity, by multiplying the amount of momentum transferred from the MCP to the gas cell with the attenuation factor $r^3/r^{3}_{\rm soft}$, where $r$ is the distance from the MCP to the cell center. Radiation MCPs are removed when they are more than $0.25\,L_{\rm box}=6.25\,\text{pc}$ away from their source sinks.

\subsection{Cluster identification}
\label{sec:cluster_membership}
Recall that sink particles represent unresolved, bound stellar groups. As the simulation progresses, the sinks become organized in viralizing clusters.
To characterize sink particle membership in clusters,
we adopt the density-based clustering algorithm
\href{http://scikit-learn.org/stable/modules/generated/sklearn.cluster.DBSCAN.html}{
{\tt DBSCAN}
}
\citep{Ester96} implemented in the {\tt scikit-learn} Python library.
The two key parameters are {\tt eps} $= 0.5\,\text{pc}$ and
{\tt n\_samples} $= 10$, which are the maximum distance between
any sinks in a cluster and the minimum number of sinks to qualify
to be a cluster, respectively.
These resolution-dependent values were chosen heuristically.
We will find that in clusters, stellar mass exceeds gas mass; therefore, in what follows we regard the center of mass of all the sinks in a cluster as the cluster's center.
\textsc{DBSCAN} is also identifies sinks not assigned to any cluster; we refer to them as `field sinks'.

\section{Results}
\label{sec:results}

%%% Basic outline of the result section.
In this section, we first characterize star formation and massive cluster properties
in the simulations.
Then we examine the impact of radiation pressure on star and cluster formation.

\begin{figure}
   \begin{center}
   \includegraphics[width=\columnwidth]{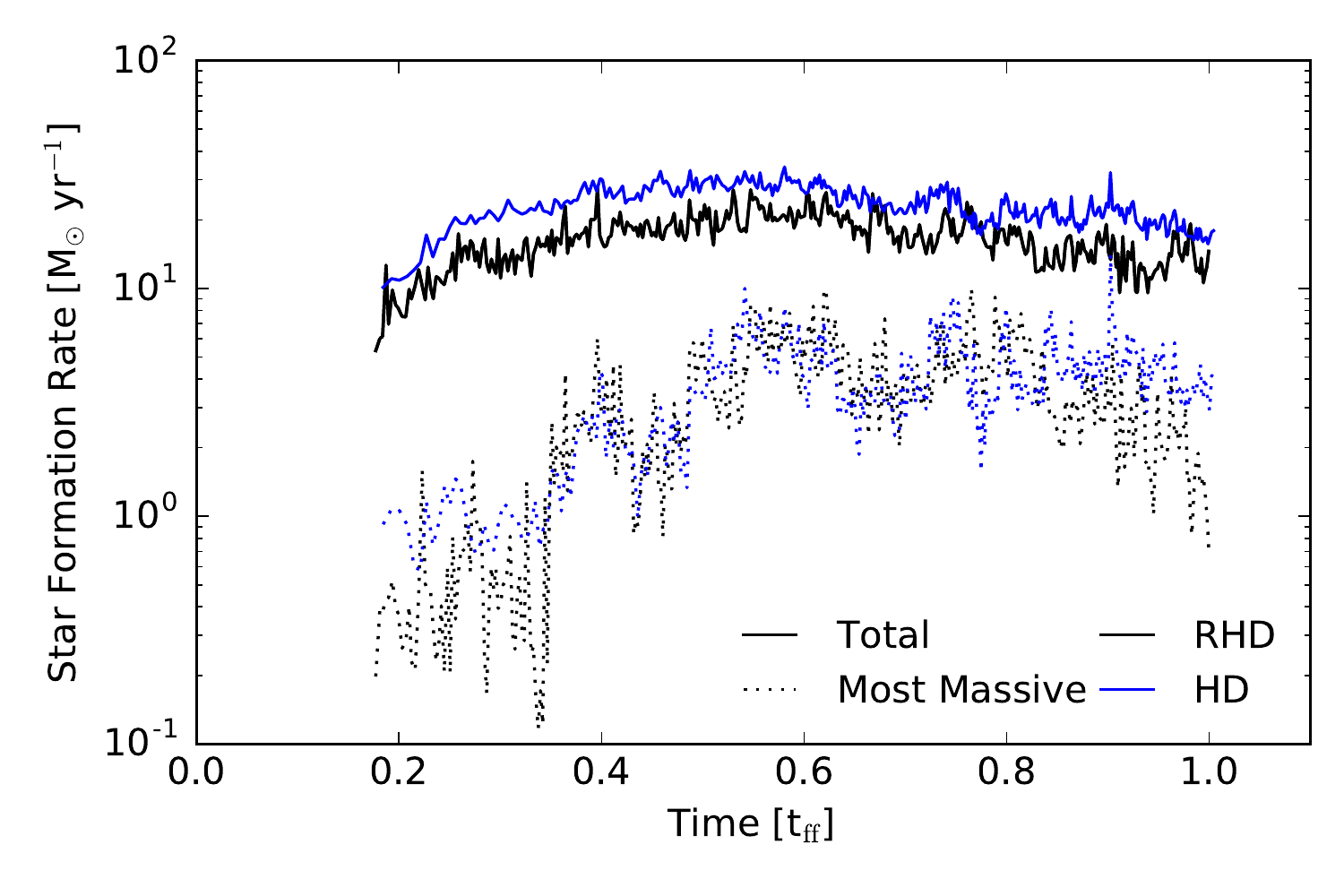}
   \end{center}
   \caption
   { Evolution of the total SFR (solid) and
     inside the most massive cluster (dotted) in the RHD (black) and
     HD (blue) run. The SFR includes the gas accretion onto existing sink particles and the creation of new sinks.
     In both simulations accretion onto existing sinks
      dominates the SFR.
    }   \label{fig:sfr}
\end{figure}

\subsection{Star formation}

The very first sink particles form shortly after gravity has been activated in sites where flow convergence compresses gas to high density.
Figure \ref{fig:dens_projs} shows the density-weighted density projections of
the entire simulation domain along the $y$-axis at
$(0.25,\,0.5,\,0.75,\,1.0)\,t_{\rm ff}$,
in the RHD and HD runs.
In both runs, the first sink particles form in the center of the densest gas clump
after $\sim5.4\,\text{kyr} \simeq 0.017\,t_{\rm ff}$ from the moment gravity was switched on.
Sink particles
subsequently form along the gas filaments as seen in the
$t =0.25\,t_{\rm ff}$ and $0.5\,t_{\rm ff}$ snapshots.
The simulations are carried out to $1\,t_{\rm ff}\sim0.3\,\text{Myr}$.
Gas flow morphology and visual sink particle clustering in the two runs remains
similar on large scales.
Figure \ref{fig:dens_projs_zoomed} shows the corresponding zoomed-in views
of the most massive clusters and their half-mass radii.
Sinks form along the filaments but a large scale gravitational field then pulls them into the nearby virialized clusters.\footnote{\citet{SafranekShrader16} describe a very similar pattern of gravitational infall, fragmentation, and virialization, but there in lower-mass protogalactic clouds formed through thermal instability.}
The collapsing structures have angular momentum and rotate.

The top panel of Figure \ref{fig:stellar_mass} shows the total stellar mass growth in both runs.
The stellar masses grow smoothly to reach $4.6\times10^{6}$ ($6.2\times10^{6}\,M_{\odot}$) in
the RHD (HD) run. These final stellar masses correspond to box-wide star formation efficiencies of 46\% (62\%). The total stellar mass in the HD run is consistently $\sim1.3$--$1.5$ times higher than in the RHD.  Section \ref{sec:roles_rad_pres} attributes this effect to radiation pressure.
Since our sinks do not merge, the number of sinks also increases smoothly
to a final total of $\sim3900$ ($\sim2600$) in the RHD (HD) run.
The lower panel of Figure \ref{fig:stellar_mass} shows the
evolution of sink particle counts.
In Section \ref{sec:cluster_prop} we ascribe the larger number of sinks in the RHD simulation to the radiation suppressing accretion onto existing sinks while leaving more gas to collapse into new sinks.
Indeed the average sink mass is systematically higher in the radiation-free
HD run.

We follow the procedure described in Section \ref{sec:cluster_membership}
to group sink particles into clusters and a separate unclustered population of field sinks.
Since each sink is itself an unresolved stellar group, the field sinks should not be construed as isolated stars, but are small clusters.
Natural compact clusters dynamically evaporate but our subgrid prescription does not model sink evaporation.
The most massive clusters acquire a significant fraction of the total stellar mass by $t = 0.5\,t_{\rm ff}$ and have stellar masses
$1.3\times10^{6}$ ($1.6\times10^{6}\,M_{\odot}$)
at the end of the RHD (HD) simulations.
The fluctuations of the measured stellar mass in the most massive cluster are an artifact of unstable cluster membership identification during cluster merging. In Figure \ref{fig:cluster_info},
we show the evolution of the cluster count and the cluster mass range.
The cluster mass ranges between
$\sim 10^{3}\,M_{\odot}$ and $\sim 10^{6}\,M_{\odot}$.
The low median cluster mass shows that most of the clusters have low masses.

\begin{figure}
   \begin{center}
   \includegraphics[width=\columnwidth]{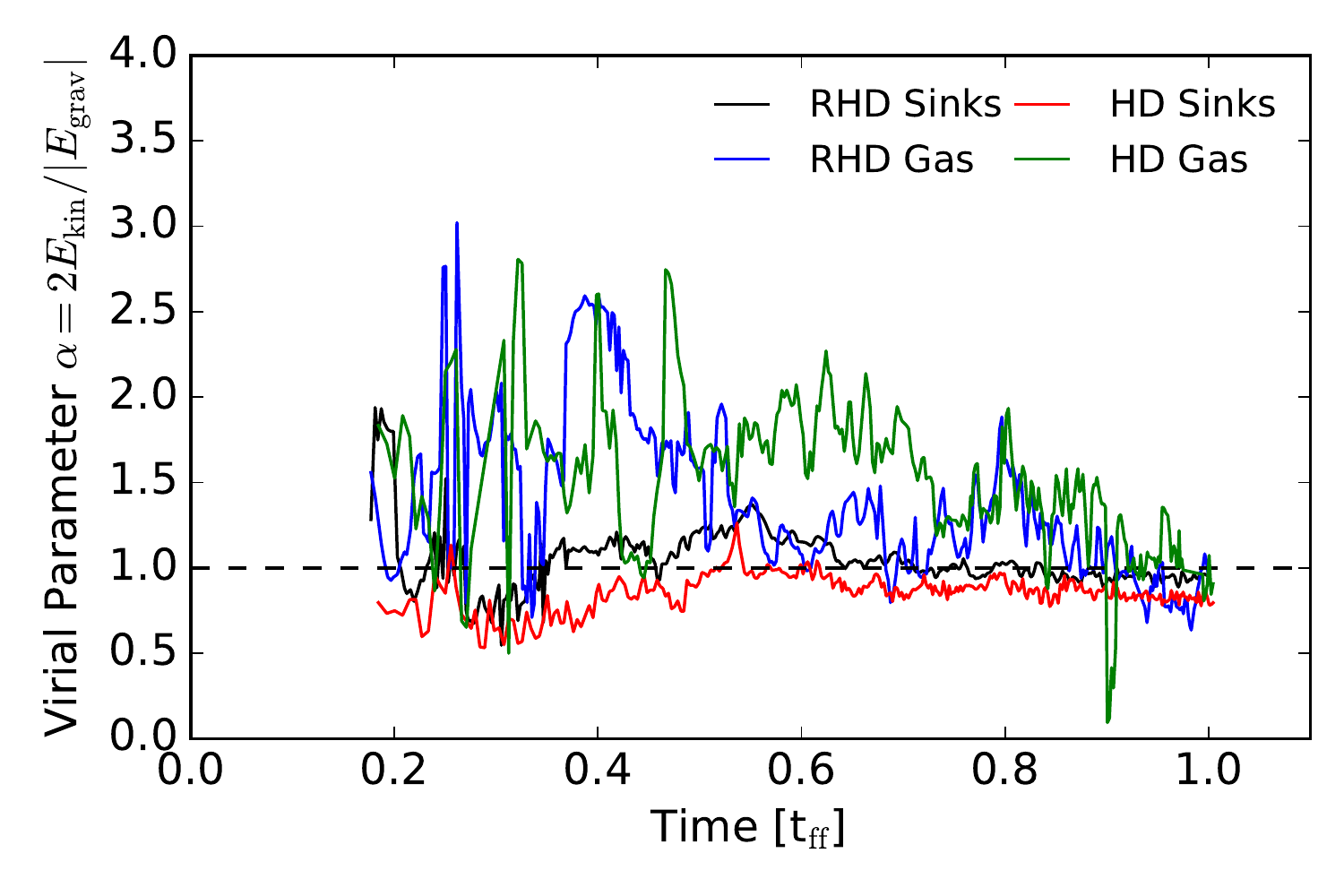}
   \end{center}
   \caption
   { Evolution of the virial parameter
     $\alpha_{\rm vir} = 2 E_{\rm kin} / |E_{\rm grav}|$
     of the stellar and gas components inside the most massive cluster in
     the two runs.
     In both runs, the star clusters represented by the sink particles
     virialize about 0.6\,$t_{\rm ff}$ after the formation of the first sink
     and maintain the virialized state while they grow in mass.
     The gas component is bound ($\alpha_{\rm vir} < 2$) to the sink cluster after about
     0.5\,$t_{\rm ff}$.
    }   \label{fig:virial_par}
\end{figure}

A comparison of the global star formation rates (SFR) in both runs is shown
in Figure \ref{fig:sfr}.
The SFRs account for the
formation of new sinks and the accretion of mass onto existing sinks.
Throughout the simulations, gas accretion on existing sinks dominates new sink
creation.
The global SFRs inside the simulation boxes increase sharply at early times to
$\ge 10\,M_{\odot}\,\textrm{yr}^{-1}$ within $\sim0.2\,t_{\rm ff}$ and peak at
$\sim 20\,M_{\odot}\,\textrm{yr}^{-1}$ and
$\sim 30\,M_{\odot}\,\textrm{yr}^{-1}$ in the RHD and HD runs, respectively.
The SFRs then decline slightly, but the strong gravitational collapse still
keeps the SFRs above 10\,$M_{\odot}\,\textrm{yr}^{-1}$ in both runs.
Figure \ref{fig:sfr} shows that although the global SFR is
$\sim$1.2--2 times higher in the non-radiative run, the SFRs inside the
most massive clusters are almost identical between the two runs.
This is the first indication that \emph{radiation is not deleterious to a massive cluster's---SSC's---ability to continue forming stars}.
If we further decompose the total SFR into the clusters vs.\ the field, we find that
similar to the trend seen in the stellar mass, clusters contribute a larger fraction to the total SFR
in the RHD run.
As sinks become incorporated in virialized clusters where they move fast and are subject to frequent two-body encounters that can scatter them to still higher velocities, an increasing fraction of sinks, especially those in clusters, ceases to accrete.
While particles prefer to accrete outside clusters where they move slowly relative to gas, it seems that it is also outside clusters that radiation pressure has the greatest impact on the local gas density and  separately suppress accretion.

\subsection{Properties of clusters}
\label{sec:cluster_prop}

In this section we analyze the properties of the clusters in the simulations. Figure \ref{fig:virial_par} shows the
virial parameter $\alpha_{\rm vir} = 2 E_{\rm kin} / |E_{\rm grav}|$,
where $E_{\rm kin}$ and $E_{\rm grav}$ are the total kinetic and gravitational
potential energy of the gas or sink component inside the half-mass radius.
Since the stellar mass largely dominates over the gas mass, we do not include the contribution from gas self-gravity in the
$E_{\rm grav}$ term.
Independent of whether radiation pressure is enabled, the stellar component
virializes quickly, at about 0.6\,$t_{\rm ff}$.
The gas component virializes later and remains gravitationally
bound to the stellar component.

\begin{figure}
   \begin{center}
   \includegraphics[width=\columnwidth]{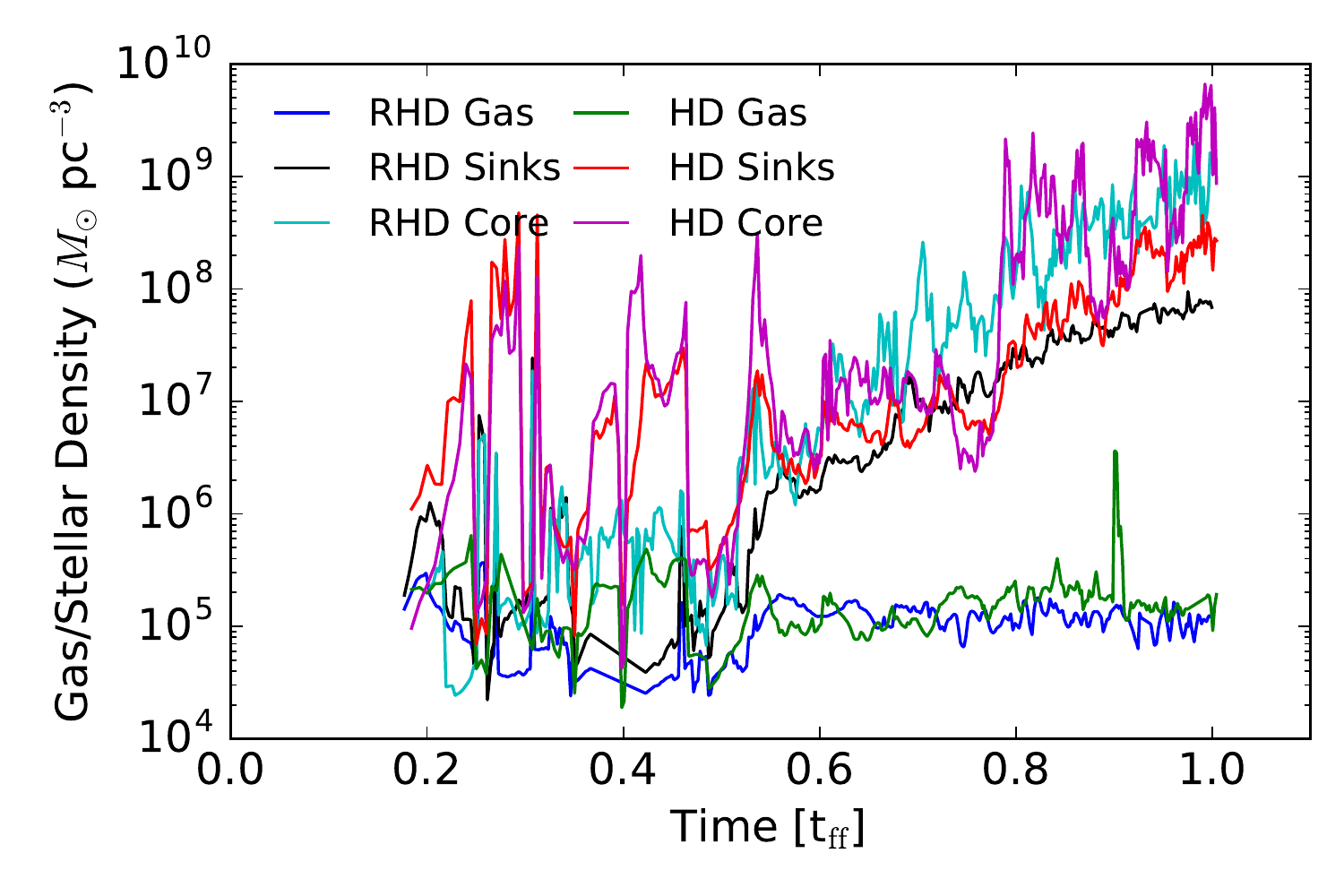}
   \end{center}
   \caption
   { Evolution of stellar and gas densities measured within the
     half-mass radius of the most massive cluster.
     Since the massive cluster is centrally-concentrated,
     we also estimate the peak
     stellar density by averaging the density of 10
     sink particles closest to the cluster's center of mass.
     Stellar mass dominates gas mass during most of the
     simulations.
     The peak stellar density reaches
     $\sim 10^{8}\,M_{\odot}\,\textrm{pc}^{-3}$ where
      runaway stellar merging can produce a very massive star.
     The gas density is near-constant at
     $1-2\times 10^{5}\,M_{\odot}\,\textrm{pc}^{-3}$ after 0.5\,$t_{\rm ff}$.
    }   \label{fig:stellar_gas_dens}
\end{figure}

Figure \ref{fig:stellar_gas_dens} plots the gas and stellar density within the half-mass radius of the most massive cluster.
By $t\simeq1\,t_{\rm ff}$ the gas density in both runs saturates
at a near-constant value $\rho_{\rm gas}\approx 10^{5}\,M_{\odot}\,\text{pc}^{-3}$.
The stellar
density reaches
$\rho_{*} \approx 6\times10^{7}\,M_{\odot}\,\textrm{pc}^{-3}$ in the RHD run
and $\rho_{*} \approx 2\times10^{8}\,M_{\odot}\,\textrm{pc}^{-3}$ in the HD run.
The increase of stellar density near the end of the simulation is
likely the result of dynamical relaxation.
The peak stellar density is lower in the RHD run.  From this we can conclude that radiation pressure does seem to influence cluster structure. However, radiation does not limit the density as proposed in the literature \citep{Hopkins10}.
Interestingly, stellar density is such that collisional formation of very massive stars (VMSs) is expected \citep[][]{MC11}.

In Figure \ref{fig:stellar_gas_dens} we further estimate the core stellar density in clusters by measuring density in the sphere enclosing 10 sink particles closest to
the cluster center.
The peak density
reaches $8\times10^{8}\,M_{\odot}\,\text{pc}^{-3}$ by the end of the simulations.
Thus, radiation pressure does not
seem to have a large impact on the final peak density.
Note that this extreme stellar density is reached very early-on in the
clusters' evolution at $t = 1\,t_{\rm ff} = 0.3\,\text{Myr}$,
well before the arrival of the first supernovae.

Since the sink particles exhibit a significant mass spread, we expect mass segregation should occur.  Mass segregation can shorten the core collapse time scale below the nominal two-body relaxation time.
\emph{Since sink particle masses are resolution-dependent quantities, the degree of mass segregation in the simulations is not predictive of mass segregation in real clusters.} Nonetheless, to understand the simulation we must quantify the effect of segregation.  We estimate it using the widely adopted minimum spanning tree (MST)
method \citep{Allison09,Olczak11,MaC11}.\footnote{For a comparison of different
mass segregation estimators see \citet{PG15}.
For a hybrid approach that quantifies mass segregation combining
the MST method and a $k$th nearest
neighbor density estimator see \citet{Yu17}.}
An MST is the minimum total edge length of an undirected acyclic graph connecting a set of sinks.
Our analysis follows the standard approach to quantify the level of mass segregation
in a star cluster using the MST ratio $\Lambda_{{\rm MSTR}, m} = \bar{l}_{{\rm random}, m}/l_{{\rm massive}, m}$,
where $l_{{\rm massive}, m}$ is the total edge length of the MST connecting the $m$ most
massive sinks in the cluster,
and $\bar{l}_{{\rm random}, m}$ is the same but averaged over 100 sets of $m$ sinks
randomly selected independent of sink mass.
When $\Lambda_{{\rm MSTR}, m} > 1+{\cal O}(1/\sqrt{m})$,
the $m$ most massive sinks are likely to be sampling a more compact region than the region sampled by the whole cluster.
We pick $m = 10$ and construct the MST using the
\href{https://docs.scipy.org/doc/scipy/reference/sparse.csgraph.html}
{\tt csgraph}
implementation of the
{\tt Scipy} library.
From $t \ge 0.5$\,$t_{\rm ff}$ onwards, we observe a mean ratio of
$\Lambda_{\rm MSTR} \sim 1.6-1.7$ in both runs, suggesting a moderate degree of mass segregation.
These values are not sensitive to the precise choice of $m$ for $10 \leq m \leq 200$.

\begin{figure}
   \begin{center}
   \includegraphics[width=\columnwidth]{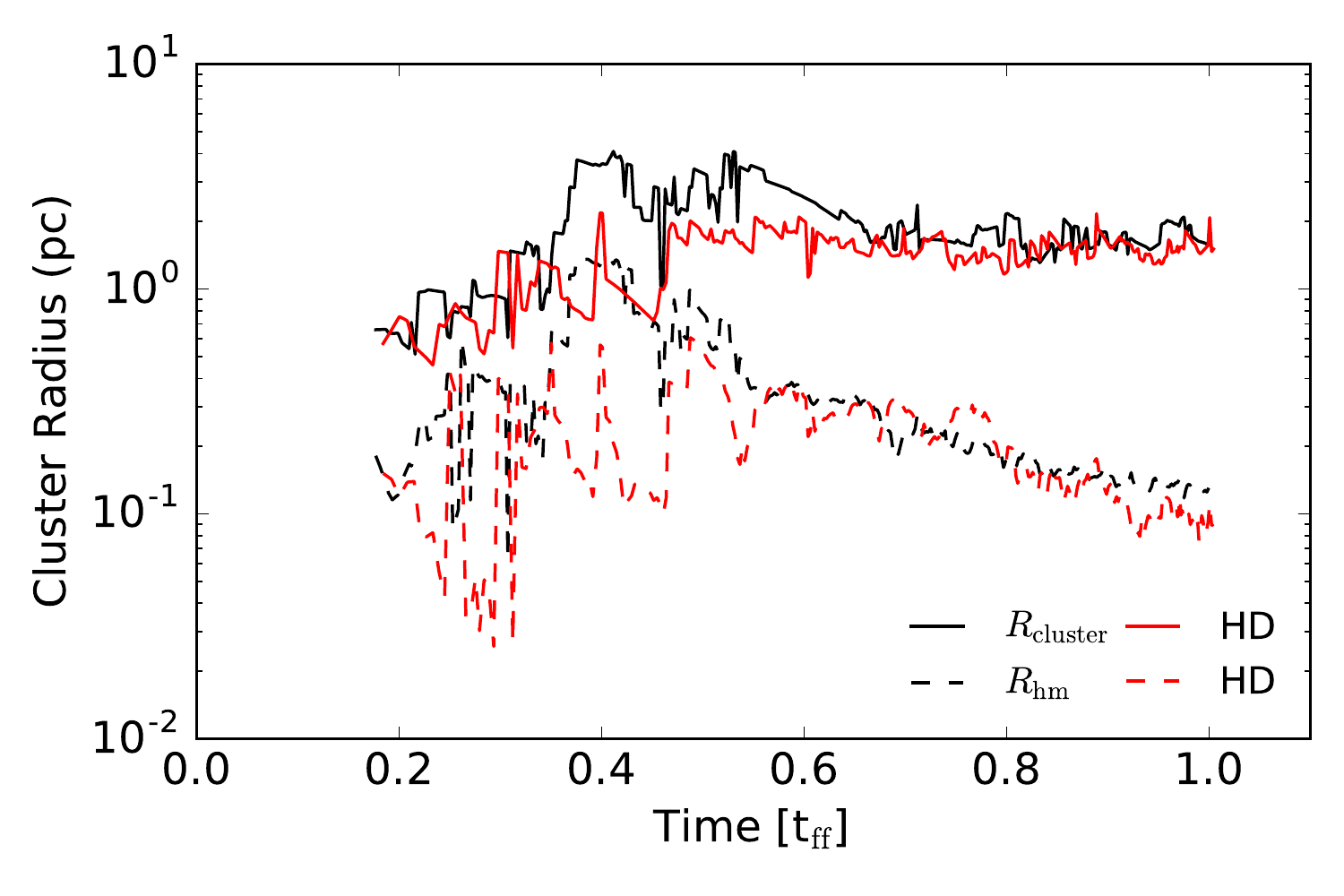}
   \end{center}
   \caption
   { Evolution of the most massive cluster radius and the half-mass radius in both runs.
    }   \label{fig:radii_time}
\end{figure}

\begin{figure*}
\begin{center}
\includegraphics[width=0.49\textwidth]{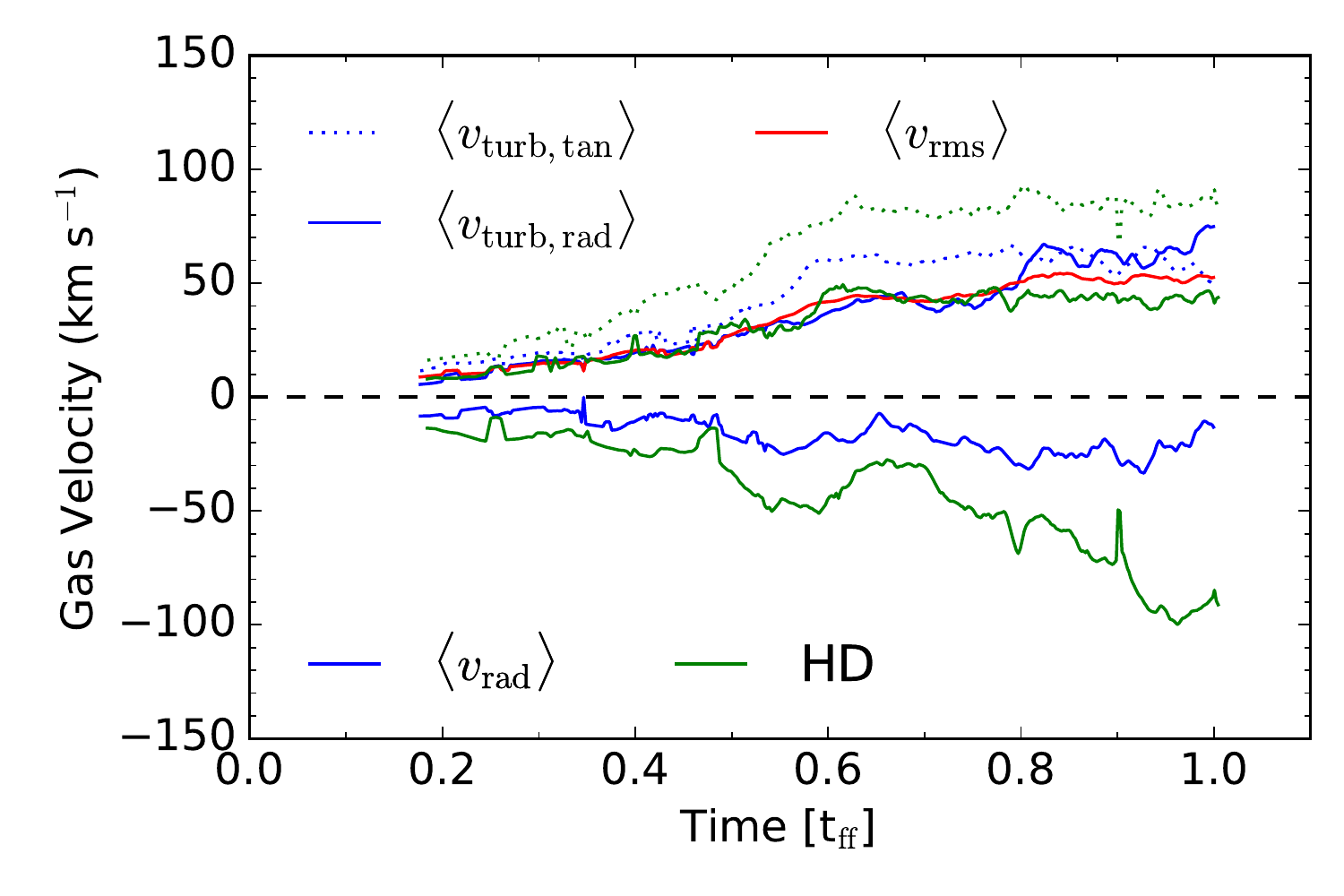}
\includegraphics[width=0.49\textwidth]{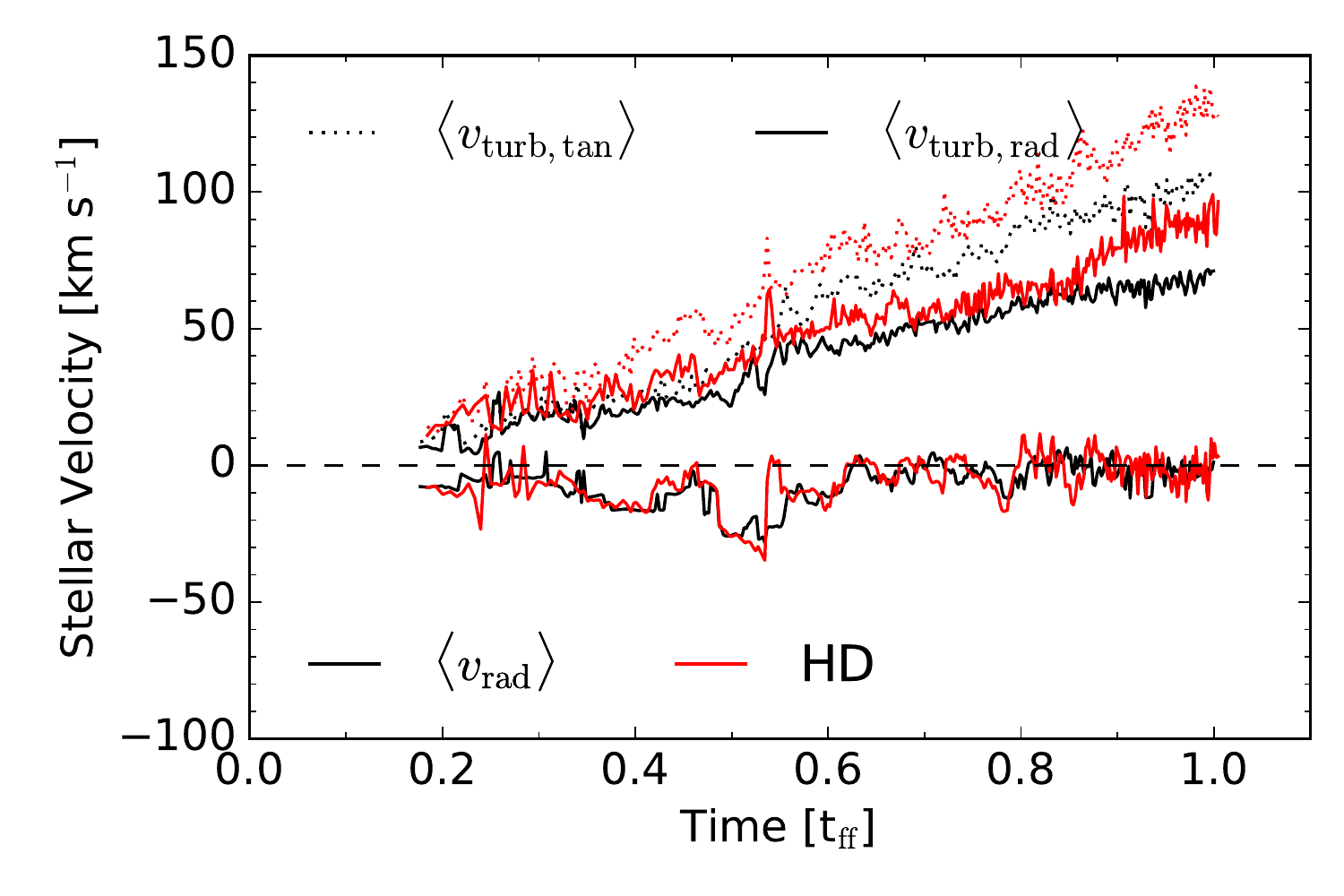}
\end{center}
\caption{
Left panel: Evolution of mass-weighted gas velocity moments in Section \ref{sec:cluster_prop}, measured inside a 1\,pc sphere centered on the most massive cluster. The red solid line is the one-dimensional velocity dispersion.
Right panel: Evolution of the corresponding particle-mass-weighted sink particle velocity moments as indicated in the legend.}
\label{fig:velocities_time}
\end{figure*}

Radio and millimeter observations provide access to the
kinematics of massive star- and cluster-forming clumps at
high angular resolution and sensitivity \citep{Herrera17,Ginsburg17,Turner17}. To enable the comparison of our modeling to observations, we investigate the velocity structure of the
gas and stellar component in the clusters.
The left panel of Figure \ref{fig:velocities_time} shows the evolution of
the mass-weighted mean radial streaming velocity and the mass-weighted root-mean-square radial and tangential turbulent gas velocities in spheres centered on the most massive star clusters
\begin{align}
  \langle {v}_{\rm rad} \rangle &= \frac{\sum_j m_{j} \mathbf{v}_{j}\cdot\hat{\mathbf{r}}}{\sum_j m_{j}},\\
   \langle {v}_{\rm turb,rad} \rangle &= \sqrt{\frac{\sum_j m_{j} |\mathbf{v}_j\cdot\hat{\mathbf{r}}-\langle v_r\rangle\hat{\mathbf{r}}|^2}{\sum_j m_{j}}},\\
  \langle {v}_{\rm turb,tan} \rangle &= \sqrt{\frac{\sum_j m_{j}( |\mathbf{v}_j|^2-|\hat{\mathbf{r}}\cdot\mathbf{v}_j|^2)}{\sum_j m_{j}}},
\end{align}
where $m_{j}$ and $\mathbf{v}_j$ are the mass and mean-cluster-motion-subtracted velocity
in cell $j$.
The sums are taken over the cells with centers within 1\,pc from the cluster center.

The HD and RHD runs differ in the evolution of the mass-weighted average radial streaming velocity $\langle v_{\rm rad} \rangle$.  The infall velocity in the radiation-free run becomes as high as $\sim 100\,\text{km}\,\text{s}^{-1}$, whereas in the run with radiation it never exceeds $\sim 30\,\text{km}\,\text{s}^{-1}$.  Since this difference cannot be explained by the radiation-free run's higher mass and central density, it seems that radiation pressure is slowing down radial infall.
The radial turbulent velocities in the two runs are
comparable until $t \sim 0.8\,t_{\rm ff}$, whereupon in the RHD run the radial turbulent velocity reaches
$\sim60$\,km\,s$^{-1}$ and the HD run remains at a somewhat lower value $\sim40$\,km\,s$^{-1}$.
The tangential velocity $\langle v_{\rm turb,tan}\rangle$
is systematically higher in the HD run.
 In Figure \ref{fig:velocities_time}, right panel, we plot the equivalent kinematical indicators, but for the sink particles.
Note that after the RHD cluster virializes at $t \sim 0.6$\,$t_{\rm ff}$,
its particle-mass-weighted mean radial streaming velocity approximately vanishes, as expected.
The velocity dispersions are larger in the HD run because the radiation-free run converts more gas into sink particles and has a larger virial mass.  We summarize properties of the final most massive cluster in the RHD run in Table \ref{tab:most_massive_cluster}.

\begin{table}
  \centering
  \begin{tabular}{lc}
  \hline \hline
  Property & Value\\
  \hline
        Stellar mass    & $1.3\times10^6\,M_\odot$ \\
        Residual gas mass & $1.4\times10^5\,M_\odot$\\
        Fiducial cluster radius & $1.6\,\text{pc}$\\
        Half-mass radius & $0.1\,\text{pc}$\\
        Stellar velocity dispersion & $73\,\text{km}\,\text{s}^{-1}$\\
        Gas velocity dispersion & $53\,\text{km}\,\text{s}^{-1}$\\
        Star formation rate & $1\,M_\odot\,\text{yr}^{-1}$\\
    \hline
  \end{tabular}
   \caption{Properties of the most massive cluster at the end of the RHD run, $0.3\,\text{Myr}$ after the beginning of star formation.}
\label{tab:most_massive_cluster}
\end{table}

\subsection{The impact of radiation pressure}
\label{sec:roles_rad_pres}

Since star formation efficiency is lower in the RHD run than
in the radiation-free HD control run, and since both simulations are artificially set isothermal, one can conclude that to a degree, radiation pressure inhibits star formation.
However, although we chose optimistically high values of dust opacity and stellar luminosity-to-mass ratio, \emph{radiation pressure did not truncate star formation}.
Even at the end of the RHD simulation, gas accretion onto sink particles
persists and new sinks are forming. Regardless of the epoch or region within the simulation, the SFR is depressed by at most a factor of 2 in the RHD compared to the HD run. Since this is at odds with certain theoretical predictions \citep{MQT10,Kim16}, in what follows we closely examine the transport of stellar radiation.

In the core of the most massive star cluster formed in the RHD simulation, gas distribution is highly inhomogeneous.  This allows radiation to leak out through
low-density channels.  The leakage reduces the coupling of the radiation pressure force with the dense clumps that contain most of the gas mass.
The radiation-matter anti-correlation in strongly inhomogeneous gas limits
radiation pressure's effect on the gas supply for star formation.
The anti-correlation was observed in previous multi-dimensional simulations of
star cluster formation \citep{SO15}, plane-parallel radiation-pressure-driven outflows \citep{KT12,KT13,Davis14,RT15,TM15,ZD17} and massive star envelopes \citep{Jiang15}.

\begin{figure}
   \begin{center}
   \includegraphics[width=\columnwidth]{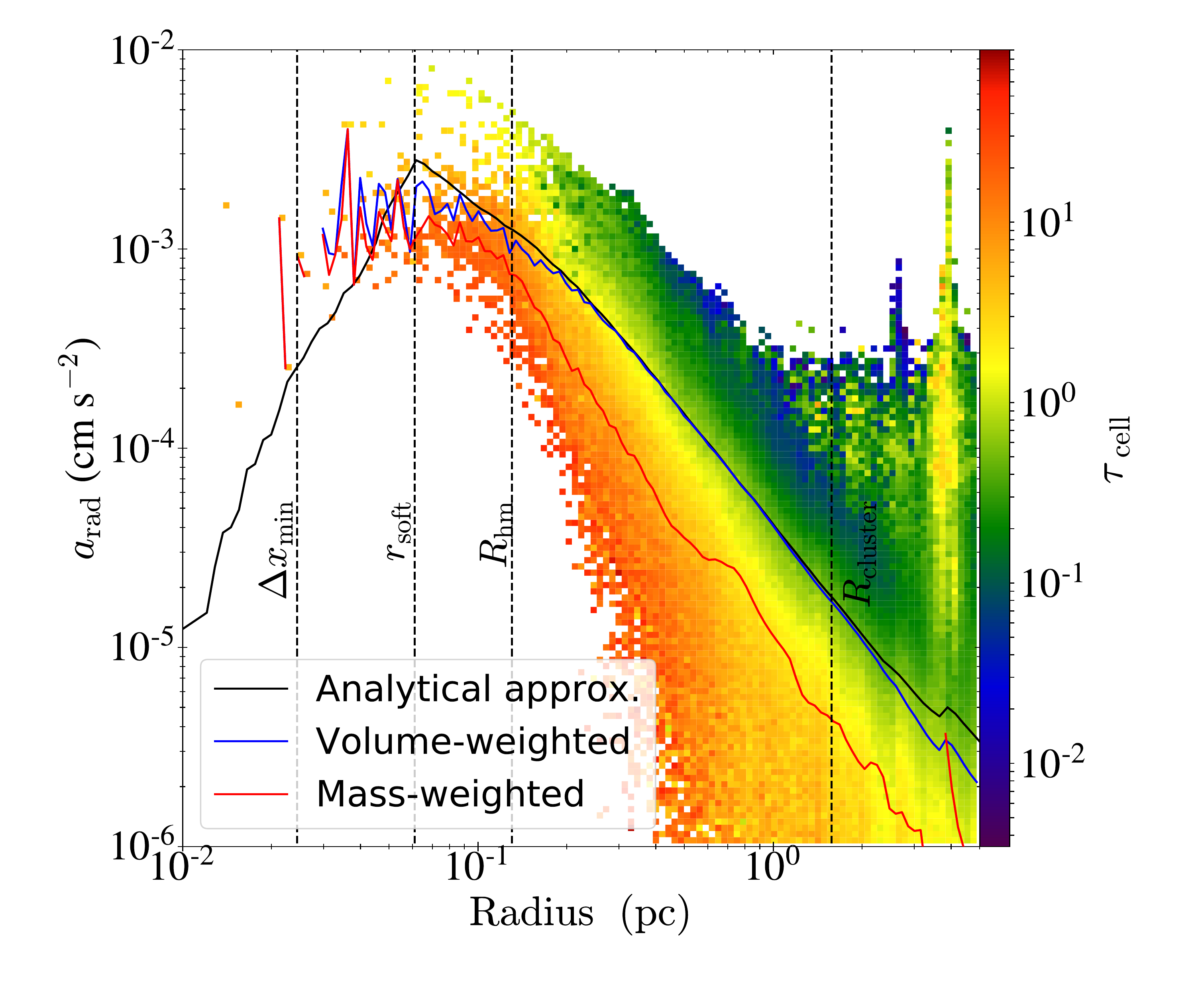}
   \end{center}
   \caption
   { Radiative acceleration and cell optical thickness as a function radius from the center of the most massive cluster at $t = t_{\rm ff}$ in the RHD run.
     The pixel color represents the average cell optical depth.
     The black solid line is the analytical acceleration assuming that stellar point sources are at the center.
     The red and blue lines are, respectively, profiles of mass- and volume-weighted radiative acceleration.
     The vertical dashed lines, from left to right, mark the minimum cell width,
     softening radius, cluster's half-mass radius, and cluster radius.
    }   \label{fig:arad_phaseplot}
\end{figure}

Figure \ref{fig:arad_phaseplot} plots volume- and mass-weighted spherically-averaged radiation pressure acceleration in the RHD run.  The spherical averaging is done around the center of the most massive cluster at $t = t_{\rm ff}$.
Also shown is the spherically symmetric analytical profile for an equivalent central source:
\begin{equation}
  a_{\rm rad, approx} (r) = \frac{\kappa_{\rm dust} \sum_{i}{L_{i}}}{4 \pi r^2 c},
\end{equation}
where the summation runs over the sink particles within $r$.
While the volume-weighted radiative acceleration closely follows
the analytical estimate,
the mass-weighted acceleration, a proxy for the actual radiation
pressure that the gas experiences, consistently lies below the analytical prediction.
To visualize the effect of radiation-matter anti-correlation, we also overlay the radial
profiles of the radiative acceleration on the color rendering of cell optical depths.
Most of the gas with $\tau_{\rm cell} > 1$ experiences a substantially reduced radiation pressure force whereas the gas with $\tau_{\rm cell} < 1$ is more strongly accelerated
than the analytical estimate.

\begin{figure}
   \begin{center}
   \includegraphics[width=\columnwidth]{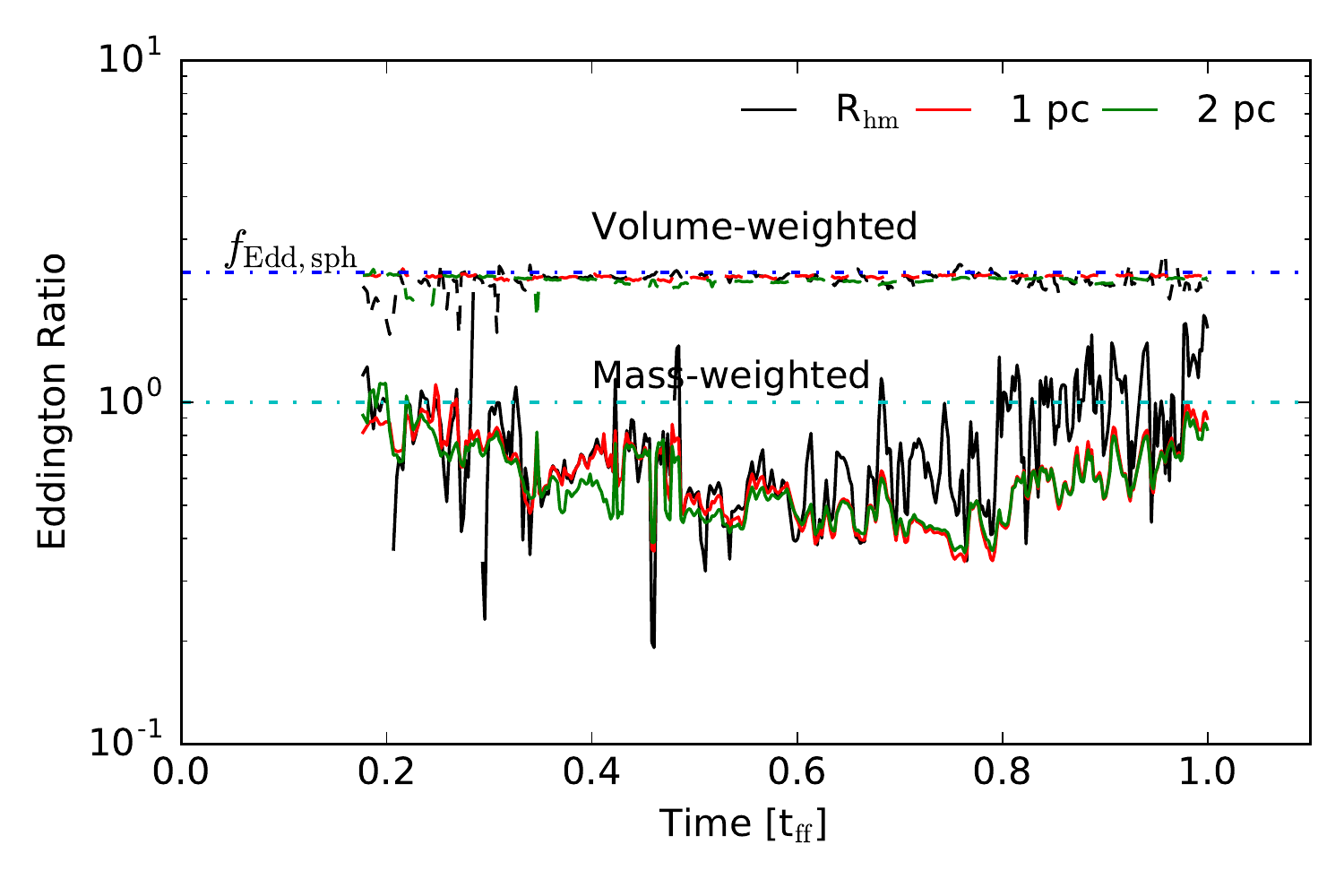}
   \end{center}
   \caption
   { Evolution of the mass-weighted (solid) and volume-weighted (dashed)
     Eddington ratio measured from the center of the most massive cluster in the RHD run.
     The ratios are calculated within spheres of three radii, the cluster's
     half-mass radius (black), 1\,pc (red), and 2\,pc (green).
     The blue dot-dashed line is the analytical value calculated assuming spherical symmetry.
     The flux-density anti-correlation reduces the mass-weighted average ratio to
     below unity throughout the entire simulation, while the volume-weighted
     value approximates the analytical estimate.
    }   \label{fig:fEdd_time}
\end{figure}

This inhomogeneity-enabled reduction in the effective radiation pressure
persists throughout the simulation.
We define mass- and volume-weighted Eddington ratio within a sphere of
radius $R$
\begin{equation}
  f_{\rm Edd}(R) = \frac{\langle a_{\rm rad} \rangle_{R}}
                        {\langle a_{\rm grav} \rangle_{R}},
\end{equation}
where the averages, weighted either by the cell mass or volume, are over the cells within the sphere.
We only include sink-gas gravitational attraction in the $a_{\rm grav}$ term;
it is a good approximation as stellar mass dominates.
Figure \ref{fig:fEdd_time} shows the evolution of the
average Eddington ratios around the most massive cluster computed within three
radii: the cluster's time-dependent half-mass radius, 1\,pc, and 2\,pc.
It shows that the radiation-matter anti-correlation renders the mass-weighted
ratio sub-Eddington whereas the volume-weighted ratio is super-Eddington
and close to the spherically-symmetric value $f_{\rm Edd, sph}$
(Section \ref{sec:sink_rad}).

\begin{figure*}
\begin{center}
\includegraphics[width=0.33\textwidth]{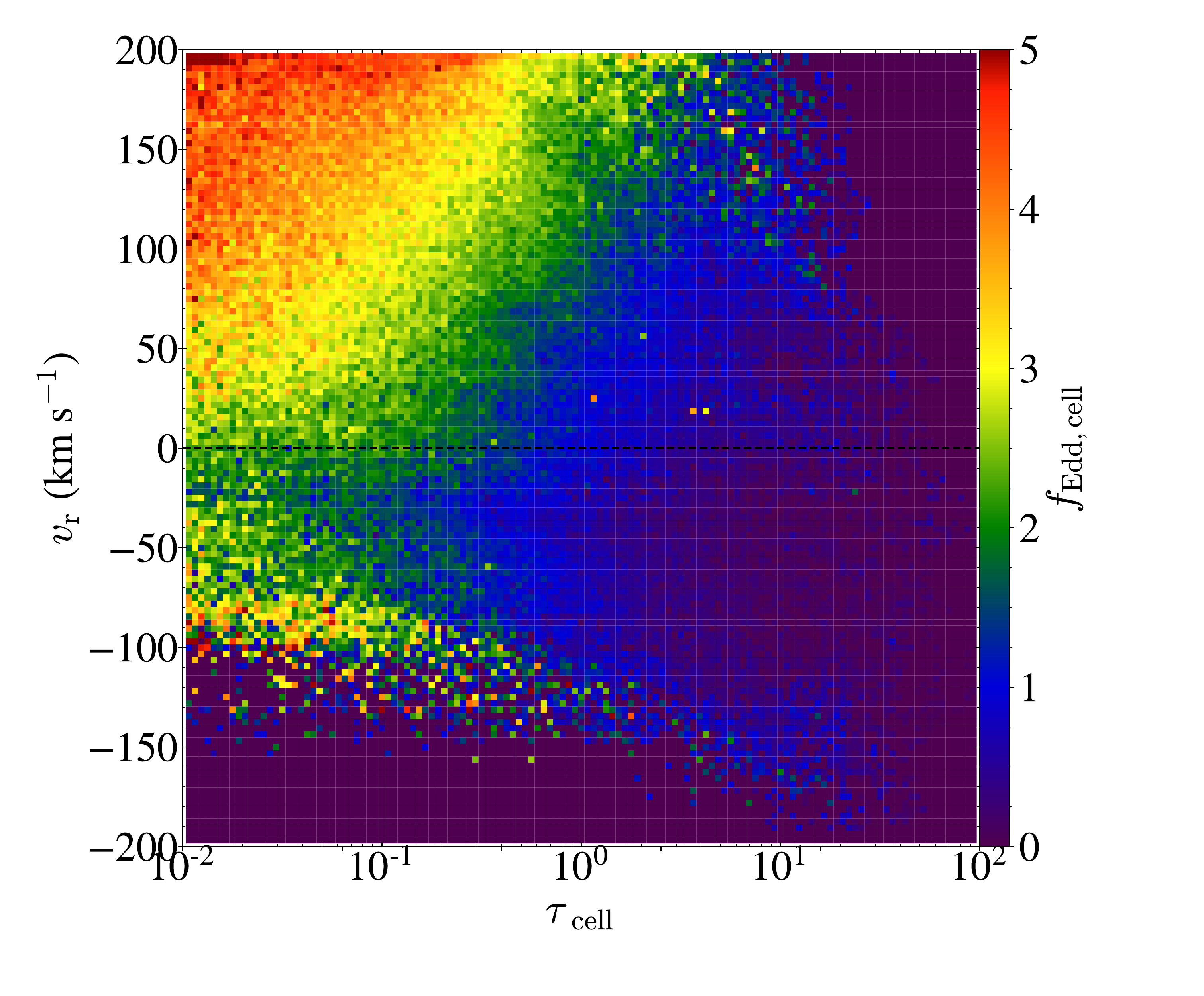}
\includegraphics[width=0.33\textwidth]{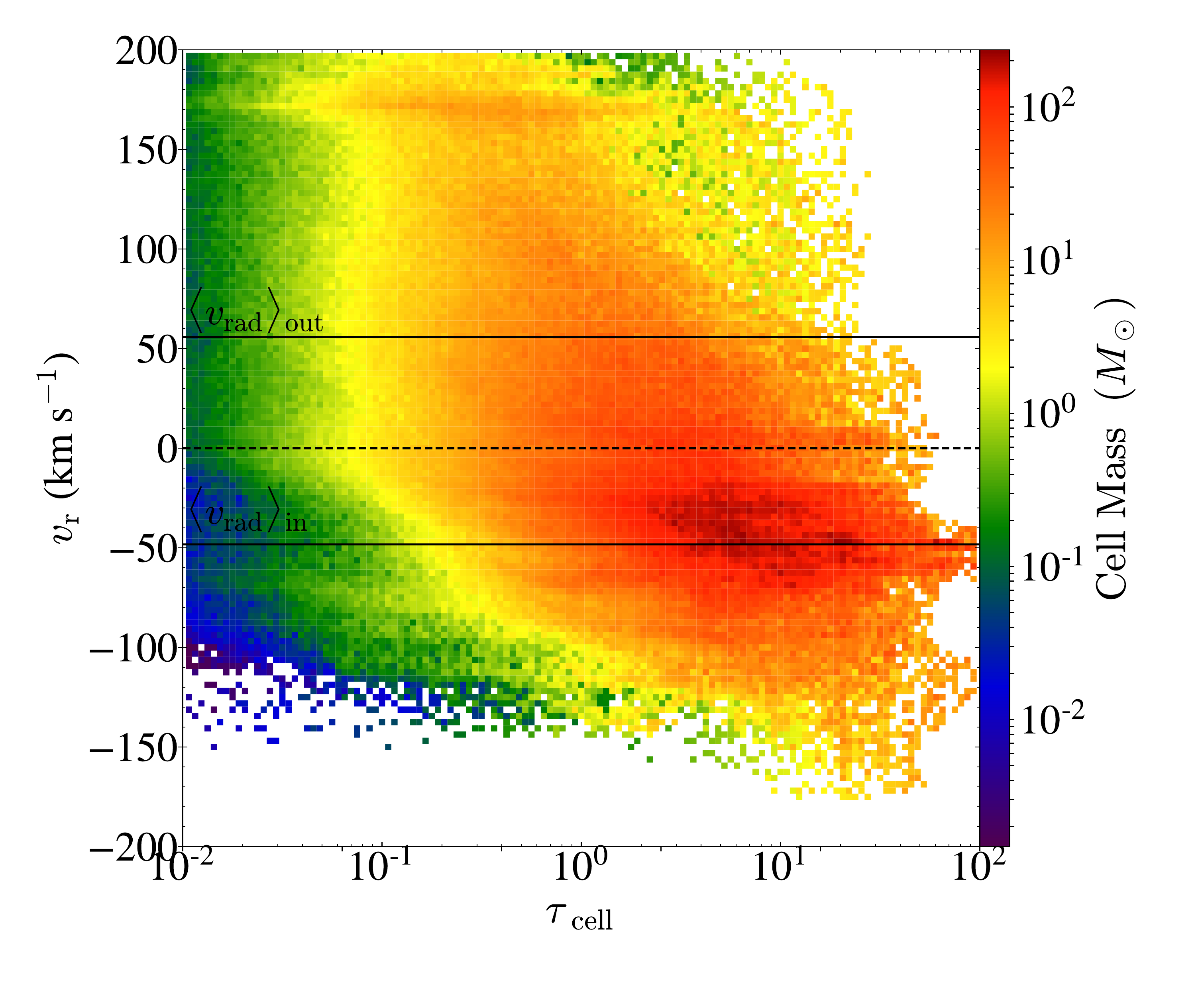}
\includegraphics[width=0.33\textwidth]{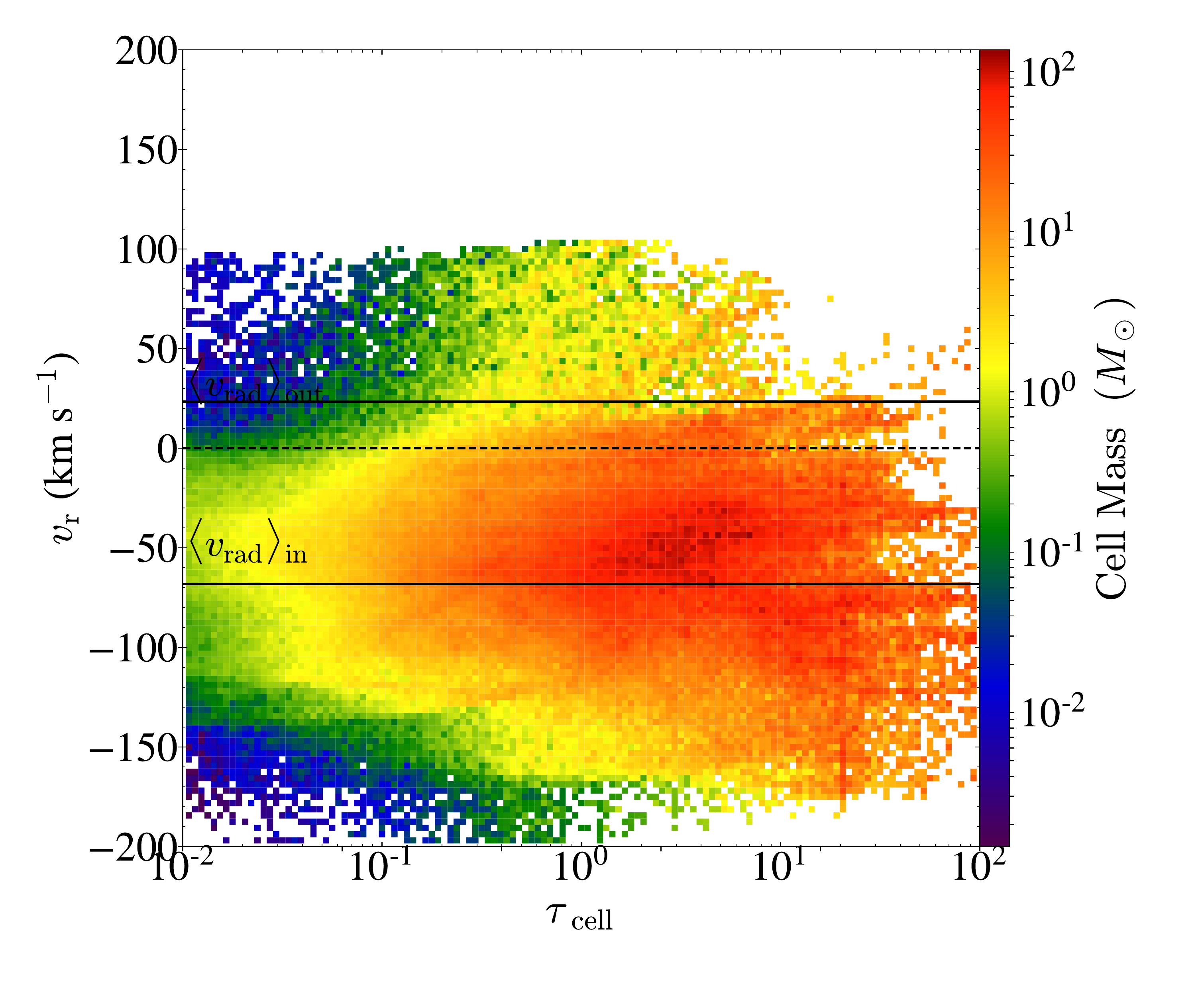}
\end{center}
\caption{The mass-weighted Eddington ratio (left) and total cell
    mass (middle and right) shown on the cell optical depth $\tau_{\rm cell}$--radial velocity $v_{\rm rad}$ plane, all at $t = t_{\rm ff}$.
    The right panel is taken from the HD run.
    Inhomogeneity-induced matter-radiation anti-correlation is clearly
    visible on the left and middle panel.
    Super-Eddington cells accelerated by radiation to high velocities
    have low optical depths $\tau_{\rm cell}$. Most of the gas mass is
    undergoing strong gravitational infall subject to weak radiative acceleration.}
\label{fig:phaseplot_tile}
\end{figure*}

For radiation pressure to temper star formation, the radiative acceleration not only
has to exceed the local gravitational
acceleration, but also has to cancel the inertia of the already infalling gas.
The left and middle panels of Figure \ref{fig:phaseplot_tile}
show the Eddington ratio $f_{\rm Edd, cell}$
and gas mass as a function of $\tau_{\rm cell}$ and radial streaming elocity $v_{\rm rad}$
for cells within 2\,pc from the most massive cluster center at $t = t_{\rm ff}$.
The radiation-matter anti-correlation is clear---almost all of the super-Eddington cells are optically thin $\tau_{\rm cell} < 1$
(top left of left panel),
and most of the optically thick gas mass with $\tau_{\rm cell} > 1$ is sub-Eddington.
In addition, a fraction of optically thin cells is being radiatively accelerated in an outflow.
The mass-weighted average radial velocity in the outflowing and inflowing cells is, respectively, 56\,km\,s$^{-1}$ and $-49$\,km\,s$^{-1}$.
The escape velocity from the cluster radius
is 90\,km\,s$^{-1}$ and thus, on average, \emph{the outflowing gas remains gravitationally bound to the cluster}.
In the right panel, we show the corresponding cell mass plot for
the radiation-free HD run.
As expected, the optically-thin, outflowing
component is absent. The bulk is in steady
gravitational infall with a slightly higher mass-weighted radial velocity $\sim 70\,\text{km}\,\text{s}^{-1}$.

\begin{figure}
   \begin{center}
   \includegraphics[width=0.5\textwidth]{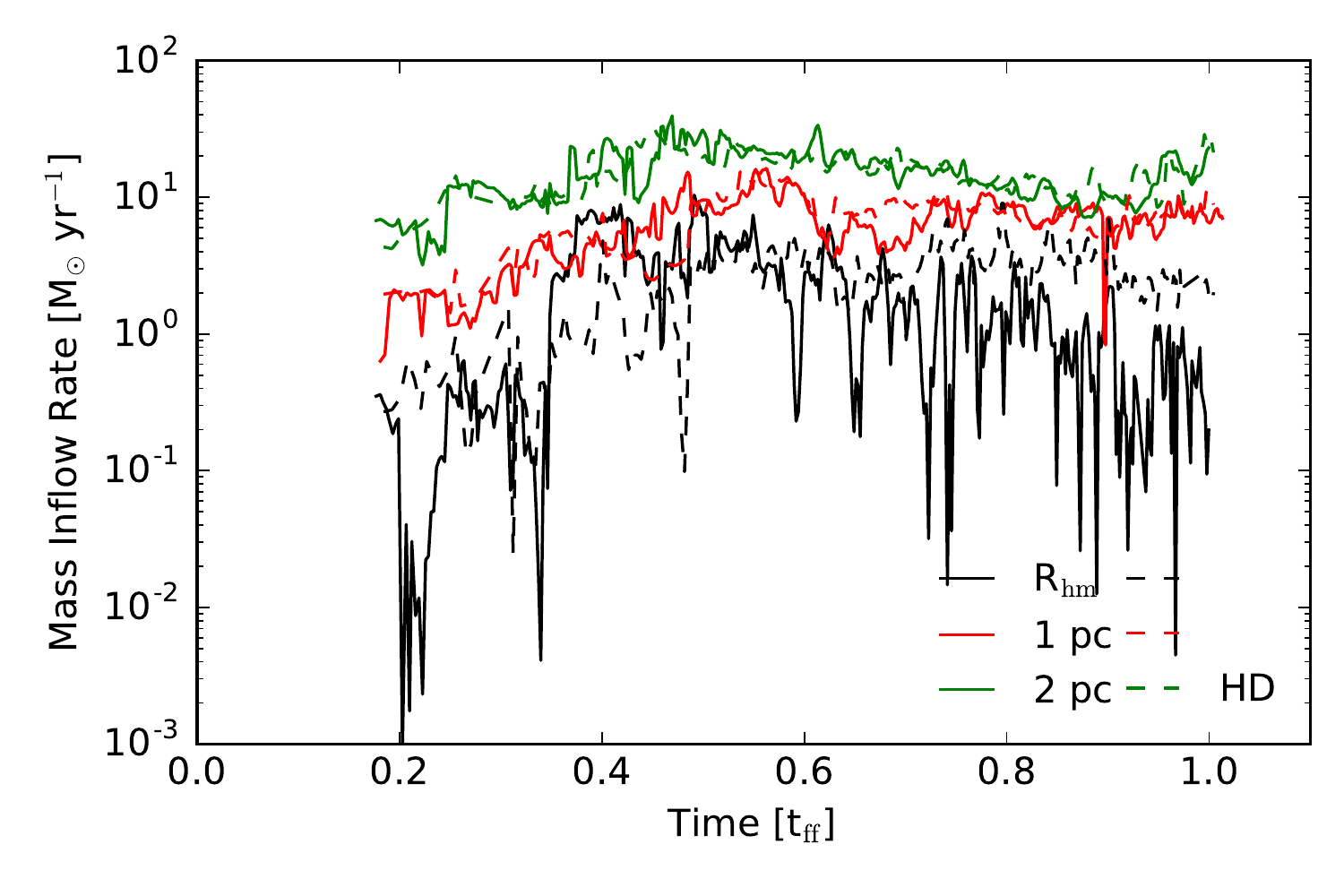}
   \end{center}
   \caption
   { Evolution of the mass accretion rates across spheres of
     different radii around the center of the most massive cluster.
     Solid and dashed lines are for the RHD and HD run, respectively.
    Radiation pressure has minor effect on the overall
     accretion rates.
    }   \label{fig:mdots_surfaces}
\end{figure}

To quantify the mass infall,
Figure \ref{fig:mdots_surfaces} compares mass accretion rates across
spheres centered on the most massive cluster.
We compute the accretion rates by averaging the mass flux in shells
\begin{equation}
  \dot{M}_{\rm acc}\left(R\right) = \frac{1}{N_{\rm cell}(R)}
                                    \sum_{j}\rho_j v_{{\rm rad},j} 4 \pi R^2,
\end{equation}
where $\rho_{j}$ and $v_{{\rm rad}, j}$ are the density and radial velocity of cell $j$,
 $N_{\rm cell}(R)$ is the number of cells inside the radial shell
$(R-\Delta x_{\rm min}/2,R+\Delta x_{\rm min}/2)$, and the sum is over the cells with centers located in the shell.
At $R= 2\,\text{pc}$, large-scale gravitational collapse channels mass toward the cluster center at high rates $\gtrsim 10\,M_{\odot}\textrm{yr}^{-1}$.
Down to $R\sim\,1\,\text{pc}$, the two runs have comparable accretion rates. At the cluster's half-mass radius the accretion rate is only slightly lower in
the RHD run.
These suggest that \emph{the overall dynamical influence of radiation pressure on cluster-scale gaseous collapse is minor}.

\begin{figure*}
   \begin{center}
   \includegraphics[width=0.45\textwidth]{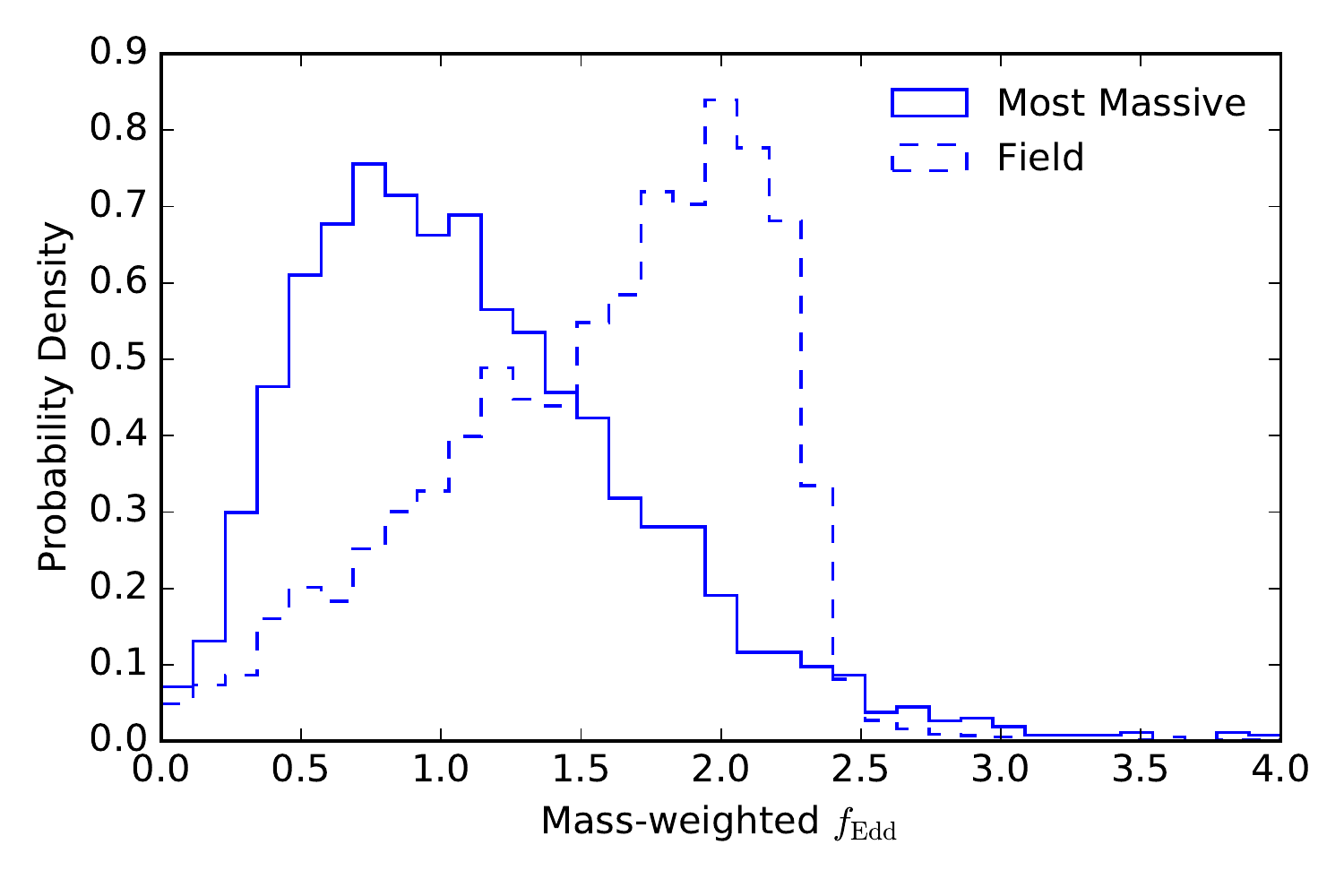}
      \includegraphics[width=0.45\textwidth]{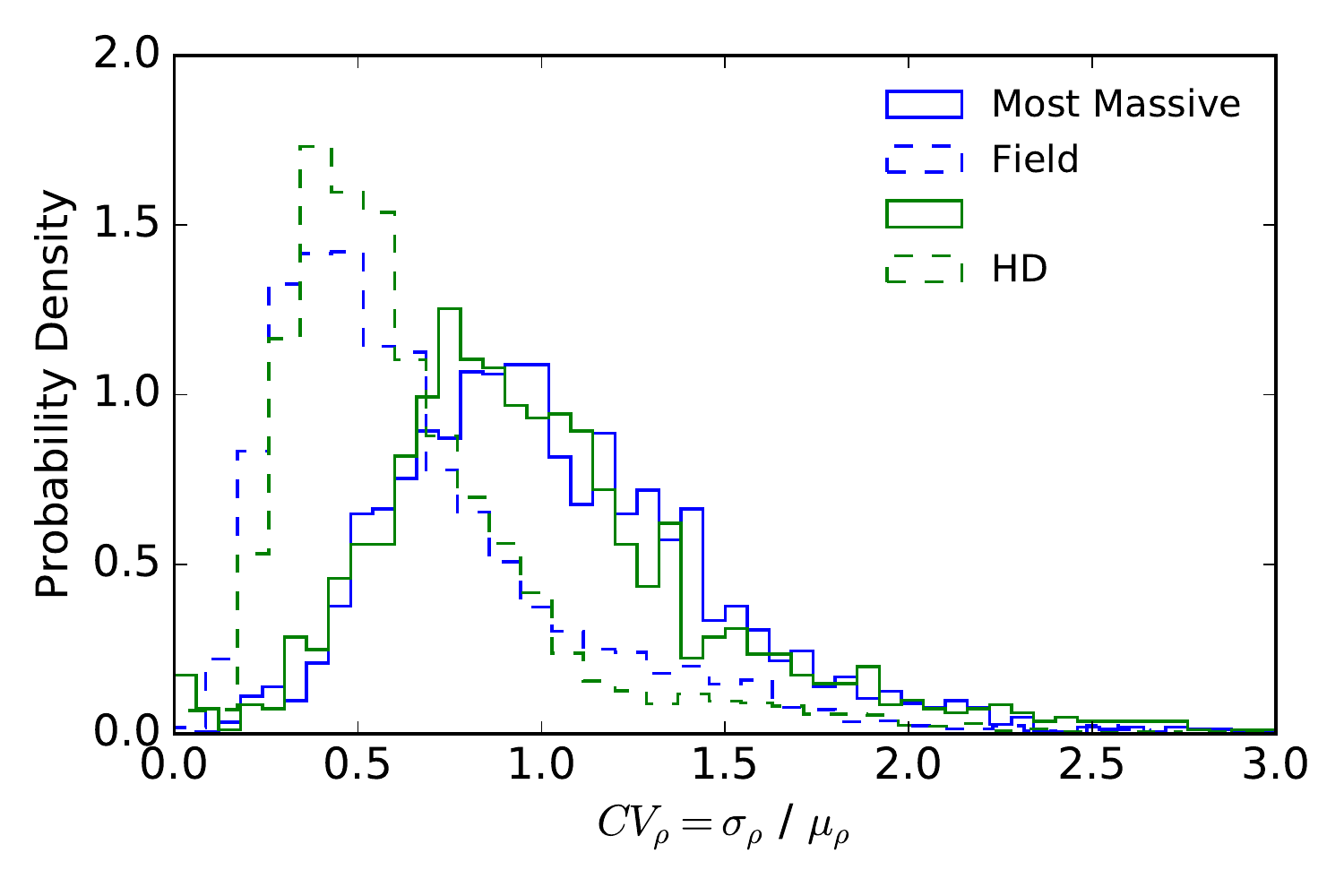}
   \end{center}
   \caption
   { Left panel: Probability distribution functions of the mass-weighted Eddington ratio
     for isolated sink particles (dashed) and those contained in the most massive
     cluster (solid). To have reliable statistics, the histograms combine data
     from 100 time steps before $t= 0.5$\,$t_{\rm ff}$,
     which span a duration of about 0.2\% $t_{\rm ff}$.
     It is much more likely for sinks to be under super-Eddington
     conditions in the field, outside any star clusters.
     Right panel: Probability density of the $\text{CV}_{\rho}$ for sinks in clusters (solid)
     and in the field (dashed).
     Sinks in the field, outside any star clusters, have more homogeneous
     local density distribution.
     The level of non-uniformity is independent of the presence of radiation suggesting that the density variation is driven by turbulent collapse.
   % Sinks in the field, outside any star clusters, have a similar level of
   % non-uniformity with and without radiation.
   % Inside the most massive cluster, however, the presence of radiation pressure
   % significantly perturb the gas distribution, suggesting that the
   % non-uniformity of gas is driven mainly by radiation.
    }   \label{fig:fEdd_hist}
\end{figure*}

Turbulent gas is organized in a network of sheets, filaments, and clumps.
These structures seed the formation and merging of star clusters with a wide range of masses.
Could the effects of radiation be more pronounced in more isolated regions that are forming lower-mass clusters?  More homogeneous gas in shallower gravitational potential wells may be more susceptible to radiation pressure.
The differential effects of radiation pressure inside and outside the most massive
star cluster can be observed in the left panel of Figure \ref{fig:fEdd_hist}.
It compares histograms of the mass-weighted average Eddington ratio
around sink particles inside the most massive cluster (solid)
and in the field, both at $t = 0.5\,t_{\rm ff}$ when the formation of clusters across the mass spectrum is pervasive.
The average ratio is computed from cells within $r_{\rm acc}$ around each
individual sink.
While 57\% of sinks inside the most massive clusters are super-Eddington $f_{\rm Edd} > 1$, that is the case for
85\% of sinks in the field regions.
Next we show this trend is the result of the greater homogeneity of gas around isolated sinks.

We quantify the non-uniformity of gas around sink particles with
the `coefficient of variation' of gas density $\text{CV}_{\rho} = \sigma_{\rho}/\mu_{\rho}$,
where $\sigma_{\rho}$ and $\mu_{\rho}$ are the standard deviation and
mean of gas density within the same $r_{\rm acc}$ spheres used to
construct the left panel of Figure~\ref{fig:fEdd_hist}.
Such a local measure of density variation is appropriate because the amount of gas
around individual sink is different.
Alternately, we could define a coefficient of variation using the line-of-sight gas surface
density by tracing rays in various directions.
We computed the ray-traced coefficient and found that it correlates tightly
with the density-based value.

We plot the distribution of $\text{CV}_{\rho}$ in the right panel of
Figure~\ref{fig:fEdd_hist}.
We see that the density variation both around field sinks and inside the most massive
cluster is very similar between the runs, and that
the density variation in the cluster is significantly higher
compared to in the field.
%Also, density variation inside the most massive cluster in the HD run is similar
%to the field density variation.
Since we expect density variation to be anti-correlated with the strength
of the radiation-matter coupling, this explains the lower peak $f_{\rm Edd}$
in the cluster than in the field in the left panel of Figure~\ref{fig:fEdd_hist}.
%But the result also suggests that \emph{radiation itself amplifies density variation inside the cluster}.
The result also suggests that
\emph{density variation inside the cluster is mainly set by the turbulent collapse, not radiation}.
In the field where sinks are only beginning to form, gravity
has not yet had the opportunity to disturb the local gas.
Inside clusters, the hierarchical collapse has had ample time to
induce inhomogeneity. The weakening of effective radiation pressure is therefore
the result of such turbulent gas rearrangement.
% induce inhomogeneities. The weakening of effective radiation pressure is therefore
% the result of such turbulent gas rearrangement.

\section{Prior art and observational implications}
\label{sec:prior_art}
We simulated a setup similar to that of \citet{SO15} but with a different
radiation transport scheme and slightly different parameters.
We reaffirm their main conclusion, that an anti-correlation between radiation
flux and gas density reduces the forcing of IR radiation pressure on
star forming gas.
Our simulations, however, sample a higher total gas mass and surface
density regime, one in which it may naively be expected that the radiation
would more easily couple to the gas.
Our simulations are initialized not from a stochastic velocity field,
but are prepared via stochastic forcing.
Our total gas mass and the mean surface and volume densities are factors of
10, $\sim5$, and $\sim2.7$ higher, respectively, than in the fiducial model of SO15.
They performed radiation transport with the M1 closure, a method that does not
correctly capture the directional structure of the radiation field close to
point sources.
Therefore they were forced to smear the point source emissivities over 1 pc
spheres. Also, we perform direct gravitational force summation for sink
particles, whereas SO15 spatially-smeared sink particles, with an
effective softening length of 0.5\,pc, onto the finite volume mesh,
and solved the Poisson subsystem for the combined gas and particle density.
They also used a more conservative threshold density for sink particle
creation.
Their simulations therefore formed substantially fewer sink particles.
From an initial gas mass of $10^{6}\,M_{\odot}$, the cluster in the fiducial model of SO15 attained a final mass of
$6\times10^{5}\,M_{\odot}$ (SFE $\sim$60\%) by
$t\sim3\,t_{\rm ff}$.
In our simulation, where the turbulent density field reached a statistical steady state by the time gravity was
switched on, the stellar mass was much larger $4.6\times10^{6}\,M_{\odot}$
(SFE $\sim46$\%) already at $t\sim1\,t_{\rm ff}$ and was continuously increasing.

% Revised.
\citet{ROS16} used the same methods as SO15 but focused instead on the direct UV non-ionizing radiation pressure.  They simulated the formation of star clusters at
lower cloud masses $\sim10^{5}\,M_{\odot}$
and densities $\sim100\,M_{\odot}\,\textrm{pc}^{-2}$.
As in the higher mass regime, they found that the presence of radiation pressure
does limit the star formation efficiency.
Furthermore, they provided an analytical
upper limit on the final star formation efficiency based on the successive
dispersal of high-surface density gas patches by the intensifying
stellar radiation from the growing clusters.
\citet{Myers14} prepared the initial conditions by driving turbulence similar to how we did, with stochastic forcing, but their forcing was with the pure solenoidal mode and the mass of the cluster-forming cloud was much lower,
$10^{3}$\,$M_{\odot}$.  \citet{Li17} extended \citet{Li15}
to simulate cluster formation in magnetized
clouds and with persistent stochastic forcing.
They modeled both radiative and outflow feedback and were able to reproduce the
local SFE of $\sim4$\%.

In contrast with these multidimensional simulations of massive cluster formation
including ours, analytic and semi-analytic models assuming artificial symmetries
have typically claimed radiation pressure was dominant and deleterious to star formation
\citep{Fall10,MQT10,Thompson15,Kim16}.  It seems clear that these divergent conclusions
can be traced to the multi-dimensional nature of the turbulent clouds.
Allowing for the matter-radiation anti-correlation that is ubiquitous in the cited multidimensional simulations and ours, the relative
strengths of various stellar feedback processes---radiation and ionized gas pressure, stellar wind momentum, supernovae, etc.---becomes undetermined in the presently available models. Not only is the forcing of gas by radiation pressure weaker without symmetry, but the gravitational confinement of ionized gas throws into question the thermalization efficiency of stellar winds and supernovae: the winds and ejecta colliding with dense ionized gas might cool and condense on a dynamical time.

Other simulations of feedback in massive cluster formation can be found in the literature.
\citet{Vazquez-Semadeni16} simulated hierarchical collapse of
filamentary structures in the collision of clouds with a total mass $\sim10^{5}$\,$M_{\odot}$ and
studied feedback from photoionization heating.
\citet{Howard16,Howard17} included both photoionization heating and UV direct radiation pressure to simulate cluster formation in a
molecular cloud with mass $10^{6}\,M_{\odot}$, close to ours.
They adopted a large initial cloud radius, 33.8\,pc, and the mean density was a factor $\sim$100 lower than ours, and
 found that feedback was ineffective.
\citet{Dale14} included ionization heating and stellar winds and found that
the dynamical impact from winds was very modest compared to photoionization.
Their highest-density model had a mean
density still a factor 4.2 below ours.
 \citet{Gavagnin17} focused on ionization feedback and followed
the collapse of a turbulent spherical cloud with mean density $\sim$14 times
lower than ours. Their accurate direct gravitational force summation
enabled them to study dynamical evaporation from the cluster.
They found that radiative feedback limited the stellar density
and this attenuated dynamical evaporation.
In their simulations, the gravitational softening radius was much smaller than
ours, at 0.5\,$\Delta x_{\rm min}$ compared to our 2.5\,$\Delta x_{\rm min}$.
The details of force softening can critically influence dynamical evolution and mass segregation \citep{PDE15}.

Observational estimates of the masses of SSCs outside the Local Group typically rely on comparing
photometric colors and luminosities with spectral synthesis models
\citep{Whitmore10,Westmoquette14}. SSCs were found to be ubiquitous in starburst and interacting galaxies.
However, such estimates are subject to uncertainties in the assumed IMFs and
the specific stellar evolutionary tracks.
Within the Local Group, spectroscopic measurement of stellar velocity dispersions
in SSCs provides a direct constraint on the cluster mass via the virial theorem.
For example, in the Antennae, \citet{Mengel02} measured velocity dispersions of
$\sim10-20\,\textrm{km}\,\textrm{s}^{-1}$ towards five clusters with sizes of 3.6--6\,pc
from which they deduced cluster masses of $6.5\times10^{5}\,M_{\odot}$--$4.7\times10^{6}\,M_{\odot}$.
\citet{Larsen04} studied a young cluster in the nearby spiral galaxy NGC 6946.
A velocity dispersion of $8.8\,\textrm{km}\,\textrm{s}^{-1}$ and a radius of 10.2\,pc
pointed to a virial mass of $1.7\times10^{6}\,M_{\odot}$.
In the nuclear regions of M82, \citet{McCrady05,McCrady07} found that the range of velocity
dispersion  in a dozen of SSCs implied a cluster mass range of
$2\times10^{5}\,M_{\odot}$--$4\times10^{6}\,M_{\odot}$.
The final masses of the most massive clusters in our simulations are close to the upper end
of the observed range, however, the stellar velocity dispersion
($73\,\textrm{km}\,\textrm{s}^{-1}$) is considerably higher.
These estimates are subject to the dependence on the internal mass distribution
in the cluster that is poorly constrained in observations.

The age of our clusters 0.3\,Myr is much smaller than the typical ages
(6--10\,Myr) of the systems observed in the optical and
near-IR, yet by 6\,Myr, the most massive and earliest to form stars are already lost to supernovae.
With high-angular-resolution instruments such as ALMA, recently we are able to
probe the massive cluster forming gas.
\citet{Johnson15} and \citet{Leroy15} observed the star-forming clouds in the
Antennae and NGC 253, respectively, and identified clouds with
masses $10^{7}\,M_{\odot}$ and length scales of 30\,pc.
The clouds in NGC253 were further observed to have high line widths of
$20-40\,\text{km}\,\text{s}^{-1}$.
For direct comparison with observations, in Figure \ref{fig:velocities_time} we show the one-dimensional velocity dispersion
$\langle v_{\rm rms} \rangle$ in the RHD run. After $\sim0.6\,t_{\rm ff}$, the velocity dispersion becomes $\sim 50\,\text{km}\,\text{s}^{-1}$.
In comparison, \citet{Turner17} measured the CO molecular line width in the super-star-cluster-forming
Cloud D1 in
the dwarf galaxy NGC 5253 to be $21.7\pm0.5\,\text{km}\,\text{s}^{-1}$. The dispersion implies a virial mass of $2.5\times10^{5}\,M_{\odot}$
at the reference radius of 2.8\,pc.
When our most massive cluster grows to this mass (at $\sim$0.4\,$t_{\rm ff}$),
its cluster radius is 3.5\,pc, and the one-dimensinoal velocity
dispersion is 20.6\,km\,s$^{-1}$, both consistent with Cloud D1.
Similar to our simulations,
\citet{Turner17} also found that the molecular gas only constituted a
small fraction of the global dynamical mass.
These suggest that Cloud D1 could indeed be a growing massive cluster.
The gas dynamics observed in our simulation supports their
proposal that the star-forming molecular clumps orbit in a deep gravitational
potential containing a high-mass stellar component.
Furthermore, the light-to-mass ratio of Cloud D1 was estimated to be
$\sim4000$\,erg\,s$^{-1}$\,g$^{-1}$, providing observational motivation for
our optimistic choice.

\section{Conclusions}
\label{sec:conclusions}

We have extended the IMC radiation transport scheme
into a hybrid IMC-DDMC scheme that is efficient enough to permit simulating the radiation-hydrodynamics of super star cluster formation in realistic turbulent molecular clouds.
We drove turbulence in $10^{7}\,M_{\odot}$ of gas and then
followed gravitational collapse and subgrid star formation. The star particles organized into hierarchically merging clusters to produce a final SSC with mass $\gtrsim 10^{6}\,M_{\odot}$ and radius
$\sim1\,\textrm{pc}$.
Our direct gravitational force summation in combination with the accurate hybrid
radiation transport scheme we implemented allowed us to resolve gas virialization and radiative forcing in the very nucleus of the SSC.

In the course of the 0.3\,Myr simulation we find that radiation pressure reduced the simulation-wide star formation efficiency by 30-35\% and star formation rate by 15-50\% , both relative to a radiation-free run.
In contrast to previous analytical arguments that hinged on idealized geometries, we found that
radiation pressure did not truncate gas infall or star formation.
At the end of the simulation massive clusters continued growing at rates $\sim1\,M_{\odot}\,\textrm{yr}^{-1}$.

Similar to \citet{SO15}, we attribute the  ineffectiveness of radiation pressure
relative to 1D models to a radiation-matter anti-correlation in the
turbulent cluster-forming gas.
The turbulent infall enhances the density inhomogeneity in and around clusters.
The gas distribution is more homogeneous around the stellar sources far from massive clusters and there the radiation accelerates gas more effectively.

The simulation produced a final peak stellar density of $\sim10^{8}\,M_{\odot}\,\textrm{pc}^{-3}$
in an SSC with a one-dimensional stellar velocity dispersion exceeding $70$\,km\,s$^{-1}$.  These conditions favor stellar collisions and runaway merging that can produce a very massive star.
We speculate that this very massive star accretes clouds condensing from the ionized gas that is
gravitationally-confined in the SSC potential.  If the very massive star avoids severe mass loss, such as it might embedded in the dense gaseous environment of the SSC nucleus, it can collapse directly into a massive black hole.

The hybrid IMC-DDMC particle-based radiation transport scheme is distinct
from but competitive with the most advanced existing moment- or characteristics-based methods.
The DDMC component of the hybrid scheme accelerates transport at high optical depths and interfaces seamlessly with the IMC scheme at low optical depths.
Our implementation exploits the native parallel infrastructure of the
\textsc{flash} hydrodynamic code to deliver excellent speedups.

\section*{Acknowledgments}
%We are grateful to the referee for very helpful comments
We thank Tiago Costa, Melvyn Davies, Jeong-Gyu Kim, Lucio Mayer, and Eve Ostriker for invaluable discussions
and insightful exchanges.
B.~T. thanks Volker Bromm for his encouragements and support during the course
of the study,
and Aaron Smith for many helpful discussions and comments throughout.
The \textsc{flash} code used in this work was developed in part by the DOE NNSA-ASC
OASCR Flash Center at the University of Chicago.
Most the data analysis was performed with of the publicly available code
{\tt yt} \citep{Turk11} and the open source Python ecosystem including
{\tt Numpy} and {\tt Scipy}.
The cluster-identifying algorithm {\tt DBSCAN} adopted in the study is
part of the Python library {\tt scikit-learn}.
We acknowledge the Texas Advanced Computing Center at
The University of Texas at Austin for providing HPC resources.
Computations were performed on the Lonestar 5 and Stampede supercomputers. This study was supported by the NSF grant AST-1413501. M.~M. thanks the Kavli Institute for Theoretical Physics where this research was supported in part by the National Science Foundation under Grant No. NSF PHY-1125915.

\markboth{Bibliography}{Bibliography}

\bibliographystyle{mnras}
\bibliography{paper}

\clearpage

\end{document}